\documentclass[fleqn, usenatbib]{mnras}
\pdfoutput=1

\usepackage{newtxtext,newtxmath}
\usepackage{amsmath}
\usepackage{xcolor}

\usepackage[T1]{fontenc}

\DeclareRobustCommand{\VAN}[3]{#2}
\let\VANthebibliography\thebibliography
\def\thebibliography{\DeclareRobustCommand{\VAN}[3]{##3}\VANthebibliography}

\usepackage{graphicx}	
\usepackage{lscape}
\usepackage{amsmath}	
\usepackage{bm}
\usepackage{float}

\title[DSFGs with Euclid]{{\it Euclid} view of the dusty star-forming galaxies at $z\gtrsim1.5$ detected in wide area sub-millimetre surveys}

\author[D. Mitra et al.]{
Dipanjan Mitra$^{1}$\thanks{E-mail: MitraD@cardiff.ac.uk},
Mattia Negrello$^{1}$, Gianfranco De Zotti$^{2}$, Zhen-Yi Cai$^{3}$
\\
$^{1}$School of Physics and Astronomy, Cardiff University, The Parade, CF24 3AA, UK\\
$^{2}$INAF, Osservatorio Astronomico di Padova, Vicolo Osservatorio 5, I-35122
Padova, Italy\\
$^{3}$Department
of Astronomy, University of Science and Technology of China, Hefei 230026, China
}

\date{Accepted XXX. Received YYY; in original form ZZZ}

\pubyear{2015}

\begin{document}
\label{firstpage}
\pagerange{\pageref{firstpage}--\pageref{lastpage}}
\maketitle

\begin{abstract}
We investigate the constraints provided by the 
{\it Euclid} space observatory on the physical properties of dusty star-forming galaxies 
(DSFGs) at $z\gtrsim1.5$ detected in wide area sub-millimetre surveys with {\it Herschel}. We adopt a physical model 
for the high-$z$ progenitors of spheroidal galaxies, which form the bulk of DSFGs at $z\gtrsim1.5$. We improve the model by combining
the output of the equations of the model with a formalism for the spectral energy distribution (SED). 
After optimising the SED parameters to reproduce the measured 
infrared luminosity function and number counts of DFSGs, we simulated a sample of DSFGs over 100\,$\hbox{deg}^2$
and then applied a $5\,\sigma$ detection limit of $37\,$mJy at 250\,$\mu$m.
We estimated the redshifts from the {\it Euclid} data and then fitted the {\it Euclid}+{\it Herschel} photometry
with the code CIGALE to extract the physical parameters. We found that $100\%$ of the \textit{Herschel} galaxies are detected in all 4 \textit{Euclid} bands above $3\,\sigma$. For $87\%$ of 
these sources the accuracy on $1 + z$ is better than 15\%. 
The sample comprises mostly massive, i.e. $\log(M_{\star}/M_{\odot})\sim10.5-12.9$, highly star--forming, i.e. $\log(\hbox{SFR}/M_{\odot}\hbox{yr}^{-1})\sim1.5-4$, dusty, i.e. $\log(M_{\rm dust}/M_{\odot})\sim7.5-9.9$, galaxies.
The measured stellar masses 
have a dispersion of  $0.19$ dex around the true value, thus  
showing that {\it Euclid} will provide reliable stellar mass estimates 
for the majority of the bright DSFGs at $z\gtrsim1.5$ detected by {\it Herschel}. 
We also explored the effect of complementing the \textit{Euclid} photometry with that from the \textit{Vera C. Rubin Observatory/LSST}.

\end{abstract}




\section{Introduction}

In the 1990s, surveys using the Submillimeter Common User Bolometer Array (SCUBA) at $850\,\mu$m on the 15\,m \textit{James Clerk Maxwell Telescope} \citep[JCMT;][]{smail_deep_1997}, for the first time detected a population of high-redshift ($z\gtrsim1.5$) galaxies which are extremely luminous at far-infrared (FIR) and submillimeter (sub-mm) wavelengths and are highly dust-obscured at optical wavelengths. These are now commonly referred to as dusty star-forming galaxies (DSFGs) or sub-millimeter galaxies (SMGs) and are the dominant star-forming galaxies at $z\gtrsim1.5$ \citep{casey_population_2012,bothwell_survey_2013,cai_hybrid_2013} and most likely to be in the early evolutionary phase of present-day elliptical galaxies \citep{ivison_herschel-atlas_2013,fu_rapid_2013,dokkum_forming_2015,simpson_scuba-2_2017}. These galaxies undergo intense star formation with star formation rates (SFRs) from a few hundreds to  thousands solar masses per year \citep{blain_submillimeter_2002,magnelli_herschel_2012,riechers_dust-obscured_2013,swinbank_alma_2014,mackenzie_scuba-2_2017,michalowski_scuba-2_2017,gullberg_alma_2019,castillo_vla_2023}. During the cosmic noon (at $z\sim 2$) \citep{madau_high-redshift_1996,lilly_canada-france_1996} when the star formation rate density (SFRD) of the Universe peaked, these galaxies contributed heavily to the overall galaxy population \citep{magnelli_evolution_2011-1,burgarella_herschel_2013,rowan-robinson_star_2016}. Therefore, the discovery of these galaxies caused a big revolution in the field of extragalactic astronomy making their study vital for a detailed understanding of the formation and evolution of galaxies which led to the development of more sensitive instruments like the SCUBA-2 \citep{holland_scuba-2_2013}, the Large Apex BOlometer CAmera \citep[LABOCA;][]{siringo_large_2009},the  AZtronomical Thermal Emission Camera \citep[AzTEC;][]{wilson_aztec_2008} and the \textit{Herschel Space Observatory} \citep[\textit{Herschel};][]{pilbratt_herschel_2010}.\\

Our understanding of the DSFGs has increased drastically thanks to \textit{Herschel} which, during its four years of operation, from 2009-2013, has mapped $\sim 1300$ sq. deg. of the sky at wavelengths in the range 100--$500\,\mu$m leading to the detection of more than a million FIR and sub-mm bright galaxies. In particular, the \textit{Herschel} Astrophysical  Terahertz Large Area Survey \citep[H-ATLAS;][]{eales_herschel_2010} and the \textit{Herschel} Multi-tiered Extragalactic Survey \citep[HerMES;][]{oliver_herschel_2012} together surveyed over $\sim 930$ sq. deg. of the sky at 250, 350 and $500\,\mu$m. Despite such surveys, our understanding of the underlying physical processes responsible for the evolution of galaxies is far from complete, but it is speculated that there will be a clearer understanding of these processes in the next decade with big projects like the \textit{Euclid} space observatory \citep{laureijs_euclid_2011}, which was launched on 1st July 2023, and the \textit{Vera C. Rubin Observatory/LSST} \citep{ivezic_lsst_2019}. The \textit{Euclid} Wide Survey \citep[EWS;][]{euclid_collaboration_euclid_2022} will observe 15\,000 sq. deg. of the sky at both optical \citep{cropper_vis_2018} and near-infrared \citep[NIR;][]{euclid_collaboration_euclid_2022-1} wavelengths, thus providing crucial constraints on the stellar mass of galaxies. On the other hand, \textit{Rubin's} LSST (Legacy Survey of Space and Time) Survey will observe 18\,000 sq. deg. of the sky in 6 bands - $u$, $g$, $r$, $i$, $z$, and $y$ - having a wavelength coverage of $0.36-1.01$ $\mu$m.

 Many sophisticated phenomenological models have been developed to study the cosmological evolution of galaxies \citep{matteucci_effect_2009, franceschini_galaxy_2010, popesso_effect_2011, bethermin_unified_2012}. These models adopt simple functional forms to describe the luminosity function of multiple galaxy populations, which are assigned different spectral energy distributions (SEDs) and varying
evolutionary properties. Active galactic nuclei (AGNs) are also considered in some of the models. 
In this paper, we adopt instead the physically motivated model by \citet[C13 hereafter]{cai_hybrid_2013} for the formation and evolution of $z\gtrsim1.5$ DSFGs, which links the star formation activity and the growth of a central supermassive black hole in a self-consistent way. 
The C13 model has proved successful in reproducing data on luminosity functions, number counts and redshift distributions over a broad wavelength range, from the mid-IR to the millimetre. We use the model to investigate how {\it Euclid} can detect and study the $z\gtrsim1.5$ proto-spheroidal galaxies discovered by the H-ATLAS survey. For this purpose, we improve on the C13 original model by implementing a more sophisticated formalism for the SED, which allows us to predict the optical/near-IR properties of these galaxies.

 The paper is organised in the following manner. In Section \ref{sec2}, we provide a brief description of the model and theory used in this paper. In Section \ref{sec3}, we present a detailed methodology of catalogue creation, deriving photometric flux densities and estimating photometric redshifts and other physical properties using SED fitting. In Section \ref{secfor}, we discuss the surveys for which we make forecasts. The results obtained are analysed and discussed in Section \ref{sec4} and the final conclusions are summarized in Section \ref{sec5}. Throughout the paper, we adopt a flat $\Lambda$-CDM cosmology with
present-day matter and baryon density (in units of the critical density), $\Omega_{m,0} = 0.3153$ and $\Omega_{b,0} = 0.0493$. We set the value of the Hubble-Lema\^itre constant to $h = H_0/100 = 0.6736$, the slope of the spectrum of primordial density perturbations to $n=0.9649$ and the normalization of the density fluctuations on a scale of $8h^{-1}$\,Mpc to $\sigma_{8} = 0.8111$ \citep{planck_collaboration_planck_2020}.

\section{Adopted model for the proto-spheroidal galaxies}
\label{sec2}
In this section, we provide a brief description of the C13 model for the proto-spheroidal galaxies and illustrate the formalism used for predicting the SED of these galaxies.

\subsection{Model Outline}

\begin{figure*}

\centering
\includegraphics[width=.5\textwidth]{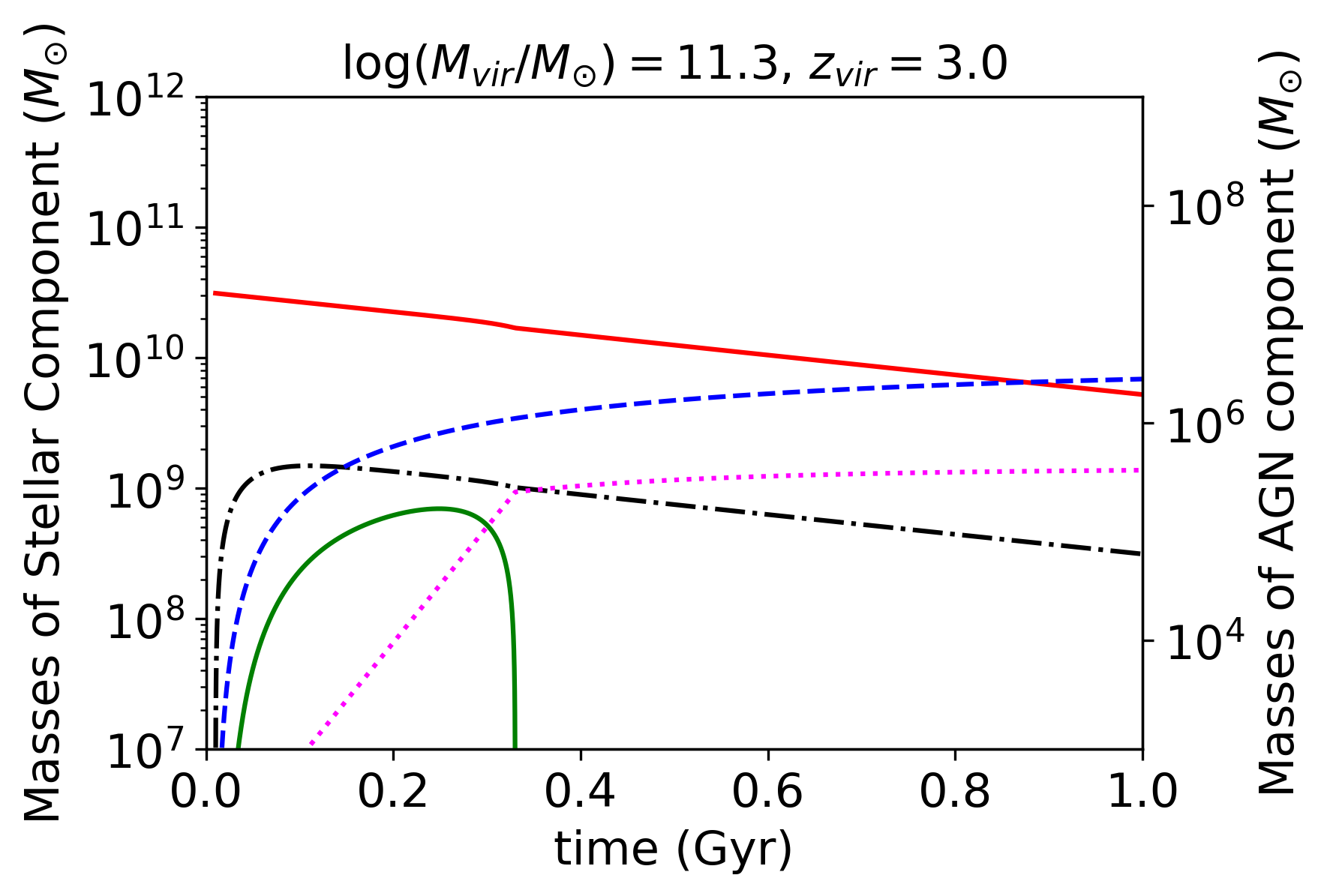}\hfill
\includegraphics[width=.5\textwidth]{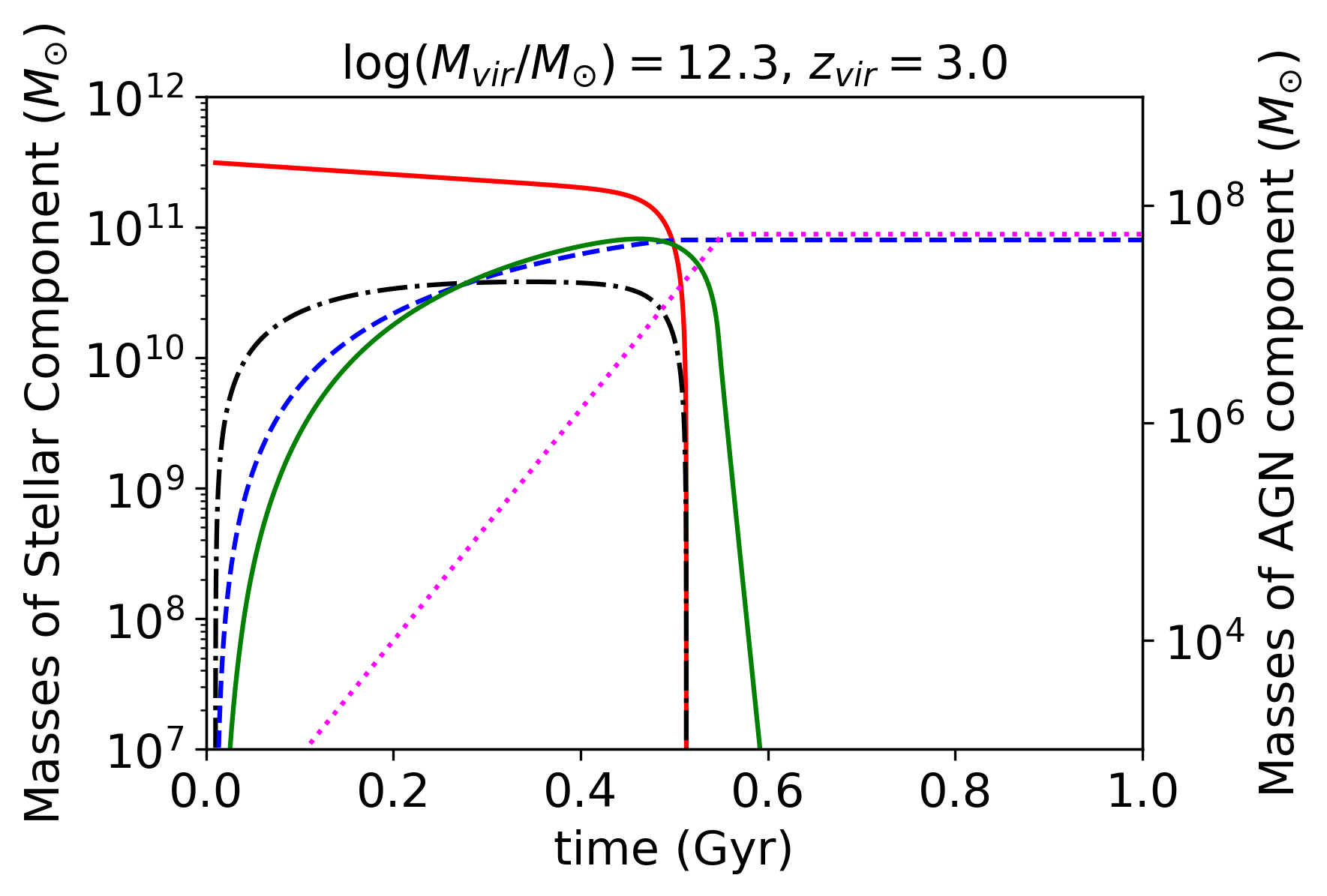}

\caption{Evolution of the properties of the stellar and AGN components of proto-spheroidal galaxies as a function of galactic age for a halo of mass $M_{\rm vir}=10^{11.3}\,M_{\odot}$ (left) and a halo of mass $M_{\rm vir} = 10^{12.3}\,M_{\odot}$ (right). Both have virialisation redshift $z_{\rm vir}=3$. In each plot, the left y-axis refers to the mass evolution of the stellar component: infalling gas (solid red line), cold gas (dot-dashed black line), and stellar mass (blue dashed line); while the right y-axis refers the mass evolution of AGN component: gas falling in the reservoir (solid green line) and super-massive black hole (SMBH) mass (dotted magenta line). We observe that
the stellar mass increases as more and more cold gas is condensed into stars, thus leading to a corresponding steady decline in the mass of the gas components. 
At the same time, the central BH mass increases as a fraction of the cold gas loses its angular momentum via the stellar radiation drag and settles down into a reservoir around the central SMBH. The BH accretes matter from the reservoir via viscous dissipation, which powers the nuclear activity. With the evolution of the galactic age, the AGN feedback sets in. Its effect can be observed in the most massive dark matter (DM) halos where the star formation activity and, consequently, the stellar mass growth, are stopped within less than 1\,Gyr from the galaxy's birth. 
}
\label{figmodelcurve}
\end{figure*}

The model by C13 is based on the fact that spheroids in the local universe (i.e. ellipticals and bulges of disc galaxies) are mostly comprised of old stellar populations that formed at redshifts $z\gtrsim1.5$. On the other hand, disc galaxies consist of comparatively younger stellar populations with luminosity-weighted age $\lesssim 7$ Gyr, indicating a formation redshift $z\lesssim 1$. Therefore, the progenitors of local spheroids (also called proto-spheroidal galaxies or proto-spheroids) are the dominant star-forming galaxies at $z\gtrsim 1.5$, whereas star formation in discs takes place mostly at $z\lesssim 1.5$. The model provides a physically motivated co-evolution of the SFR of the proto-spheroids along with the supermassive black hole at their centres. The star formation history of these proto-spheroids is calculated using a set of equations considering the evolution of gas phases and of the AGN, including cooling, condensation to form stars, accretion into the supermassive black hole, and feedback from both the stellar component and the AGN (see Figure \ref{figmodelcurve}). The model depicts two ways of dark matter halo formation, i.e. a fast collapse of the halo triggering star formation, followed by a slow growth of the halo outskirts, which have little effect on the inner region of the halo. The first phase is the one driving the build-up of most of the stellar mass in these objects.
Both the star formation and the growth of the super-massive black hole at the centre are controlled by feedback mechanisms from the supernovae and the AGN. The AGN feedback is effective in quenching the star formation in the most massive proto-spheroids, while SFR is mainly affected by SN feedback in less massive ones \citep{granato_physical_2004,scannapieco_effects_2008,rosas-guevara_supermassive_2016,rosito_role_2021}. A detailed explanation of the equations governing the evolution of both the stellar and the AGN components is given in Appendix \ref{appendixA}. 

The bolometric luminosity function of the proto-spheroidal galaxies, i.e. the volume density of galaxies per unit interval of luminosity,  is calculated by convolving the halo formation rate function with the galaxy luminosity distribution predicated by the equations of the model, i.e.
\begin{eqnarray}
    \Phi(\log L,z) & = & \int_{M_{\rm vir}^{\rm min}}^{M_{\rm vir}^{\rm max}}\int_{z}^{z_{\rm vir}^{\rm max}}\left|\frac{dt}{dz_{\rm vir}}\right|\frac{dN_{\rm ST}}{dt}  \nonumber \\   
    &~ & P(\log L,z;M_{\rm vir},z_{\rm vir})  dz_{\rm vir}dM_{\rm vir},
\end{eqnarray}
where $N_{\rm ST}$ is the halo mass function, for which the model adopts the \cite{sheth_large-scale_1999} approximation, while $P(\log L,z;M_{\rm vir},z_{\rm vir})$ is the luminosity distribution function of galaxies at redshift $z$ with virialisation mass and redshift given by $M_{\rm vir}$ and $z_{\rm vir}$ respectively. The distribution function is a log-normal distribution given by
\begin{equation}
    P[\log L|\log \bar{L}]d\log L = \frac{e^{-\frac{\log^2\left(L/\bar{L}\right)}{2\sigma^2}}}{\sqrt{2\pi\sigma^2}}d\log L,
\end{equation}
where $\sigma$ is the dispersion around the mean luminosity, $\bar{L}$, which is computed using the formalism illustrated in Sec.~\ref{sedmodel} for the stellar component. The AGN luminosity is computed using Eq.\,(\ref{eqnagn}). The total luminosity of a galaxy is the sum of the stellar luminosity and the AGN luminosity. The dispersion is considered to reflect the uncertainties in the adopted values of the main parameters that are involved in the equations of the model. 



\subsection{SED modelling}
\label{sedmodel}

The SED of a galaxy is an observable that reflects the properties of the stars, dust and gas in the galaxy. As such, it can be used to extract important information. In fact, while the UV/optical/near-IR part of the SED depends on the star formation history, metallicity and dust extinction, the far-IR/sub-mm/mm portion of the SED reveals the star formation activity occurring in the dust-enshrouded birth clouds and can be used to constrain the properties of the dust itself. 

C13 adopted a single representative SED for the whole population of high-$z$ proto-spheroidal galaxies, i.e. the SED of the well-studied $z=2.3$ lensed galaxy SMM\,J2135-0102.
That choice was dictated by the need of speeding up the computation of the relevant statistical properties on the galaxies, such as number counts and redshift distributions. Also, C13 were mainly focused on observables at far-IR/sub-mm/mm wavelengths
where the SED depends on a very limited number of free parameters, e.g. the dust temperature and the dust emissivity index. In fact, with the SED of SMM\,J2135-0102, 
C13 did achieve a good qualitative fit 
to the multi-wavelength (from the mid-IR to millimeter waves) data on number counts, both global and per redshift slices \citep[e.g.][]{bethermin_unified_2012}. Moreover, the model made a successful prediction of the sub-mm \citep{negrello_detection_2010,negrello_herschel-atlas_2017} and mm \citep{vieira_extragalactic_2010,cai_high-z_2020,cai_interpreting_2022} number counts of strongly lensed galaxies.


However, the use of a single SED does limit the applicability of the model to the investigation and the forecast of the statistical properties of dusty proto-spheroidal galaxies at optical/near-IR wavelengths, where the SED is much more complex than at longer wavelengths. Therefore, we have decided to improve the C13 model by implementing an SED modelling that reflects such complexity. The SED formalism is based on that put forward by \cite{da_cunha_simple_2008}, with a few modifications, and it is described in this section.

\subsubsection{Stellar component and dust attenuation}\label{SED_stars_and_attenuation}

 The stellar emission spectrum of the proto-spheroidal galaxies is calculated by coupling the star formation history (SFH, hereafter), as obtained from the equations of the model (Eq.\,(\ref{eqnsfr})) for any given value of $M_{\rm vir}$ and $z_{\rm vir}$, with some single stellar population (SSP) synthesis templates. There are several SSP templates available in literature \citep{bruzual_stellar_2003,jimenez_hipparcos_1998,buzzoni_statistical_1993}. Here we use the SSP templates by \cite{bruzual_stellar_2003}, which are provided at wavelengths from $91$ Å to $160$ $\mu$m and between ages $1\times10^5$\,yr and $2\times10^{10}$\,yr, for metallicities $Z=0.008$, $0.02$ and $0.05$ 
 and three IMFs, i.e. Salpeter, Korupa and Chabrier. We adopt the SSP template corresponding to $Z_{\odot}$ and a Chabrier IMF. Dust attenuation is modelled using the attenuation law by \cite{charlot_simple_2000}. It takes into account attenuation by both birth clouds (BC) and the interstellar medium (ISM), represented by different power laws. This also helps in differentiating between the emission from the young stars still embedded in their dust-enshrouded birth clouds and the old stars located in the ISM, respectively, which is important for IR galaxies \citep{malek_help_2018, buat_dust_2018}.




The luminosity emitted by a galaxy per unit interval of wavelength at time $t$ is given by \citep{charlot_simple_2000}
\begin{equation}
    L_{\lambda}(t) = \int_0^{t}\psi(t-t')S_{\lambda}(t')e^{-\hat{\tau_{\lambda}}(t')}dt'
    \label{eqnattst}
\end{equation}
where $\psi(t-t')$ is the SFR at $(t-t')$, and $S_{\lambda}(t')$ is the luminosity per unit interval of wavelength and per unit mass of an SSP at $t'$. The function $\tau_{\lambda}(t')$ represents the absorption optical depth of dust at the time $t'$ and is defined as
\begin{equation}
\hat{\tau_{\lambda}}(t')=
    \begin{cases}
        \hat{\tau_{\lambda}}^{\rm BC}+\hat{\tau_{\lambda}}^{\rm ISM} & \text{for } t' \leq t_0,\\
        \hat{\tau_{\lambda}}^{\rm ISM} & \text{for } t' > t_0
    \end{cases}
\end{equation}
$t_0$ being the time scale of BC dissipation, while $\hat{\tau_{\lambda}}^{\rm BC}$ and $\hat{\tau_{\lambda}}^{\rm ISM}$ are the absorption optical depths of the dust in the BCs and in the ambient ISM, respectively, modelled as
\begin{equation}
    \hat{\tau_{\lambda}}^{\rm BC} = (1-\mu)\hat{\tau_{V}}\left(\frac{\lambda}{5500\AA}\right)^{-1.3}
\end{equation}
and
\begin{equation}
    \hat{\tau_{\lambda}}^{\rm ISM} = \mu\hat{\tau_{V}}\left(\frac{\lambda}{5500\AA}\right)^{-0.7}.
\end{equation}
Here, $\hat{\tau_{V}}$ is the total effective V-band absorption optical depth of the dust in the BCs and $\mu = \frac{\hat{\tau_{V}}^{\rm ISM}}{\hat{\tau_{V}}^{\rm BC}+\hat{\tau_{V}}^{\rm ISM}}$ is the fraction contributed by the dust in the ISM. 

\begin{figure}

\centering
\includegraphics[width=.45\textwidth]{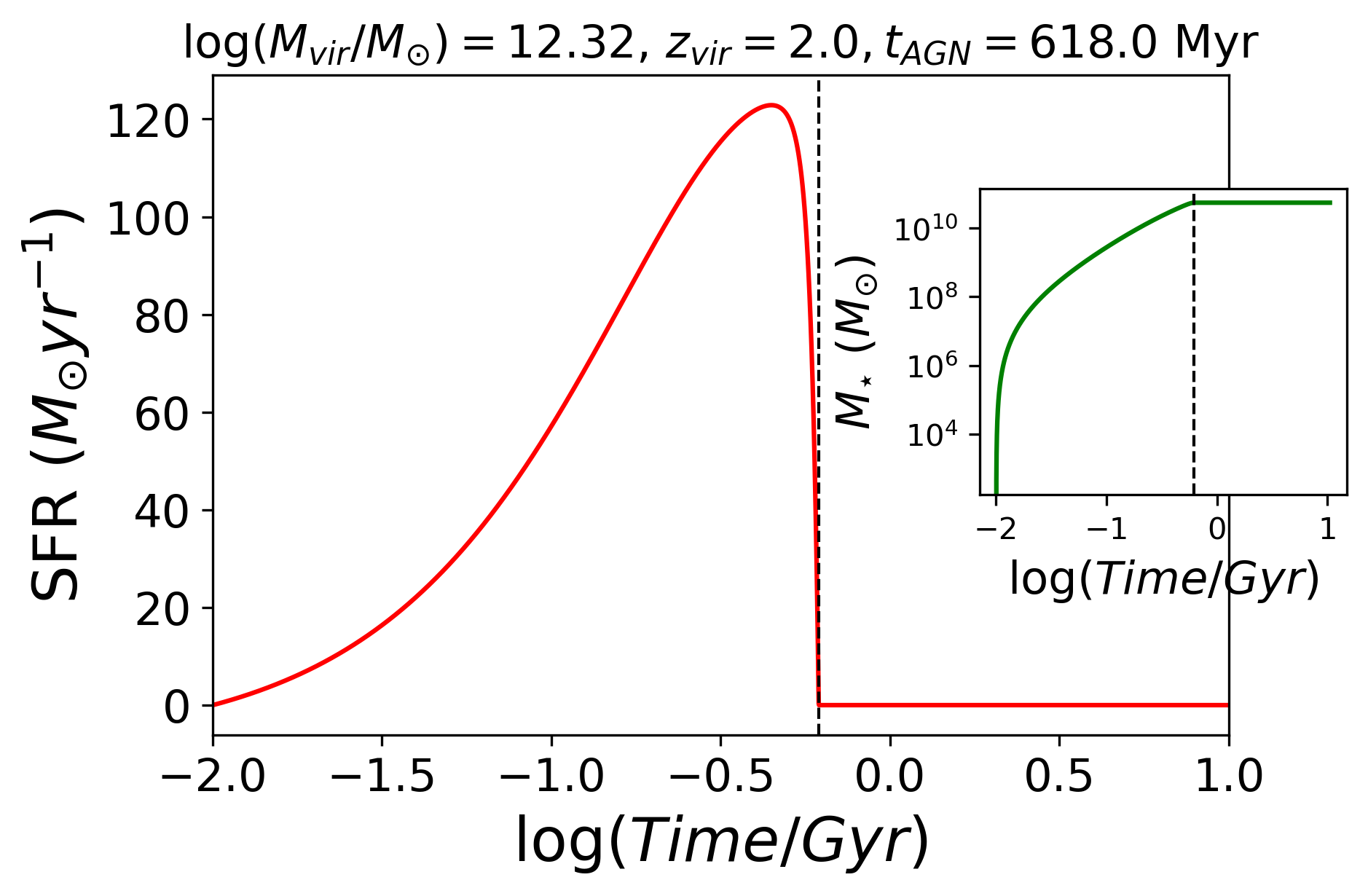}

\caption{SFR as a function of the galaxy age in an object forming in a halo of mass $M_{\rm vir}=10^{12.32}\,M_{\odot}$ virialised at redshift $z_{\rm vir}=2$. The inset plot shows the evolution of the stellar mass with galaxy age. The black dashed line marks the time ($t_{\rm AGN}$) when AGN feedback quenches SF in this galaxy and the SFR drops to zero. For this galaxy, $t_{\rm AGN}=618.0$ Myr. As there is no more SF going on, the stellar mass remains constant afterwards as illustrated by the stellar mass vs. time plot.}
\label{figagnquench}

\end{figure}

Taking into consideration the AGN feedback which quenches star formation (see Figure \ref{figagnquench}), we modify and re-define $\tau_{\lambda}(t')$ as

%
%

%
\begin{equation}
\hat{\tau_{\lambda}}(t')=
    \begin{cases}
        \hat{\tau_{\lambda}}^{\rm BC}+\hat{\tau_{\lambda}}^{\rm ISM} & \text{for } t' \leq \min(t_0, t_{\rm AGN}),\\
        \hat{\tau_{\lambda}}^{\rm ISM} & \text{for } t' > \min(t_0, t_{\rm AGN}),\\
        0 & \text{for } t' > t_{\rm AGN}
    \end{cases}
\end{equation}
where $t_{\rm AGN}$ is the time at which the AGN feedback has completely stopped the star formation. 
In practice, when this happens, 
all the gas and dust in the galaxy have been removed by the AGN feedback so that  $\hat{\tau_{\lambda}}(t')=0$. \\


The amount of starlight absorbed by dust present in the BCs and in the ISM is reprocessed and re-radiated at infrared wavelengths. The dust luminosity is thus given by
\begin{equation}
    L^{\rm dust}_{\rm total}(t) = L^{\rm dust}_{\rm BC}(t)+L^{\rm dust}_{\rm ISM}(t)
    \label{dustlum}
\end{equation}
where
\begin{equation}
    L^{\rm dust}_{\rm BC}(t) = \int_{0}^{\infty}(1-e^{-\hat{\tau_{\lambda}}^{\rm BC}})d\lambda\int_{0}^{t_0}\psi(t-t')S_{\lambda}(t')dt'
    \label{dustbc}
\end{equation}
and
\begin{equation}
    L^{\rm dust}_{\rm ISM}(t) = \int_{0}^{\infty}(1-e^{-\hat{\tau_{\lambda}}^{\rm ISM}})d\lambda\int_{t_0}^{t}\psi(t-t')S_{\lambda}(t')dt'
    \label{dustism}
\end{equation}
denote the amount of starlight absorbed and re-radiated by dust in the BCs and in the ISM, respectively.
For later use in the analysis, we also define the fraction of IR luminosity coming from the dust in the ISM as
\begin{equation}
    f_{\mu} = \frac{L^{\rm dust}_{\rm ISM}(t)}{L^{\rm dust}_{\rm total}(t)},
\end{equation}
Clearly, this fraction depends on $\hat{\tau_{V}}$, $\mu$ and the SFH.

\subsubsection{Infrared emission by dust}\label{SED_dust}

Here, we briefly describe the way in which we compute the infrared part of the SED, which is associated with the dust emission.
We assume three main dust components: Polycyclic Aromatic Hydrocarbons (PAHs), which produce strong emission in the near-infrared (NIR) at wavelengths of about $3-20$ $\mu$m, small dust grains (size $<0.01$ $\mu$m), which are responsible for the hot mid-infrared (MIR) emission, and large dust grains (size $\sim 0.01-0.25$ $\mu$m) which are in thermal equilibrium with the radiation field and produce the colder FIR emission.
\begin{enumerate}
    \item \textit{PAH emission:} in star-forming galaxies, strong emission features at $3.3$, $6.2$, $7.7$, $8.6$, $11.3$ and $12.7$ $\mu$m are observed. They are attributed to emission from PAH molecules. In the model, the PAH emission is represented by the spectrum of the photo-dissociation region (PDR) in the star-forming region M17 SW from ISOCAM observation of \citet{cesarsky_infrared_1996} and extracted by \cite{madden_ism_2006}. In order to include the $3.3$ $\mu$m PAH emission feature in the template spectra we extend it blueward beyond $5$ $\mu$m using a Lorentzian profile by \cite{verstraete_aromatic_2001}. The SED of the PAHs in the model is calculated as
    \begin{equation}
        l_{\lambda,\rm PAH} = L_{\lambda,M17}\left(\int_{0}^{\infty}L_{\lambda,M17}d\lambda\right)^{-1}
    \end{equation}
    where, $L_{\lambda,M17}$ is the template spectra used here.
    \item \textit{MIR emission:} Besides the emission from PAH molecules, the spectra of star-forming galaxies consist of a MIR continuum emission due to the heating of small dust grains to very high temperatures. This part of the spectra is computed using a 'greybody' function as
    \begin{equation}
        l_{\lambda,T_{\rm dust}} = \kappa_{\lambda}B_{\lambda}(T_{\rm dust})\left(\int_{0}^{\infty}\kappa_{\lambda}B_{\lambda}(T_{\rm dust})d\lambda\right)^{-1}
        \label{greybody}
    \end{equation}
    where, $B_{\lambda}(T_{\rm dust})$ is the Planck function at temperature $T_{\rm dust}$ and $\kappa_{\lambda}$ is the dust mass absorption coefficient of the form
    \begin{equation}
    \label{dabs}
        \kappa_{\lambda}\propto\lambda^{-\beta}
    \end{equation}
    where $\beta$ is the dust emissivity index. Studies show that $\beta\approx1$ for carbonaceous grains (radiating energy in the MIR) while $\beta\approx1.5-2$ for silicate grains \citep{hildebrand_determination_1983, draine_optical_1984}, radiating most of the energy in the far-IR/submm wavelengths. In the model, the MIR continuum emission is characterised by the sum of two greybody functions at temperatures 130 K and 250 K respectively, having equal contribution to the IR luminosity. Mathematically, 
    \begin{equation}
        l_{\lambda,\rm MIR} = \left(l_{\lambda,130 K}+l_{\lambda,250 K}\right)\left(\int_{0}^{\infty}(l_{\lambda,130 K}+l_{\lambda,250 K})d\lambda\right)^{-1}
    \end{equation}
    where $l_{\lambda,130 K}$ and $l_{\lambda,250 K}$ are calculated using Eq.\,(\ref{greybody}). The value of $\beta$ for the MIR emission is set to 1.
    
    \item \textit{Dust grains in equilibrium:} the FIR portion of the SED comprises emission from bigger dust grains which are in thermal equilibrium and have comparatively low temperatures. The model takes into account two types of grains: warm grains which can be present in BCs and in the ISM, with temperatures $T_{w,BC}$ and $T_{w,ISM}$ respectively; and cold grains which are found in the ISM at temperatures $T_{c,ISM}$. The emissions are computed using greybody functions (Eq.\,(\ref{greybody})) with $\beta=1.5$ and $2.0$ respectively. 
    
    The total IR emission from BCs writes
    \begin{equation}
\begin{split}
L_{\lambda,BC}^{\rm dust} = (\xi_{\rm BC}^{\rm PAH}l_{\lambda,\rm PAH}+\xi_{\rm BC}^{\rm MIR}l_{\lambda,\rm MIR}+\xi_{W}^{\rm BC}l_{\lambda,T_{w,\rm BC}})\times\\ (1-f_{\mu})L^{\rm dust}_{\rm total}(t)
\end{split}
\end{equation}
    satisfying the condition
    \begin{equation}
         \xi_{\rm BC}^{\rm PAH}+\xi_{\rm BC}^{\rm MIR}+\xi_{W}^{\rm BC}=1
    \label{relationbc}
    \end{equation}
    Likewise, the ISM emission can be computed as
    \begin{equation}
    \begin{split}
        L_{\lambda,\rm ISM}^{\rm dust} = (\xi_{\rm ISM}^{\rm PAH}l_{\lambda,\rm PAH}+\xi_{\rm ISM}^{\rm MIR}l_{\lambda,\rm MIR}+\xi_{W}^{\rm ISM}l_{\lambda,T_{w,\rm ISM}}+\\ \xi_{C}^{\rm ISM}l_{\lambda,T_{c,\rm ISM}})\times f_{\mu}L^{\rm dust}_{\rm total}(t)
        \end{split}
    \end{equation}
    satisfying the condition
    \begin{equation}
         \xi_{\rm ISM}^{\rm PAH}+\xi_{\rm ISM}^{\rm MIR}+\xi_{W}^{\rm ISM}+\xi_{C}^{\rm ISM}=1
    \label{relationism}     
    \end{equation}
\end{enumerate}

\begin{figure*}
    \centering
      \includegraphics[width=15cm,height=7cm]{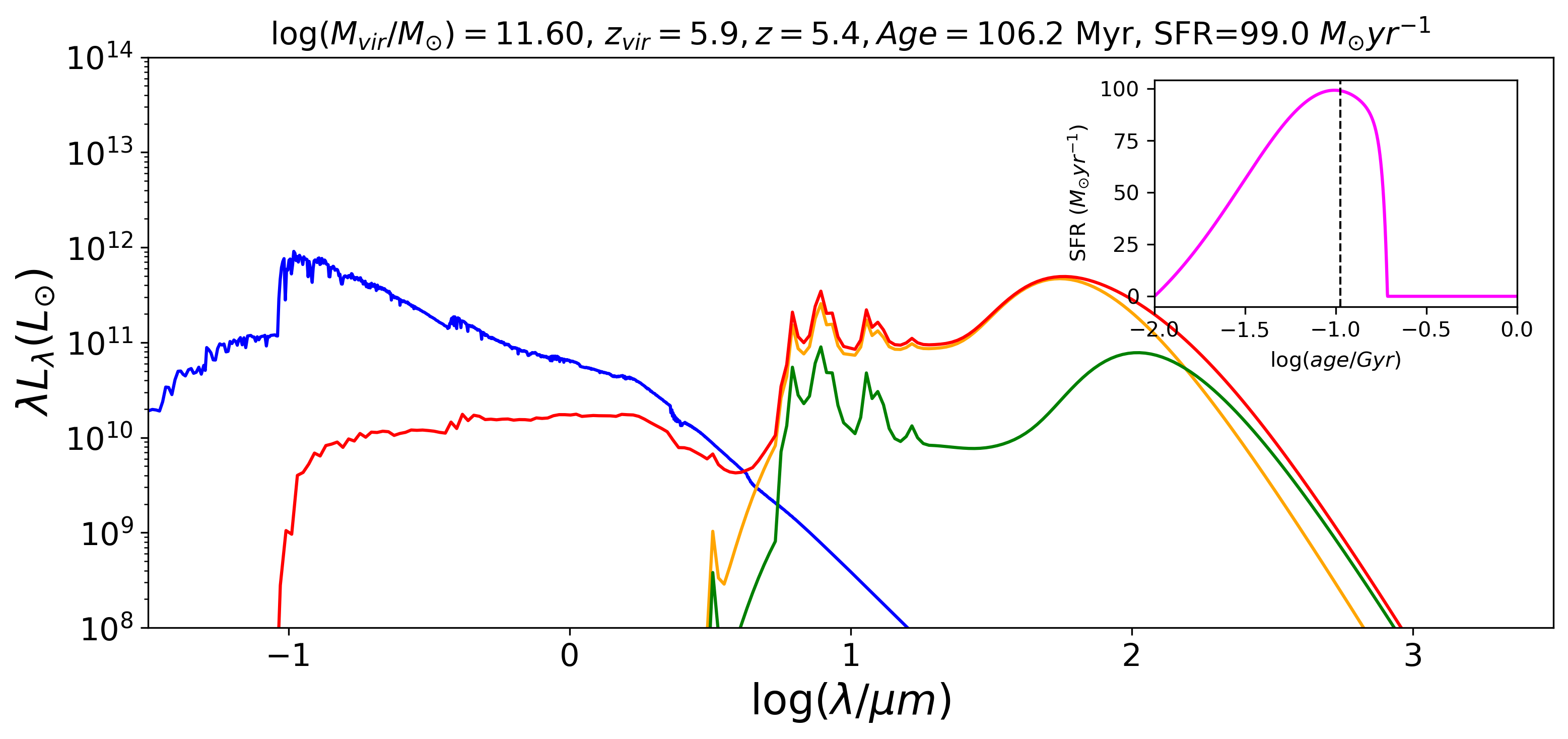}
      \caption{Example of the SED of a proto-spheroidal galaxy (red curve) with halo mass of $M_{\rm vir}=10^{11.60}\,M_{\odot}$ and formation redshift $z_{\rm vir}=5.9$. The SED accounts for the effect of the suppression of the stellar emission in the UV/optical/near-IR due to dust absorption and for the reprocessing at longer wavelengths of the starlight that has been absorbed by dust. The galaxy is observed at $z=5.4$ at an age of $106.2$ Myr when the SFR is $\sim99\,M_{\odot}\hbox{yr}^{-1}$. The evolution of the SFR with time is shown in the inset plot where the dashed black line marks the age of the galaxy at the time of observation. The blue line shows the unattenuated stellar spectrum. The IR emission from the BCs is shown in orange while that from the ISM is shown in green.}
      \label{figsed}
\end{figure*}

The adopted values of the fractions mentioned in Eqs. (\ref{relationbc}) and (\ref{relationism}) are discussed in Section \ref{sec33}. Fig.\,\ref{figsed} shows an example of the different components that make up the modelled SED of a proto-spheroidal galaxy.

\subsubsection{Emission associated with the AGN}
\label{agnsed}

\begin{figure}

\centering
\includegraphics[width=.45\textwidth]{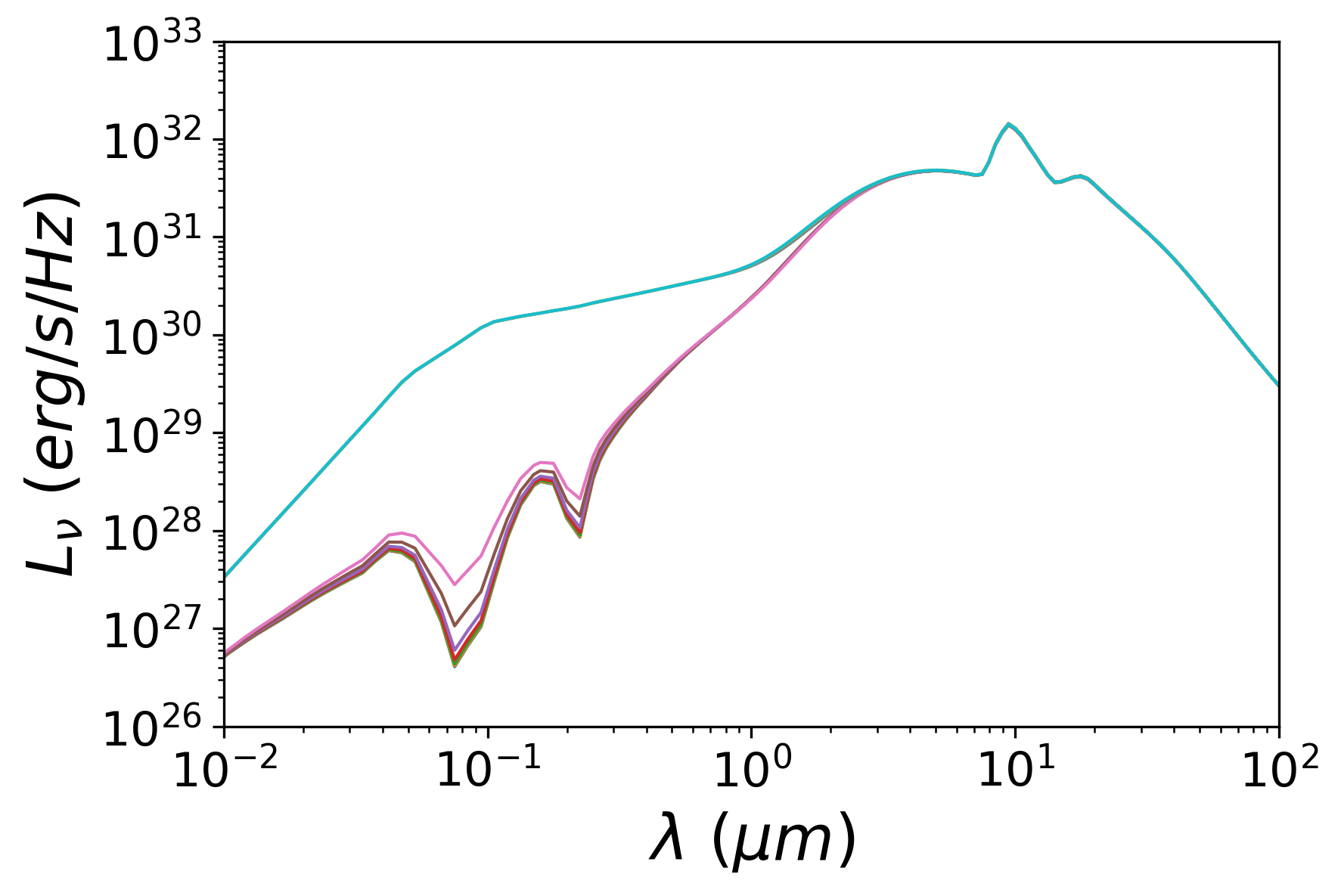}
\caption{AGN SEDs as a function of wavelength for 10 different viewing angles from $0^{\circ}$ (cyan, topmost curve) to $90^{\circ}$ (blue, lowest curve) with $\tau_{\rm eq}(9.7)=0.1$ and $\Theta=140^{\circ}$.}
\label{figagnsed}
\end{figure}

The SED of the AGN component is modelled using the smooth torus model introduced by \citet[][hereafter F06]{fritz_revisiting_2006}. We considered a smooth torus model with the ratio of maximum to minimum radii $R_{\rm max}/R_{\rm min}$ = 10 to 100. The torus amplitude angle, $\Theta$, has values of $40^{\circ}$, $100^{\circ}$ and $140^{\circ}$. The variation of the gas density inside the torus along the radial and angular direction is given by the equation $\rho(r,\theta) = \alpha r^{\beta}e^{-\gamma|\cos(\theta)|}$, where $\beta=0$ and $\gamma=0$ mean constant density. The model assumes that the luminosity of the central power source of the torus is described by a broken power law as $\lambda L_{\lambda}(\lambda) = L_{0}\lambda^\xi$ (erg/s), where $\xi = 1.2$ if $0.001\mu{\rm m}<\lambda<0.03\mu{\rm m}$, $\xi=0$ if $0.03\mu{\rm m}<\lambda<0.125\mu{\rm m}$ and $\xi=-0.5$ if $0.125\mu{\rm m} <\lambda<20\mu{\rm m}$. The equatorial optical depth at $9.7\,\mu$m, $\tau_{\rm eq}(9.7)$, ranges from 0.1 to 10. The model considers 10 viewing angles ranging from $0^{\circ}$ (type 1 AGN) to $90^{\circ}$ (type 2 AGN). For more details about the model we refer the reader to F06. 

In our simulations, as explained in Sec.\,\ref{sec3}, we associate to each simulated galaxy an AGN SED by randomly sampling the available range of values of the SED parameters. We then normalise the SED so that the resulting bolometric AGN luminosity equals the one obtained from Eq.\,(\ref{eqnagn}). 

Some of the light emitted from the AGN gets absorbed by the dust in the ambient ISM. So, the AGN luminosity emitted per unit wavelength at time $t$ is given by
\begin{equation}
    L_{\lambda}^{\rm AGN}(t) = \int_{0}^{t}L_{\lambda}e^{-\hat{\tau}_{\lambda,\rm AGN}(t')}dt',
    \label{eqnattagn}
\end{equation}
where $L_{\lambda}$ is the luminosity per unit wavelength interval from the template. We define
\begin{equation}
  \hat{\tau}_{\lambda,\rm AGN}(t')=
    \begin{cases}
        \hat{\tau_{\lambda}}^{\rm ISM} & \text{for } t' < t_{\rm AGN},\\
        0 & \text{for } t' > t_{\rm AGN}
    \end{cases}  
\end{equation}
The total AGN luminosity absorbed by dust in the ISM is then given by
\begin{equation}
    L^{\rm dust,AGN}_{\rm ISM}(t) = \int_{0}^{\infty}\int_{0}^{t}(1-e^{-\hat{\tau}_{\lambda,\rm AGN}(t')})L_{\lambda}dt'd\lambda .
\end{equation}
Therefore, the total dust luminosity of the galaxy is redefined as
\begin{equation}
    L^{\rm dust}_{\rm total}(t) = L^{\rm dust}_{\rm BC}(t)+L^{\rm dust}_{\rm ISM}(t)+L^{\rm dust,AGN}_{\rm ISM}(t).
    \label{eqntotdustlum}
\end{equation}
and the fraction of the dust luminosity coming from the ISM is now
\begin{equation}
    f_{\mu} = \frac{L^{\rm dust}_{\rm ISM}(t)+L^{\rm dust,AGN}_{\rm ISM}(t)}{L^{\rm dust}_{\rm total}(t)}.
\end{equation}
The absorbed AGN luminosity is reprocessed by dust and emitted at IR wavelengths. Likewise, the total IR emission from the BC and the ISM is now re-defined as
 \begin{equation}
\begin{split}
L_{\lambda,BC}^{\rm dust} = (\xi_{\rm BC}^{\rm PAH}l_{\lambda,\rm PAH}+\xi_{\rm BC}^{\rm MIR}l_{\lambda,\rm MIR}+\xi_{W}^{\rm BC}l_{\lambda,T_{w,\rm BC}})\times\\ (1-f_{\mu})(L^{\rm dust}_{\rm BC}(t)+L^{\rm dust}_{\rm ISM}(t)+L^{\rm dust,AGN}_{\rm ISM}(t)),
\end{split}
\end{equation}
 and 
\begin{equation}
    \begin{split}
        L_{\lambda,\rm ISM}^{\rm dust} = (\xi_{\rm ISM}^{\rm PAH}l_{\lambda,\rm PAH}+\xi_{\rm ISM}^{\rm MIR}l_{\lambda,\rm MIR}+\xi_{W}^{\rm ISM}l_{\lambda,T_{w,\rm ISM}}+\\ \xi_{C}^{\rm ISM}l_{\lambda,T_{c,\rm ISM}})\times f_{\mu}(L^{\rm dust}_{\rm BC}(t)+L^{\rm dust}_{\rm ISM}(t)+L^{\rm dust,AGN}_{\rm ISM}(t)).
        \end{split}
    \end{equation}
respectively.

\begin{figure*}
    \centering
      \includegraphics[width=18cm,height=7cm]{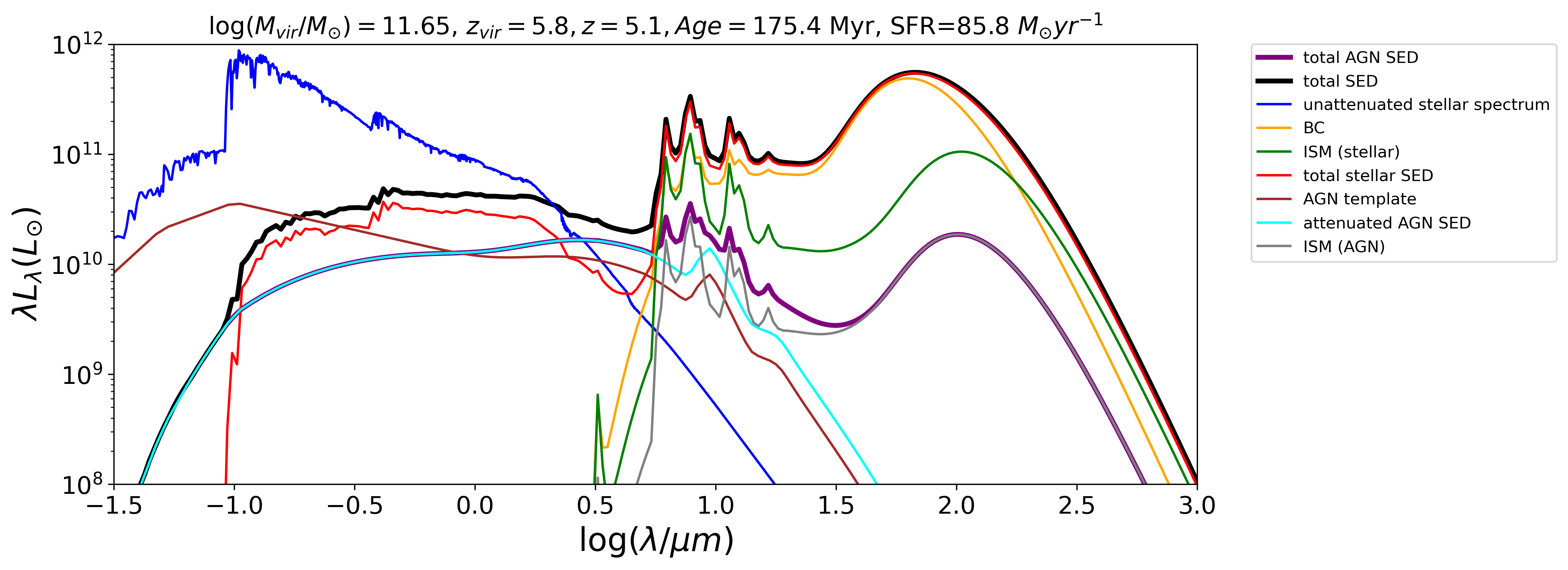}
      \caption{Example of the SED (stellar + AGN components) of a proto-spheroidal galaxy (black curve) with halo mass of $M_{\rm vir}=10^{11.65}\,M_{\odot}$ and formation redshift $z_{\rm vir}=5.8$. The galaxy is observed at $z=5.1$ at an age of $175.4$ Myr when the SFR is $\sim 86\,M_{\odot}\hbox{yr}^{-1}$. The blue curve is the unattenuated stellar spectrum, while the red curve shows the starlight attenuated by dust in the UV/optical/near-IR and reprocessed by the same dust at longer wavelengths. The IR emission from the BCs is shown in orange while that from the ISM is shown in green. 
      The unattenuated SED of the AGN component is shown in brown while the attenuated AGN SED is shown in cyan. The grey curve depicts the IR emission from the ISM due to the absorption of AGN luminosity by ISM dust. The purple curve shows the total AGN SED.  The AGN SED is typical of a Type 2 AGN ($\theta=90^\circ$) with density profile $\rho(r) = \alpha r^{-0.5}$, $\tau_{\rm eq}(9.7)=10$, $R_{\rm max}/R_{\rm min}=60$ and $\Theta=100^\circ$.}
      \label{figsedagn}
\end{figure*}

Fig.\,\ref{figagnsed} shows the F06 AGN template SED as a function of wavelength for different viewing angles (from $0^{\circ}$ to $90^{\circ}$) and a constant density profile of the gas in the torus. An example of the SED of a proto-spheroidal galaxy with a visible AGN component is shown in Fig.\,\ref{figsedagn}. 

Apart from the dusty torus and the dust in the ISM, "polar dust" can also contribute to the emission from the AGN. This dust extends in the polar direction of the dusty torus over parsec scales 
\citep{honig_dust_2013,yang_dust_2020}. 
Its emission peaks in the mid-IR \citep[see Figure 8]{honig_dust_2013}, and it is negligible at the wavelengths of interest here, i.e. UV/optical/near-IR wavelengths. Hence, we do not include the effect of polar dust in our model.

\subsubsection{Dust mass estimation}

To estimate the dust mass in galaxies, we follow the prescriptions of \cite{da_cunha_simple_2008}, which are based on the formula by \cite{hildebrand_determination_1983}. For dust grains in thermal equilibrium at the temperature $T_d$ the dust mass ($M_d$) is given by
\begin{equation}
    L_{\lambda}^{T_d} = 4\pi M_d \kappa_{\lambda}B_{\lambda}(T_d),
\end{equation}
where $L_{\lambda}^d$ is the infrared luminosity of the dust grains, while $B_{\lambda}(T_d)$ and $\kappa_{\lambda}$ are defined in Eq. (\ref{greybody}) and (\ref{dabs}), respectively. After normalising $\kappa_\lambda$ using $\kappa_{850\,\mu m}=0.77$ $\hbox{g}^{-1}\,\hbox{cm}^2$, the mass contribution of warm dust in BC and ISM is calculated as
\begin{eqnarray}
    M_{\rm dust,\rm W}^{\rm BC} = \xi_{\rm W}^{\rm BC}(1-f_{\mu})L_{\rm total}^{\rm dust}\left(4\pi\int_{0}^{\infty}\kappa_{\lambda}B_{\lambda}(T_{w,\rm BC})d\lambda\right)^{-1}
\end{eqnarray}
and
\begin{eqnarray}
    M_{\rm dust,\rm W}^{\rm ISM} = \xi_{\rm W}^{\rm ISM}f_{\mu}L_{\rm total}^{\rm dust}\left(4\pi\int_{0}^{\infty}\kappa_{\lambda}B_{\lambda}(T_{w,\rm ISM})d\lambda\right)^{-1}
\end{eqnarray}
respectively, and that from the cold dust in the ISM is calculated as
\begin{eqnarray}
    M_{\rm dust,\rm C}^{\rm ISM} = \xi_{\rm C}^{\rm ISM}f_{\mu}L_{\rm total}^{\rm dust}\left(4\pi\int_{0}^{\infty}\kappa_{\lambda}B_{\lambda}(T_{c,\rm ISM})d\lambda\right)^{-1}
\end{eqnarray}
Taking into account the contribution from stochastically heated dust grains and considering only a small contribution of a few per cent from PAH grains \citep{draine_infrared_2007}, the total dust mass of a galaxy is estimated as
\begin{eqnarray}
    M_{\rm dust}^{\rm total} \approx 1.1\left(M_{\rm dust,\rm W}^{\rm BC}+M_{\rm dust,\rm W}^{\rm ISM}+M_{\rm dust,\rm C}^{\rm ISM}\right)
\end{eqnarray}




\section{Building a simulated sample of proto-spheroidal galaxies}
\label{sec3}

Here we describe the way in which we implement the formalism presented in the previous section to generate a sample of simulated proto-spheroidal galaxies. The sample will then be used for forecast studies.

\subsection{Step 1: sampling the halo formation rate function}

We start by randomly sampling the halo formation rate function (Eq.\,(\ref{eqnHFR})) in both halo virialization mass and virialization redshift, within the intervals $z_{\rm vir}\in[1.5,12]$ and $\log(M_{\rm vir}/M_{\odot})\in[11.3,13.3]$, respectively. 
The total number of randomly sampled pairs of $M_{\rm vir}$ and $z_{\rm vir}$ is dictated by the adopted survey area, which, in turn, determines the sampled volume at any given virialization redshift.
For each simulated pair of values of $M_{\rm vir}$ and $z_{\rm vir}$ we then solve the C13 equations to compute the SFH as a function of the galaxy age, with the birth of the galaxy set at the corresponding virialization redshift.


\subsection{Step 2: generating the UV/optical/near-IR part of the SED}

At this point we apply the SED formalism illustrated in Sec.\,\ref{SED_stars_and_attenuation} to compute the SED of the dust-attenuated stellar component (Eq.\,(\ref{eqnattst})) as well as the SED of the dust-attenuated AGN (Eq.\,(\ref{eqnattagn})) at any given age of the galaxy. The main parameters affecting the stellar component are the age of the BCs, $t_{0}$, and the total effective V-band absorption optical depth of the dust in the BCs, $\hat{\tau}_{V}$, and in the ISM, $\hat{\tau}_{V}^{\rm ISM}$.
For them, we adopt the values measured by \citet[R14 hereafter]{rowlands_herschel-atlas_2014}, who performed SED fitting on a sample of 29 250\,$\mu$m-selected high-redshift DSFGs from H-ATLAS, using MAGPHYS \citep{da_cunha_magphys_2011}. They obtained a good fit by allowing $t_{0}$ to vary uniformly in logarithmic space between $10^7-10^8$ years, while for $\hat{\tau_{V}}$ and $\hat{\tau}_{V}^{\rm ISM}$ they derived mean values of $5.1\pm0.6$ and $1.0\pm0.1$, respectively. 
In practice, we take Gaussian priors for $\hat{\tau_{V}^{\rm BC}}$ and $\hat{\tau_{V}^{\rm ISM}}$ with mean 5.1 and 1.0, respectively. The standard deviation is taken as 0.6 and 0.1, respectively, for all the simulated objects, and the value of $\log(t_{0}/{\rm yr})$ is randomly assigned to each object within the range 7 to 8.

\begin{figure}
\centering
\includegraphics[width=.45\textwidth]{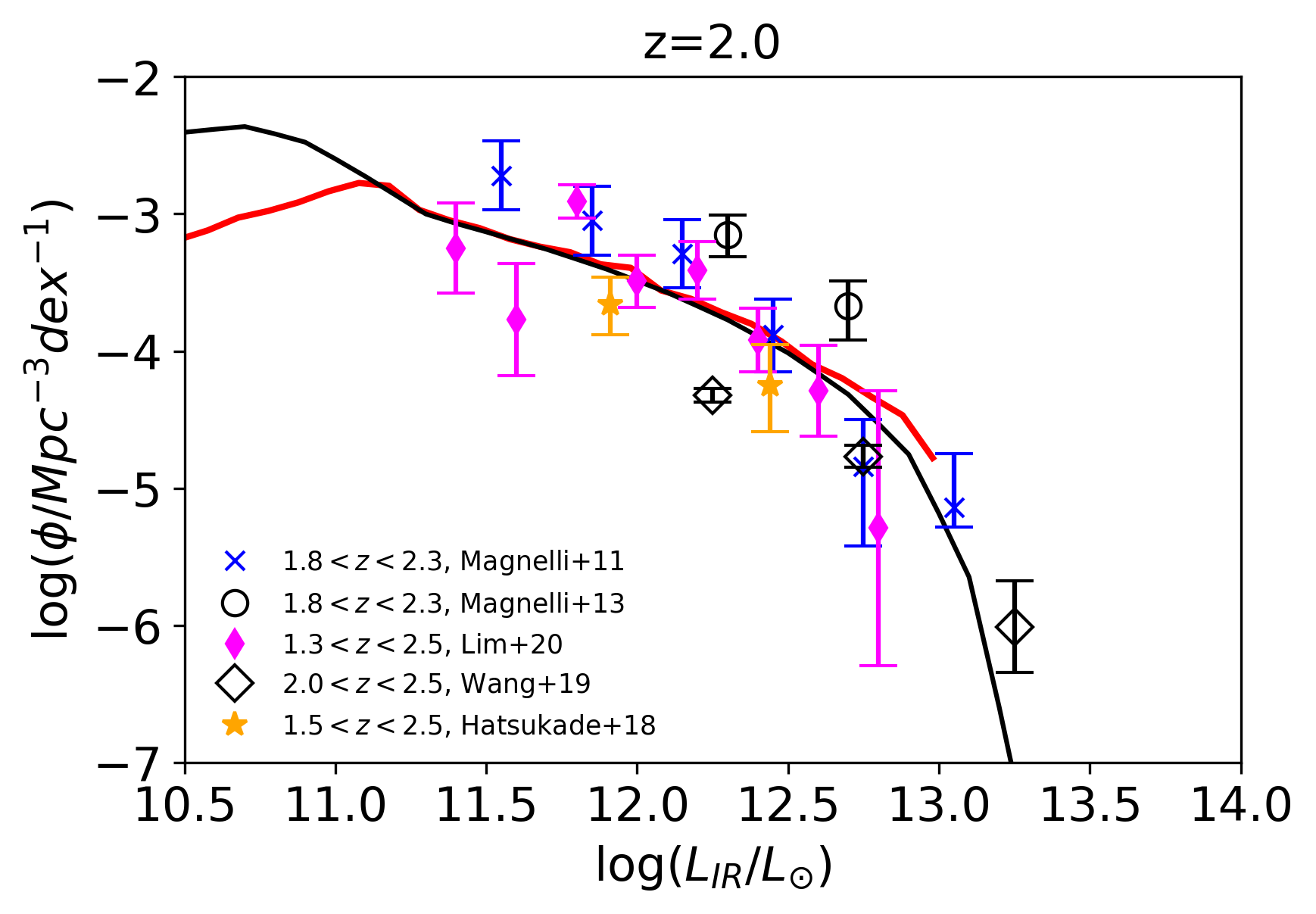}
\includegraphics[width=.45\textwidth]{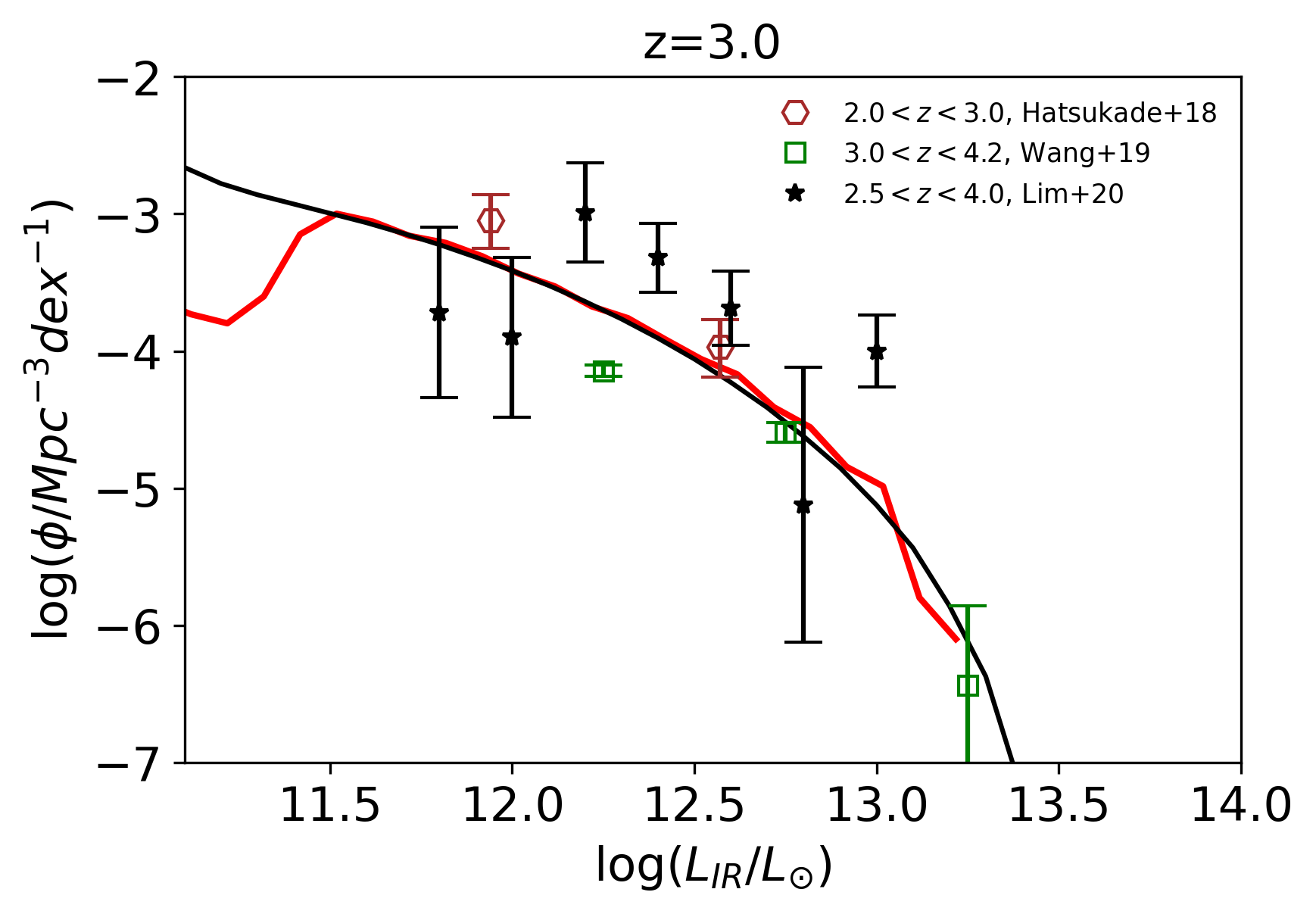}
\includegraphics[width=.45\textwidth]{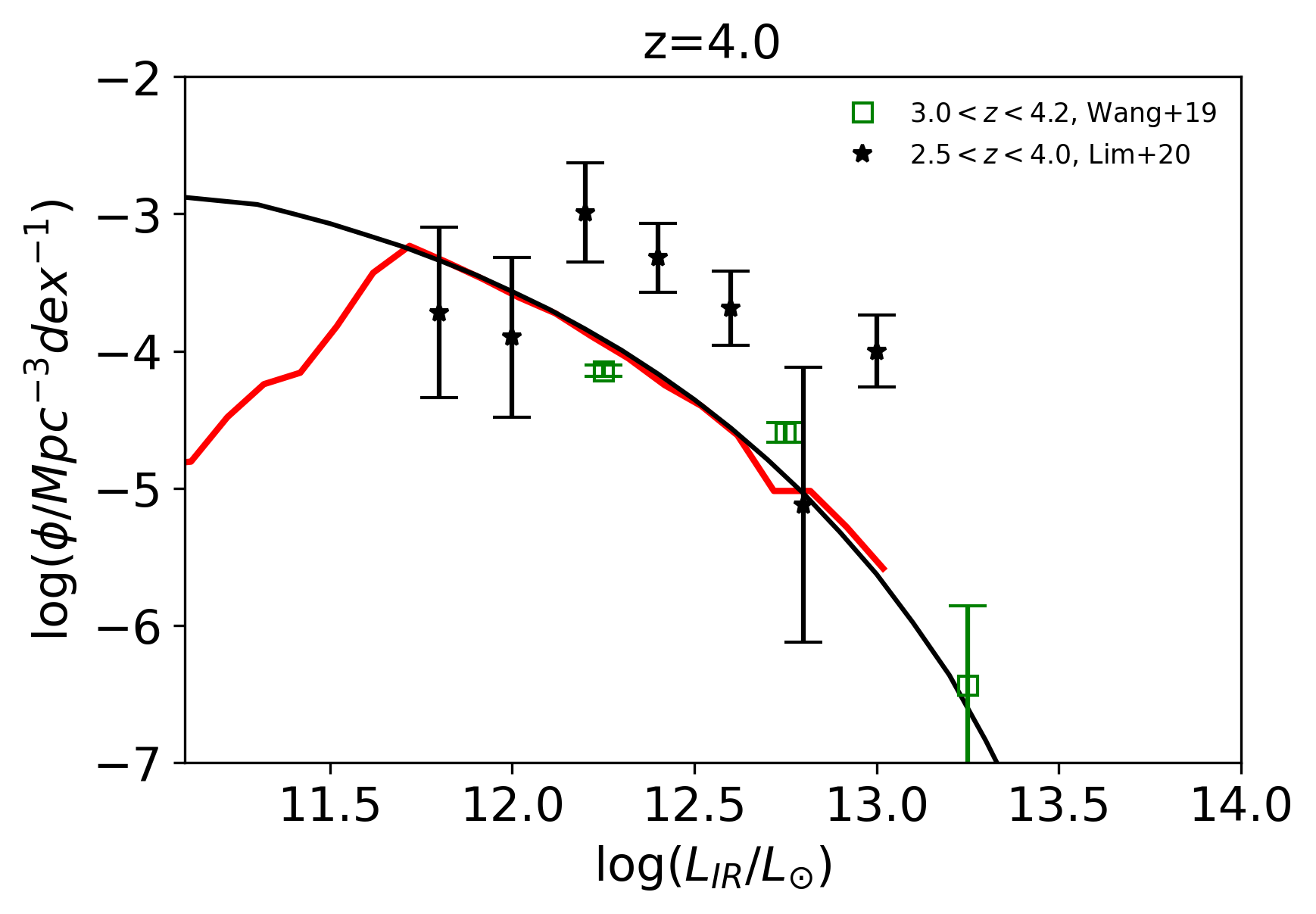}
\caption{Comparison between the theoretical LF (black), the simulated curves (red) and observational data at different redshifts. The data points are taken from  \citet{magnelli_evolution_2011}, \citet{magnelli_deepest_2013}, \citet{hatsukade_alma_2018}, \citet{wang_multi-wavelength_2019} and \citet{lim_scuba-2_2020}.}
\label{figcomplfresult}
\end{figure}

Before moving on, we checked that the infrared luminosity function (LF) of the simulated sample is consistent with current measurements. For a given redshift, we first computed the age of each object at that redshift and then used Eq.\,(\ref{eqntotdustlum}) to compute the corresponding dust luminosity, $L_{d}$. Because the latter represents the amount of energy re-emitted in the far-IR by dust, we took $L_{d}$ to be the total infrared luminosity, $L_{\rm IR}$. We then binned the objects in $L_{\rm IR}$ to generate a simulated infrared luminosity function to be compared with the data.

The results are illustrated in Fig.\,\ref{figcomplfresult} for $z=2$, 3, and 4. The simulated LF (red curve) shows a very good agreement with the C13 model (black curve) at moderate to bright infrared luminosities, i.e. above $L_{\rm IR}\gtrsim10^{11.5}\,L_{\odot}$. However, it lies significantly below the prediction from the C13 model at low luminosities. This is a consequence of the way C13 computed the infrared luminosity. Indeed, they calculated $L_{\rm IR}$ from the SFR of the objects using the relation by \citet[K98 hereafter]{kennicutt_star_1998}. However, that relation only applies during the early dust-obscured phase of the evolution of galaxies, when the far-IR emission is mainly associated with BCs where the dust is heated by young massive stars (R14). It does not hold for more evolved, low-luminosity, galaxies where the ISM contributes a significant fraction of the dust emission. 

\begin{figure}
    \centering
      \includegraphics[width=8.8cm,height=6.5cm]{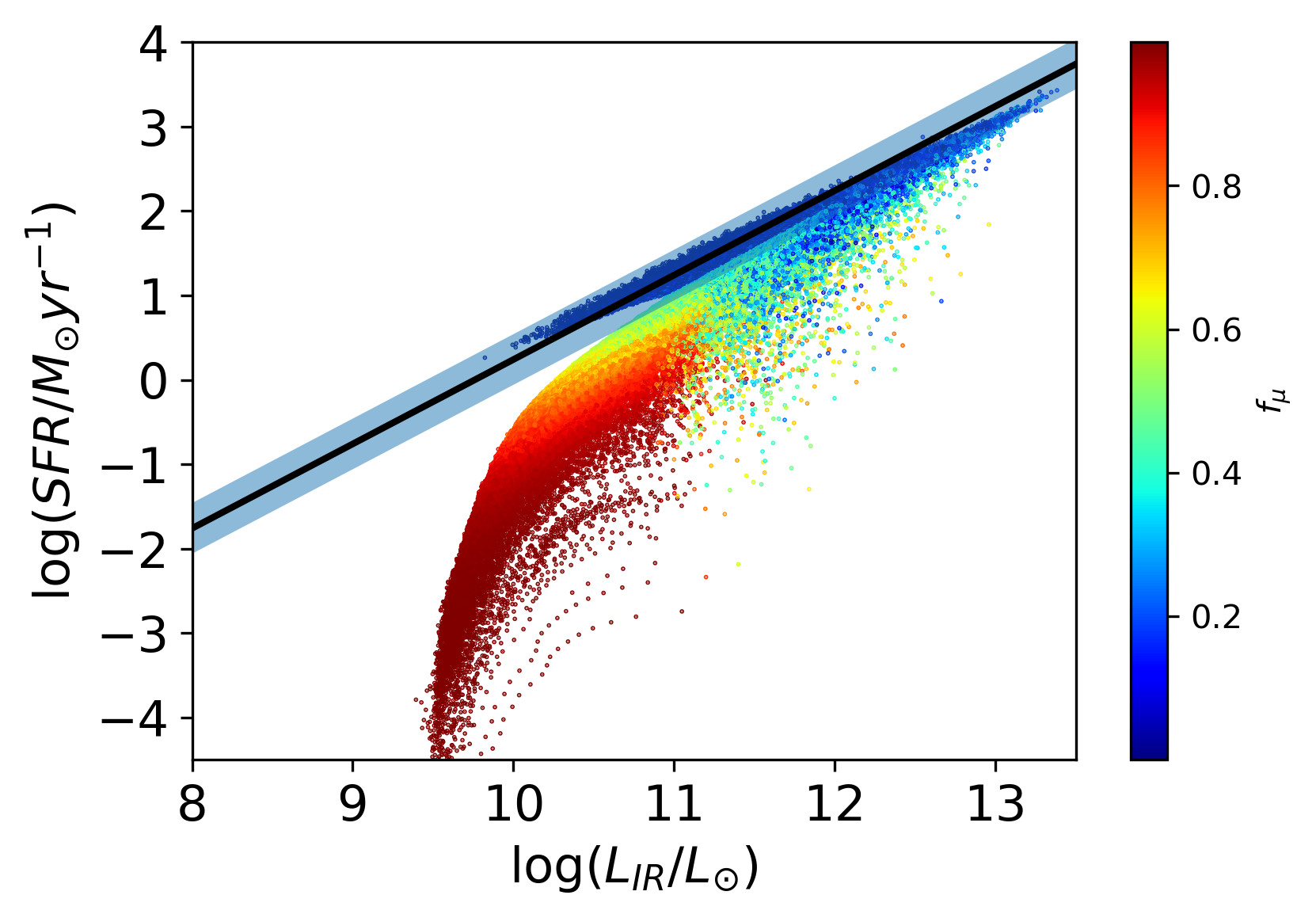}
      \caption{Correlation between SFR and total IR luminosity, $L_{\rm IR}$, for a sample of simulated galaxies at $z=2$. $L_{\rm IR}$ is calculated using Eq.\,(\ref{dustlum}). Data points are colour-coded in accordance with the value of $f_{\mu}$. The black line shows the relation between SFR and total IR luminosity (integrated from 8 to $1000\,\mu$m) from K98 with a $30\%$ dispersion shown by the shaded blue area. Galaxies with a low value of  $f_{\mu}$ lie on the K98 relation. However, for galaxies where the ISM contributes significantly to the dust luminosity (i.e. those with high values of $f_{\mu}$) the K98 relation leads to an underestimate of $L_{\rm IR}$.}
      \label{figdsfrld}
\end{figure}

This is illustrated in Fig.\,\ref{figdsfrld} where the SFR of our simulated galaxies is shown as a function of their infrared luminosity, computed using Eq.\,(\ref{dustlum}), and colour-coded according to the value of $f_{\mu}$. A high value of $f_{\mu}$ corresponds to a dominant contribution of the ISM to the dust luminosity. Below $\log(L_{\rm IR}/L_{\odot})\simeq11$, where $f_{\mu}>0.5$, there is a clear deviation from the K98 relation, with the latter significantly underestimating the infrared luminosity for any given SFR. As a result, more objects with lower $L_{\rm IR}$ are expected when using the K98 relation. This is the reason why our simulated LF curve lies below the C13 one at lower luminosities.

\begin{figure}

\centering
\includegraphics[width=.47\textwidth]{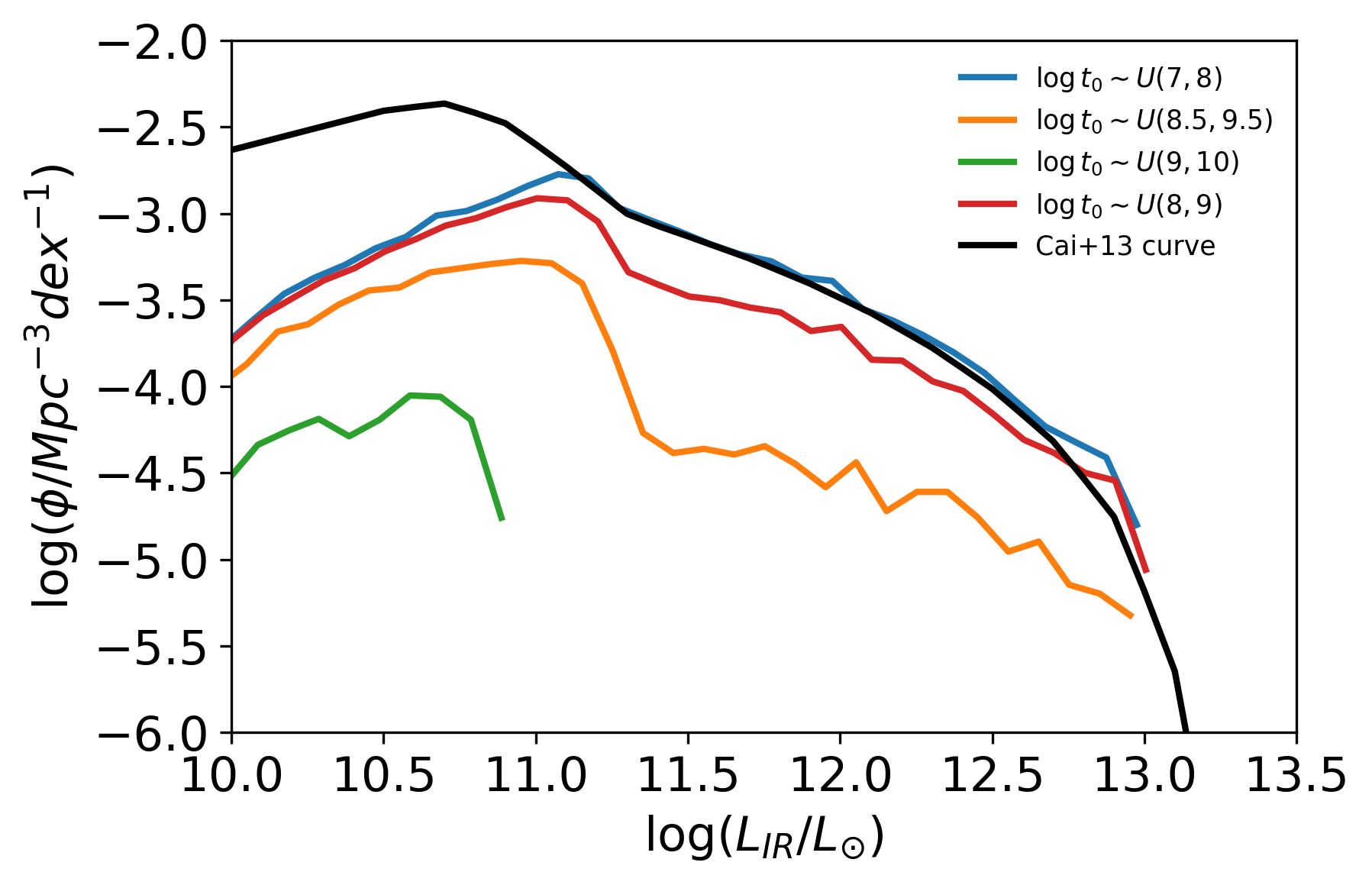}\hfill
\includegraphics[width=.47\textwidth]{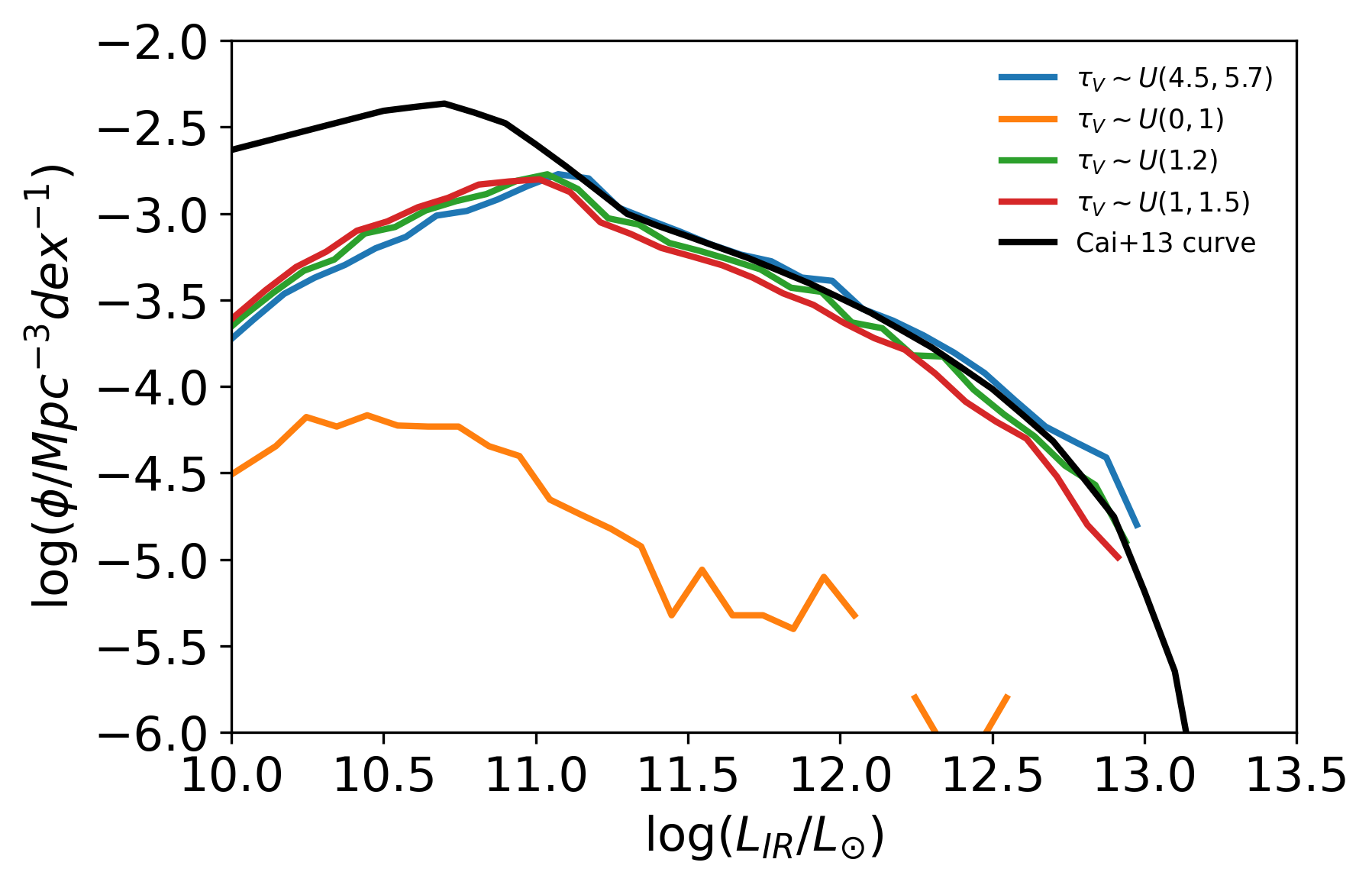}\hfill
\includegraphics[width=.47\textwidth]{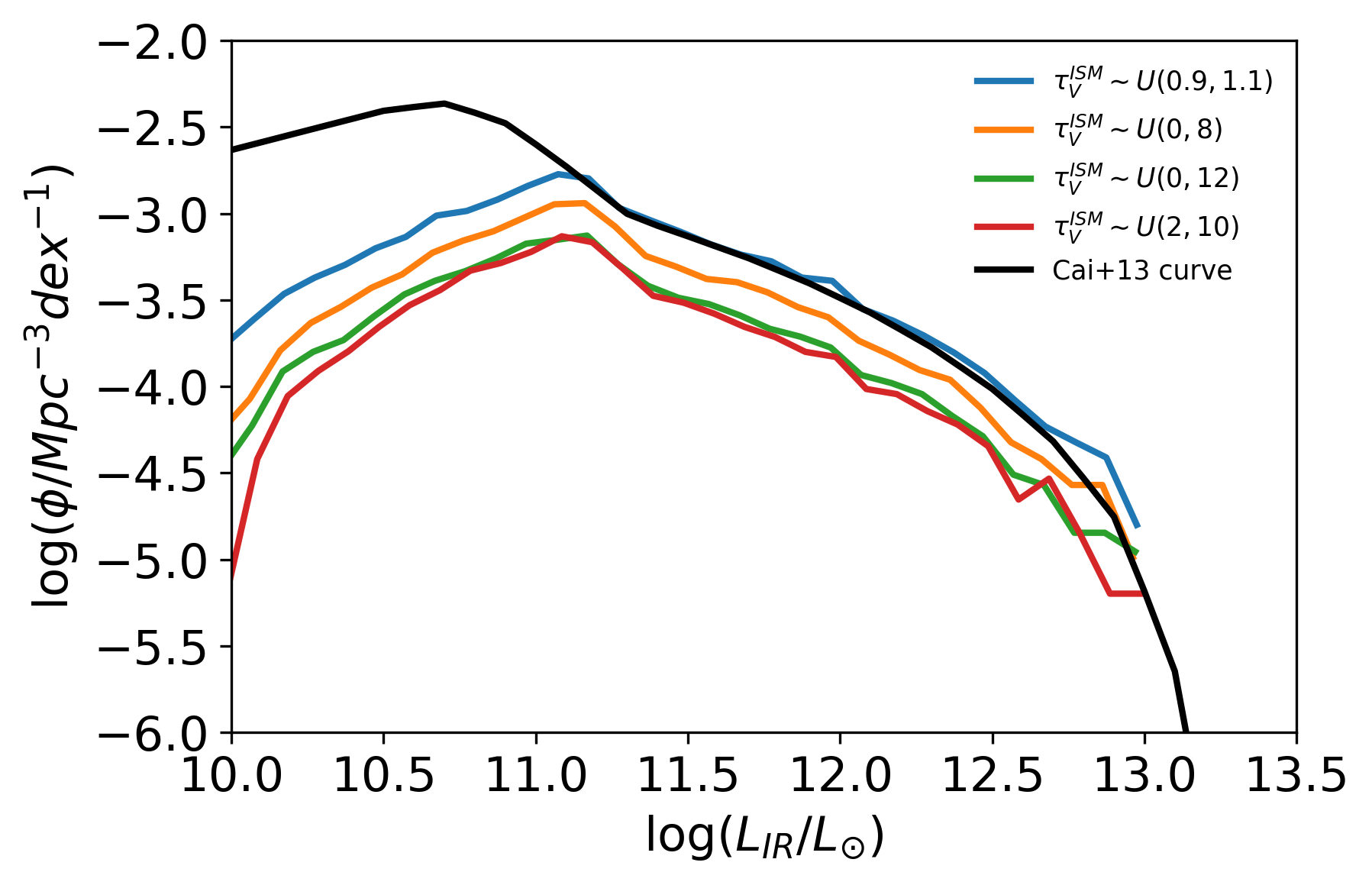}

\caption{Effect of changing the values of $t_{0}$, $\hat{\tau_{V}}$ and $\hat{\tau_{V}^{\rm ISM}}$ on the shape of the total IR LF at $z=2$. In all panels the black curve represents the theoretical curve obtained by C13 and the blue curve represents the simulated LF obtained by using the optimal values of the parameters: $\log t_{0}\in U(7,8)$, $\tau_{V}\in U(4.5,5.7)$ and $\tau_{V}^{\rm ISM}\in U(0.9,1.1)$. {\it Top panel}: varying $t_0$ uniformly in logarithmic scale in the ranges 8--9 (red), 8.5--9.5 (orange) and 9--10 (green), respectively. {\it Middle panel}: varying $\tau_V$ uniformly in the ranges 0--1 (orange), 1--2 (green), and 1--1.5 (red), respectively. {\it Bottom panel}: varying $\tau_V^{\rm ISM}$ uniformly in the ranges 0--8 (orange), 0--12 (green), and 2--10 (red), respectively.}
\label{figlfcheck}
\end{figure}

The shape of the LFs is highly dependent on the values of the parameters that determine $L_{d}$ via Eq.\,(\ref{dustlum}), especially of $t_{0}$, $\tau_{V}$ and $\tau_{V}^{\rm ISM}$. 
Figure\,\ref{figlfcheck} shows how the IR LF at $z=2$ is affected by changes in the values of the above parameters. 
The parameters are changed one at a time while keeping the others fixed at their prescribed values. It is evident that a deviation from the R14 values produces LFs that lie below the reference LF by C13 (black curve). 

By increasing the dissipation timescale of the BC, stars are allowed to reside inside the BC for a longer time. One would expect that a prolonged dust absorption of UV/optical light from the stars will lead to a higher IR luminosity, hence to an increased infrared LF. But, in reality, that is not the case. The massive stars, which contribute the bulk of the IR luminosity, have significantly evolved by the time the BC dissipates and hence, have reduced luminosity. So, there are fewer UV/optical photons being emitted from them to be absorbed and re-emitted at IR wavelengths. Moreover, by increasing the value of $t_{0}$ and by getting it closer to $t_{\rm AGN}$, the BCs are eventually destroyed by the AGN feedback. This effect also leads to a lower IR luminosity. 

Similarly, by increasing $\tau_{V}^{\rm ISM}$ while keeping $\tau_{V}$ constant (which is equivalent to decreasing $\tau_{V}^{\rm BC}$), a lower fraction of the starlight is re-radiated in the far-IR, leading to a decrease in the number of galaxies for any given $L_{\rm IR}$ compared to the reference case.


\begin{figure}

\centering
\includegraphics[width=.47\textwidth]{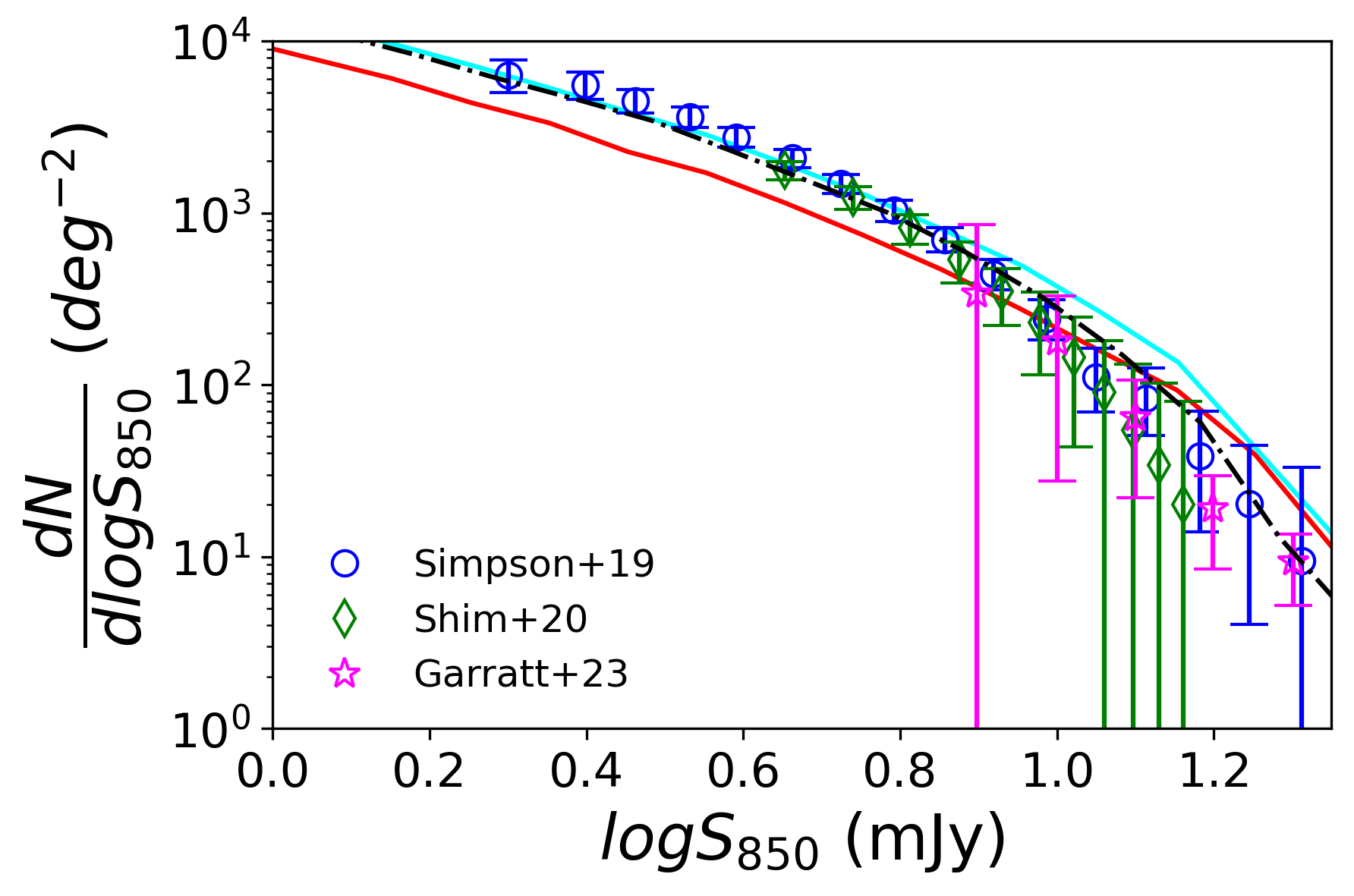}\hfill
\includegraphics[width=.47\textwidth]{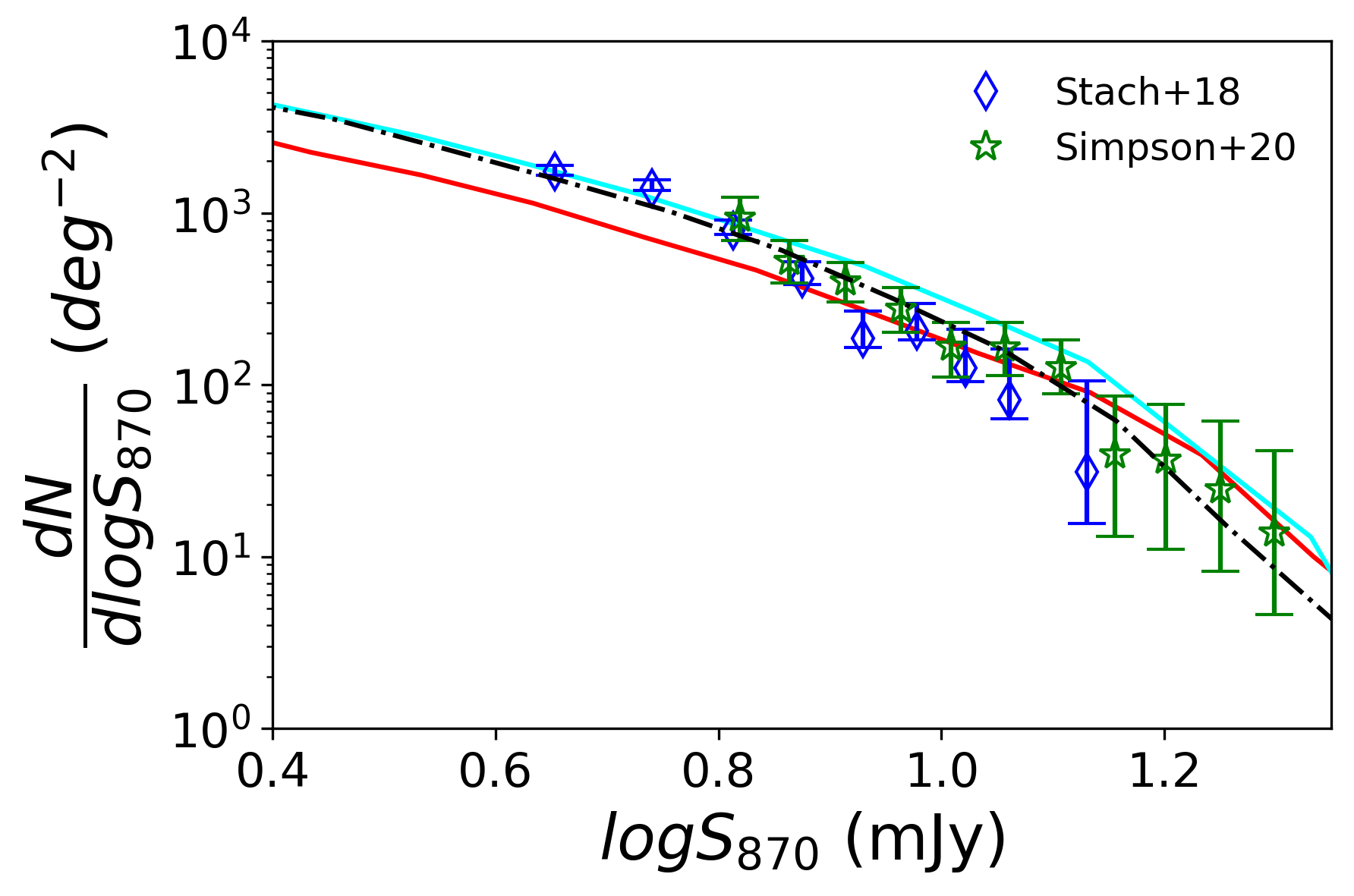}\hfill
\includegraphics[width=.47\textwidth]{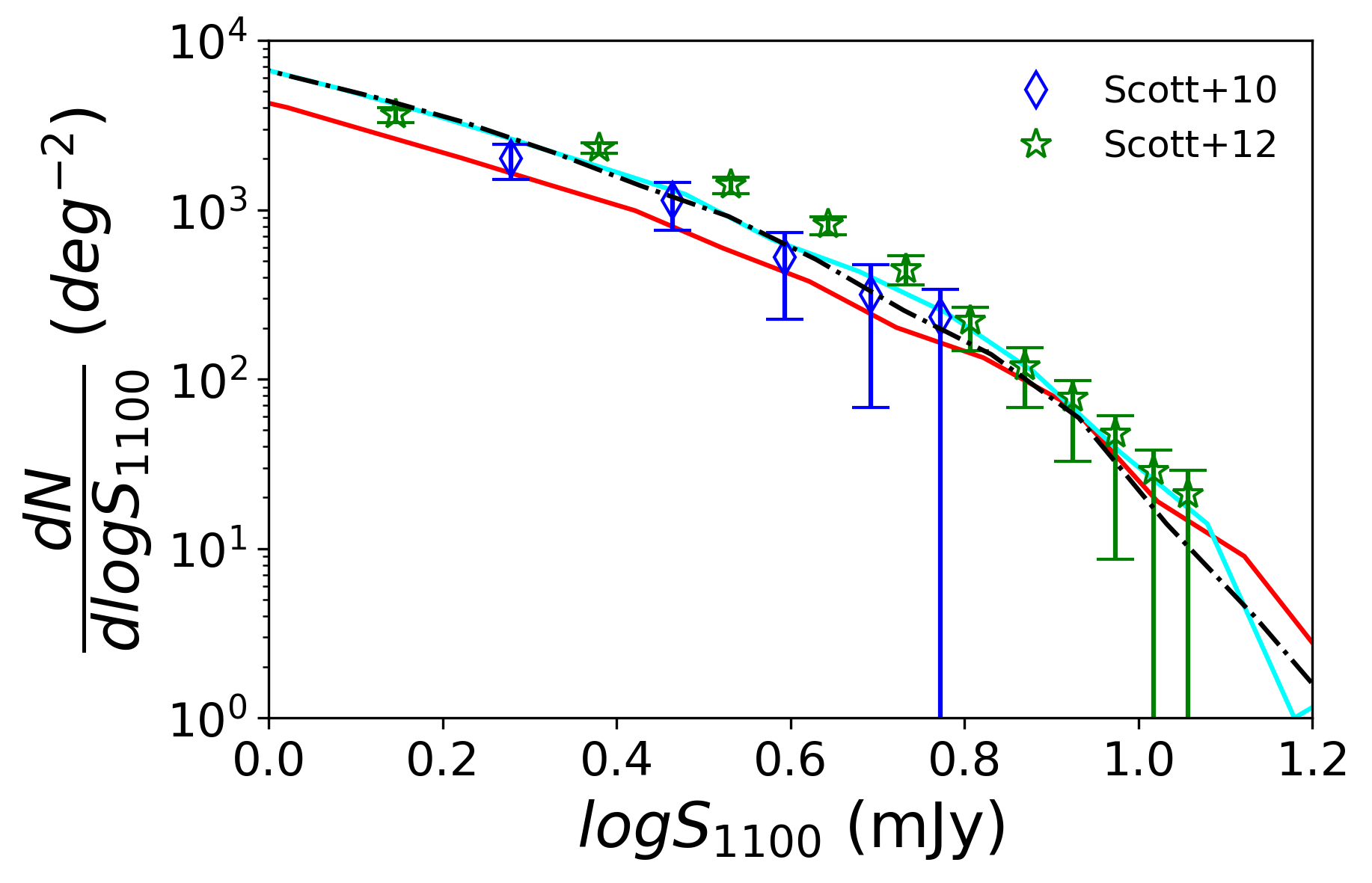}

\caption{Comparison between the simulated differential number counts (red curve) with the C13 model (dashed black dots) and the observed data at $850\,\mu$m, $870\,\mu$m and $1100\,\mu$m. 
The cyan curve shows the number counts obtained by modifying the parameter values by trial and errpr. The new curve fits the faint end but produces an excess of objects at the brightest luminosities. The data points are taken from \citet{simpson_east_2019}, \citet{shim_nepsc2_2020},\citet{garratt_scuba-2_2023},  \citet{stach_alma_2018}, \citet{simpson_alma_2020}, \citet{scott_deep_2010}, and \citet{scott_source_2012}.}
\label{figdnc}
\end{figure}

\subsection{Step 3: generating the far-IR/sub-mm/mm part of the SED}
\label{sec33}

\setlength{\tabcolsep}{4pt} 
\renewcommand{\arraystretch}{1.0} 
\begin{table}
\caption{Information about the measured differential number counts at 850\,$\mu$m, 870\,$\mu$m and 1100\,$\mu$m shown in Figs\,\ref{figdnc}-\ref{figdncfinal}.}
\label{tabdnc}
\begin{tabular}{lccc}
\hline\hline
Wavelength & Instrument & Field & Reference\\
$\hspace{3ex}(\mu$m) &  &  &\\
\hline
850 & SCUBA-2 & S2COSMOS & \cite{simpson_east_2019}\\
850 & SCUBA-2 & NEPSC2 & \cite{shim_nepsc2_2020}\\
850 & SCUBA-2 & S2LXS XMM-LSS & \cite{garratt_scuba-2_2023}\\
870 & SCUBA-2 & S2CLS UDS & \cite{stach_alma_2018}\\
870 & SCUBA-2 & S2COSMOS & \cite{simpson_alma_2020}\\
1100 & ASTE/AzTEC & GOODS-S & \cite{scott_deep_2010}\\
1100 & JCMT/AzTEC & GOODS-N & \cite{scott_source_2012}\\
\hline
\end{tabular}
\end{table}

For each simulated object we compute the far-IR/sub-mm/mm part of the SED by implementing the formalism illustrated in Sec.\,\ref{SED_dust}, and setting the values of the relevant parameters to those proposed by \cite{da_cunha_alma_2015}. The indices $\beta$ for warm and cold dust grains were fixed at 1.5 and 2 respectively. The temperature of the warm dust grains in the BCs, $T_{w,BC}$, is allowed to vary uniformly in the interval 30\,K to 80\,K, while that of the ISM, $T_{w,ISM}$, is fixed at 45\,K. The temperature of the cold grains in the ISM, $T_{c,ISM}$, is taken from a uniform distribution between 20\,K to 40\,K. In the BCs, $\xi_{\rm BC}^{\rm PAH}$ and $\xi_{\rm BC}^{\rm MIR}$ are allowed to vary uniformly in the ranges $0-0.1$ and $0.1-0.2$, respectively. For each choice of both $\xi_{\rm BC}^{\rm PAH}$ and $\xi_{\rm BC}^{\rm MIR}$, $\xi_{W}^{\rm BC}$ follows from Eq.\,(\ref{relationbc}). For the ISM, we extract $\xi_{C}^{\rm ISM}$ from a uniform distribution between $0.5$ and $1$. The other parameters are set to $\xi_{\rm ISM}^{\rm PAH}=0.550(1-\xi_{C}^{\rm ISM})$, $\xi_{\rm ISM}^{\rm MIR}=0.275(1-\xi_{C}^{\rm ISM})$, and $\xi_{W}^{\rm ISM}=0.175(1-\xi_{C}^{\rm ISM})$ so that Eq.\,(\ref{relationism}) is satisfied. 

With this parameter set, we generated differential number counts at
$850\,\mu$m, $870\,\mu$m and $1100\,\mu$m. We did that by first computing the objects' flux density at several redshifts from $z=1$ to $z=8$ in steps of 0.1. Each object was randomly assigned a redshift within the corresponding bin assuming a uniform distribution. We then computed the number of objects per unit interval of redshift and per unit interval of flux density at each redshift step. Finally, we integrated the redshift-dependent number counts to produce the differential number counts. The result is shown by the red curve in Fig.\,\ref{figdnc}, where it is compared with the C13 number counts model curve (black line) of the proto-spheroids and with existing data (see also Table\,\ref{tabdnc}).



We notice that the simulated number counts significantly underestimate the observed number counts and the C13 curve at low flux densities, i.e. $\lesssim6\,$mJy, while producing, at the same time, an excess of bright objects. This result indicates that 
the assumed values for the parameters that describe the far-IR/sub-mm/mm SED of the proto-spheroids need to be revised.
In principle, the best-fit values of the SED parameters could be found using a minimum $\chi^{2}$ approach on the number counts. However, that method, in this specific case, is very time consuming. Therefore, we opted instead, for a trial and error approach, which led to the following changes in the value of some of the SED parameters.
%
%
We assigned a higher value to the dust emissivity index of the warm dust grains, raising it from $\beta=1.5$ to $\beta=2$, while we kept the one of the cold dust grains to its original value of $\beta=2$. 
The temperatures $T_{w, BC}$ and $T_{c,ISM}$ were allowed to vary uniformly within the intervals 30-45\,K and 20-30\,K, respectively. The value of $T_{w,ISM}$ was left unchanged. 
By using these updated values for the SED parameters we achieved a very good agreement with the observed number counts at the faint end, as illustrated by the solid cyan curve in Fig.\,\ref{figdnc}. However, we were still left with an excess of objects at bright flux densities.

\begin{figure}
    \centering
      \includegraphics[width=8.6cm,height=6.0cm]{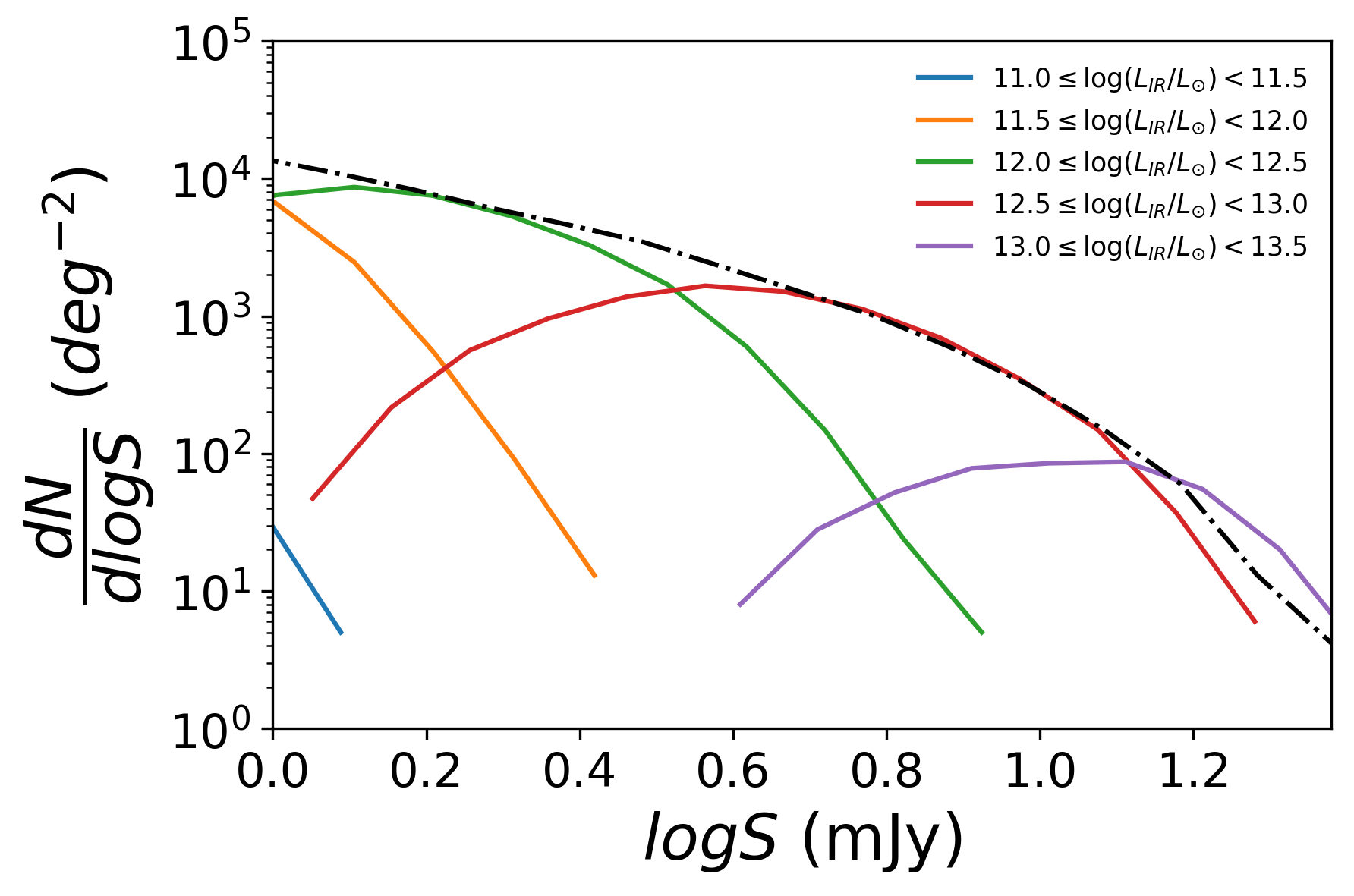}
      \caption{Simulated differential number counts at $850\,\mu$m, with contributions from  intervals of total infrared luminosity. The dot-dashed black curve represents the model number counts by C13.}
      \label{figdncirlbin}
\end{figure}

\begin{figure}
\centering
\includegraphics[width=.47\textwidth]{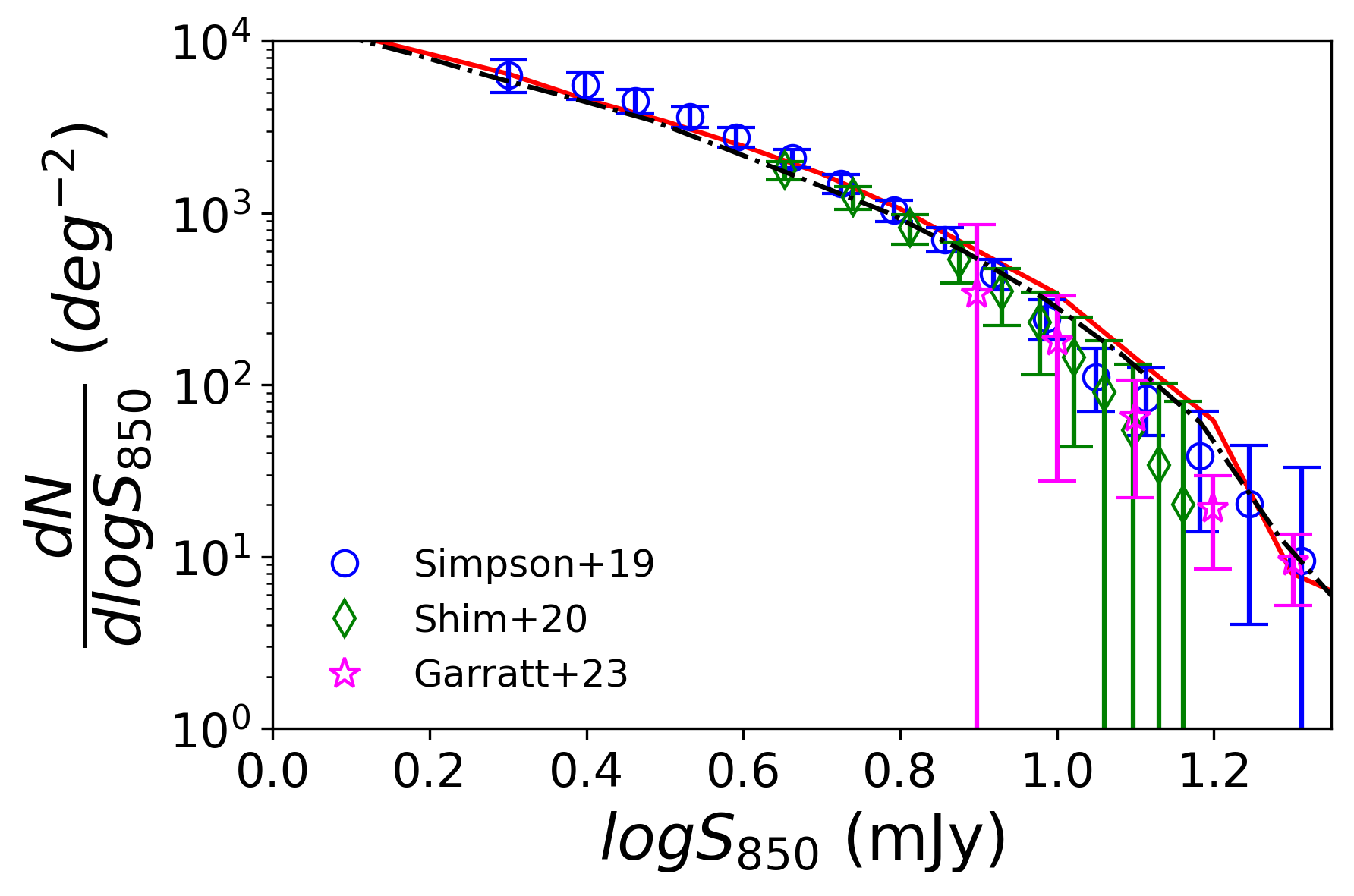}\hfill
\includegraphics[width=.47\textwidth]{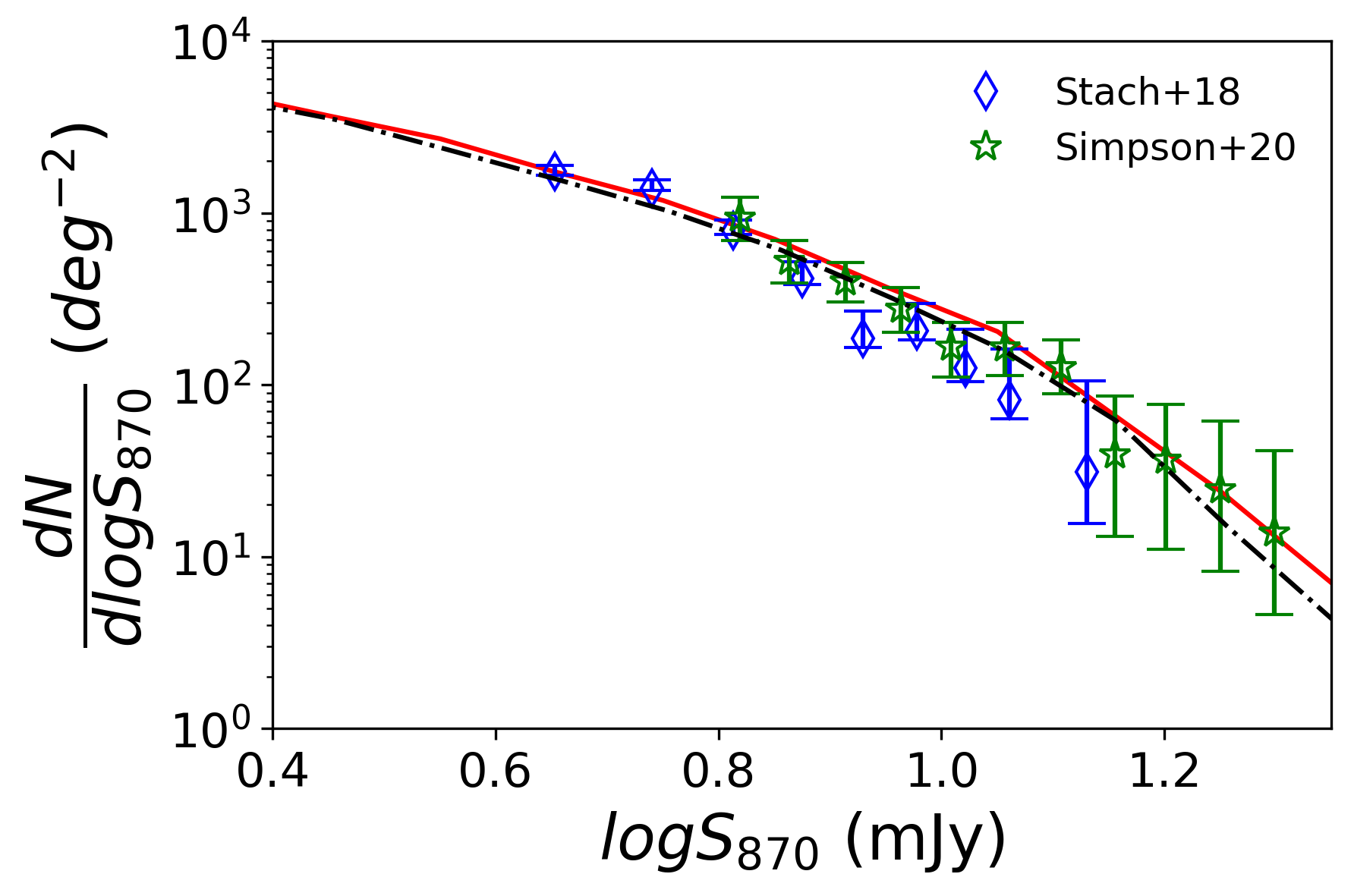}\hfill
\includegraphics[width=.47\textwidth]{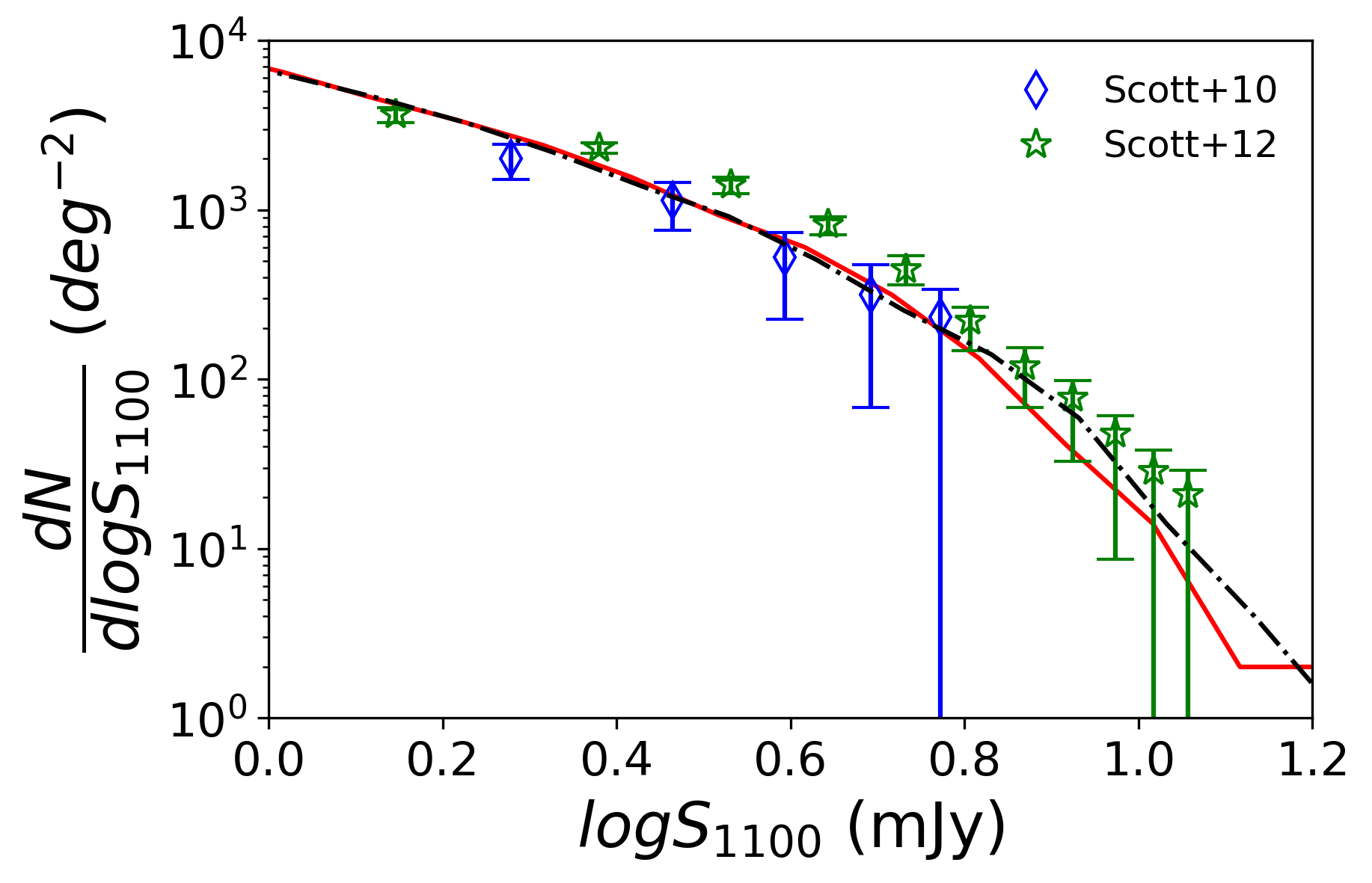}
\caption{Final version of the simulated differential number counts (solid red curve) compared to the C13 model prediction (dot-dashed black curve) and to observed data at $850\,\mu$m, $870\,\mu$m,and $1100\,\mu$m. The data points are taken from \citet{simpson_east_2019}, \citet{shim_nepsc2_2020}, \citet{garratt_scuba-2_2023},  \citet{stach_alma_2018}, \citet{simpson_alma_2020}, \citet{scott_deep_2010}, \citet{scott_source_2012}.}
\label{figdncfinal}
\end{figure}

Figure\,\ref{figdncirlbin} shows that, as expected, the bright tail of the number counts is made up by the intrinsically brightest objects, i.e. those with total infrared luminosity $L_{\rm IR}\gtrsim10^{13}\,L_{\odot}$. Therefore, we assigned a separate set of SED parameter values to these very bright objects. Because the dust temperature increases with increasing infrared luminosity \citep{hwang_evolution_2010-1}, we let $T_{w, BC}$ and $T_{c,ISM}$ to vary uniformly in the range 30-70\,K and 20-40\,K, respectively. At the same time, we also decreased the dust emissivity index of the warm dust grains to $\beta=1.8$. 
%
%
With these modifications, we obtained the number counts shown by the red solid curve in Fig.\,\ref{figdncfinal}. The agreement with the measured number counts is now satisfactory across the whole range of flux densities probed by the data.


\subsection{Step 4: generating the final sample}

With the adopted values of the SED parameters, which reproduce both the infrared luminosity function and number counts of proto-spheroidal galaxies, we generate the final catalogue of simulated objects used for forecast studies. 
We simulate proto-spheroidal galaxies over an area of $100\,$deg$^{2}$, which is of the same order of the wide area surveys conducted with {\it Herschel}.


\begin{figure}

\centering
\includegraphics[width=.47\textwidth]{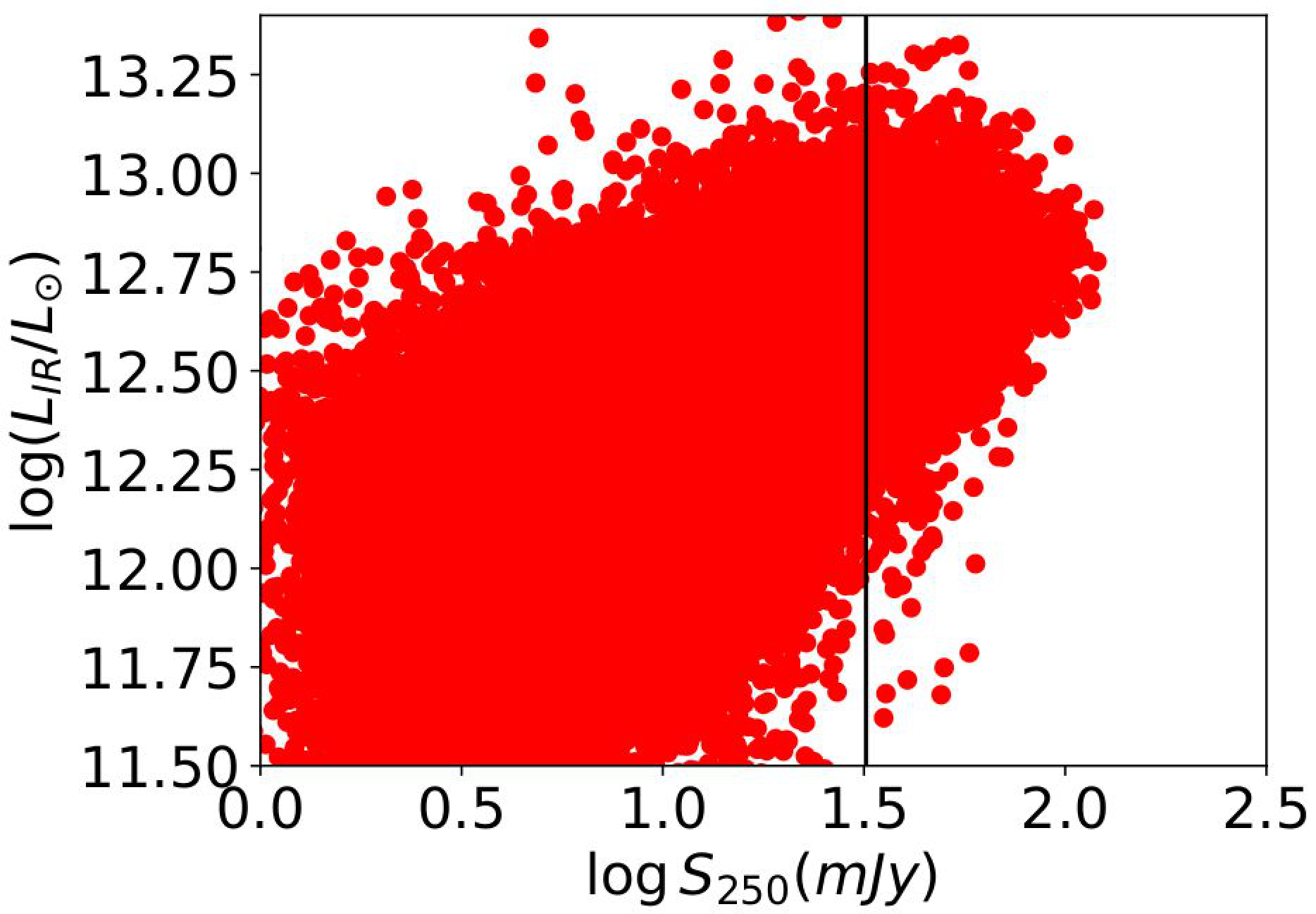}\hfill
\includegraphics[width=.47\textwidth]{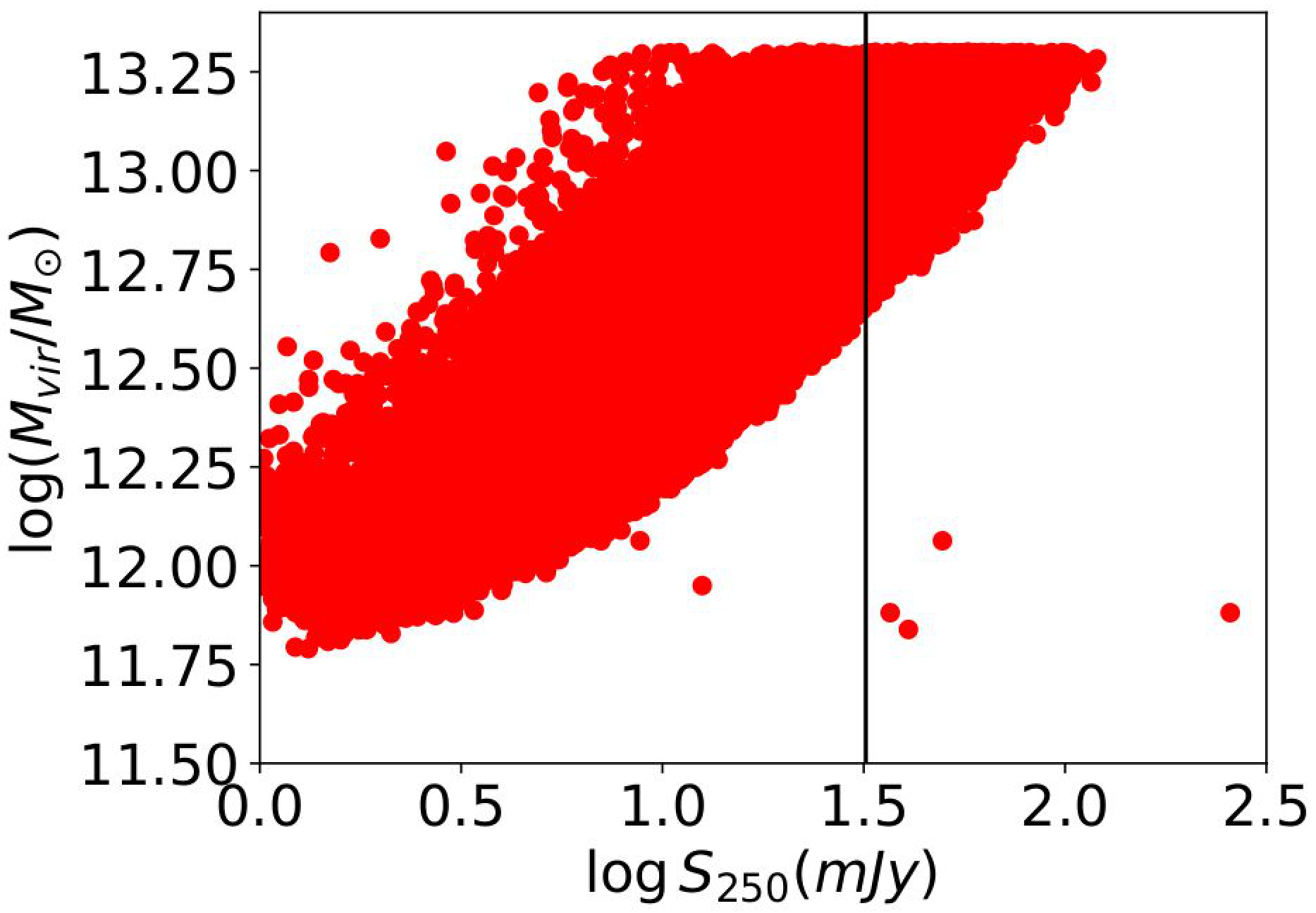}

\caption{Distribution of infrared luminosity, $L_{\rm IR}$, and virial mass, $M_{\rm vir}$, of the simulated proto-spheroidal galaxies as a function of their 250\,$\mu$m flux density, $S_{250}$. The vertical black line denotes the {\it Herschel}-ATLAS $5\,\sigma$ flux density limit of 37\,mJy at $250\,\mu$m.}
\label{figlimits}
\end{figure}

\section{Forecasts}
\label{secfor}

 We investigate the detectability of {\it Herschel}-selected sources in the {\it Euclid} wide area survey. To increase the wavelength coverage, we also test how many of these \textit{Euclid} sources are detected by the \textit{Rubin} observatory. We first provide a brief description of the {\it Herschel}, {\it Euclid} and \textit{Rubin's} wide-area surveys and then present the results of the analysis.

\subsection{Surveys}

\subsubsection{The Herschel-ATLAS}

The {\it Herschel-}ATLAS  \citep[H-ATLAS;][]{eales_herschel_2010} is a survey of 600\,deg$^2$ conducted in 5 photometric bands (100 $\mu$m, 160$\,\mu$m, 250$\,\mu$m, 350$\,\mu$m and 500$\,\mu$m), using the Photoconductor Array Camera and Spectrometer (PACS) and the Spectral and Photometric Imaging Receiver (SPIRE) cameras. The H-ATLAS observed fields in the northern and southern hemispheres and on the celestial equator. The fields were chosen to minimise the dust emission from the Milky Way. The observed fields are as follows:
\begin{enumerate}
    \item A field with an area of 150\,deg$^2$ close to the North Galactic Pole (NGP).
    \item Three equatorial fields each of approximate area of 56\,deg$^2$ coinciding with the previously surveyed fields in the Galactic and Mass Assembly (GAMA) spectroscopic survey.
    \item Two fields having a total area of 250\,$\deg^2$, close to the South Galactic Pole (SGP). 
\end{enumerate}

Because of the higher sensitivity to high-z sources of SPIRE compared to PACS, hereafter we will only focus on the SPIRE wavelengths. The $1\sigma$ noise for source detection at 250$\,\mu$m, 350$\,\mu$m and 500$\,\mu$m, is 7.4\,mJy, 9.4\,mJy and 10.2\,mJy, respectively.

\subsubsection{Euclid wide area survey}

The {\it Euclid} Wide Survey \citep[EWS;][]{euclid_collaboration_euclid_2022} will map 15000 sq. deg. of the sky from the Sun-Earth Lagrange point L2, using all four of its filters - visible ($I_E$) and NIR ($Y_E$, $J_E$ and $H_E$). The survey will have a $5\,\sigma$ depth of 26.2, 24.3, 24.5 and 24.4 mag for $I_E$, $Y_E$, $J_E$, and $H_E$, respectively. 

\subsubsection{Rubin's Legacy Survey of Space and Time (LSST)}

The \textit{Vera C. Rubin Obervatory}  \citep[previously known as Large Synoptic Survey Telescope;][]{ivezic_lsst_2019}, which is under construction in Cerro Pachón in Northern Chile, is planned to conduct a survey named \textit{Legacy Survey of Space and Time (LSST)} covering approximately 18\,000 sq. deg. of the sky in the optical in 6 bands - $u$, $g$, $r$, $i$, $z$, $y$. The survey will have a $5\,\sigma$ depth of 23.8, 24.5, 24.03, 23.41, 22.74 and 22.96 mag respectively.\\

A list of the \textit{Euclid} and \textit{Rubin} filters along with their central wavelengths and bandwidths is provided in Table \ref{tabnircam}.

\setlength{\tabcolsep}{11pt} 
\renewcommand{\arraystretch}{1.0} 
\begin{table}
\caption{Euclid filters along with their central wavelengths and bandwidths.}
\label{tabnircam}
\begin{tabular}{lcc}
\hline\hline
Filter & Central Wavelength & Bandwidth\\
&$\hspace{1ex}(\mu m)$ &$\hspace{1ex}(\mu m)$\\
\hline
$I_{E}$/VIS/Euclid & 0.731 & 0.378\\
$Y_{E}$/NISP/Euclid & 1.081 & 0.262\\
$J_{E}$/NISP/Euclid & 1.342 & 0.398\\
$H_{E}$/NISP/Euclid & 1.772 & 0.498\\
\textit{u}/Rubin & 0.358 &  0.068\\
\textit{g}/Rubin & 0.478 &  0.133\\
\textit{r}/Rubin & 0.623 &  0.127\\
\textit{i}/Rubin & 0.755 &  0.119\\
\textit{z}/Rubin & 0.871 &  0.101\\
\textit{y}/Rubin & 1.013 &  0.162\\
\hline
\end{tabular}
\end{table}

\subsection{Simulation set-up}


A simulated source is said to be detected by \textit{Herschel} if it has a flux density above the $5\,\sigma$ limit of 37\,mJy at $250\,\mu$m, and to be detected in one of the {\it Euclid} bands if it has a flux density higher than the $3\,\sigma$ limit at that band, i.e. 26.75, 24.85, 25.05, 24.95 mag for $I_E$, $Y_E$, $J_E$ and $H_E$, respectively. Moreover, we say that an \textit{Euclid} source is detected by \textit{LSST} if it has a flux density $>3\,\sigma$ in at least 4 out of the 6 bands.


To speed up the calculation, we simulated objects with a minimum virialization halo mass of $\log(M_{\rm vir}/M_{\odot})=12.5$, which corresponds to an infrared luminosity of $\log(L_{\rm IR}/L_{\odot})=12$, as illustrated in Fig.\,\ref{figlimits}. In fact, simulated sources with  $S_{250}=37\,$mJy have a minimum halo mass of $10^{12.5}\, M_{\odot}$ and a corresponding minimum IR luminosity of $10^{12}\, L_{\odot}$.

The simulated catalogue comprised 458\,994 galaxies prior to the application of any selection criterion. Approximately 33$\%$ of them (i.e. 151\,720 objects) have $S_{250}> 37\,$mJy. All of these \textit{Herschel} detected galaxies are also detected in all the four \textit{Euclid} bands above $3\,\sigma$.  Among these sources, 140\,524 have a 350$\,\mu$m flux density above 3\,$\sigma$ and 30\,337 are detected above $3\,\sigma$ at both 350$\,\mu$m and 500$\,\mu$m. Approximately $75\%$ of the \textit{Euclid} detected \textit{Herschel} galaxies are also detected by \textit{Rubin/LSST} above $3\,\sigma$ in at least 4 out of the 6 bands.

\subsection{Photometric redshifts}

The first thing needed to extract the intrinsic properties of galaxies is an estimate of the redshift. When spectra are not available, an estimate of the redshift can be obtained from the photometry. Because we are dealing with {\it Herschel} sources detected by {\it Euclid}, we can use either the optical/near-IR photometry from {\it Euclid} or the sub-mm {\it Herschel} photometry, or a combination of both to estimate the redshift.

\begin{figure}
    \centering
      \includegraphics[width=.47\textwidth]{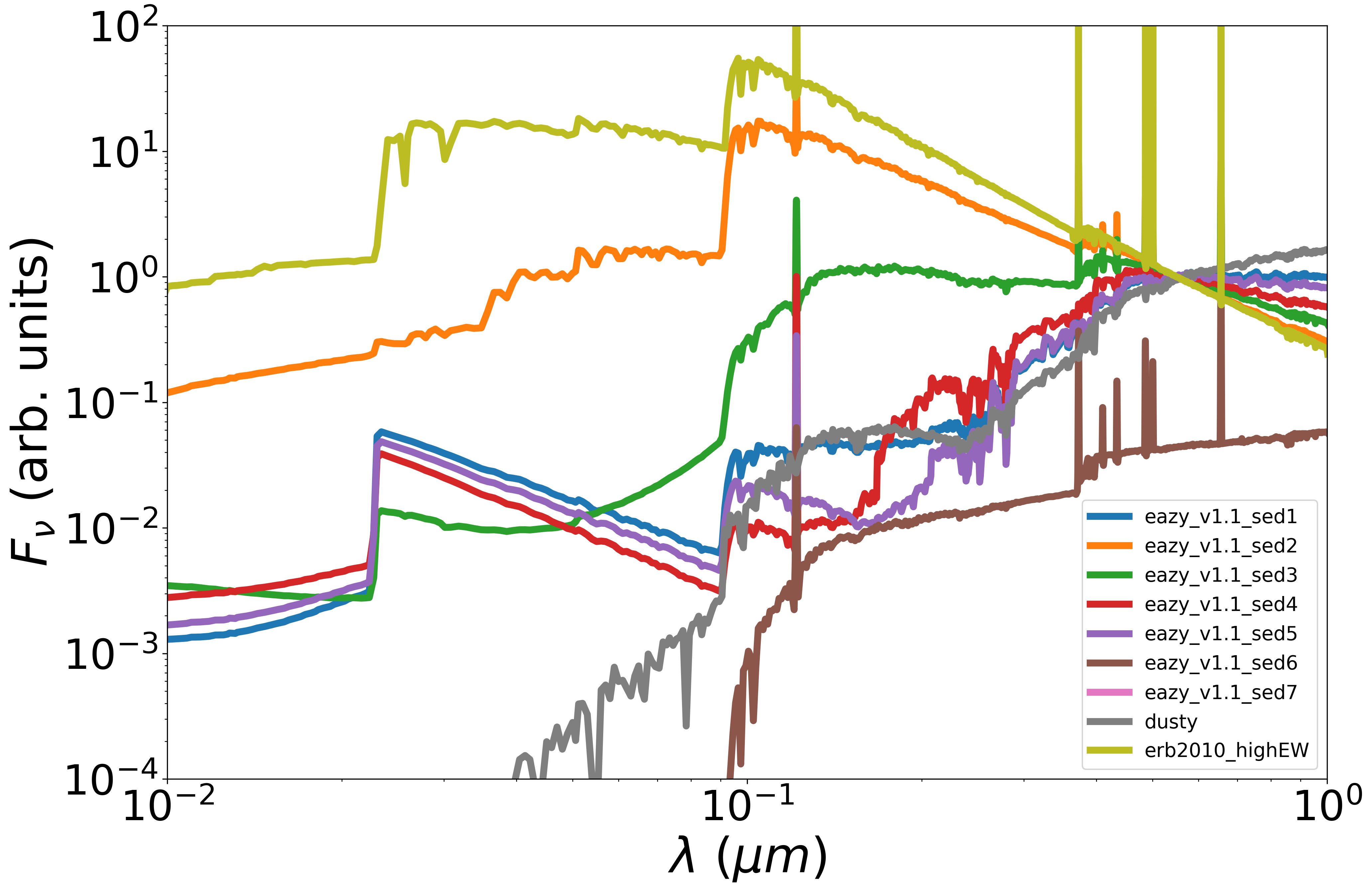}
      \caption{The set of EAZY SED templates used in this work.}
      \label{figeazysed}
\end{figure}

\subsubsection{Redshifts from the optical/near-IR Euclid photometry}


To estimate the photo-$z$ from the {\it Euclid} data we used the photometric redshift estimation code EAZY \citep{brammer_eazy_2008}, which is designed to work with UV/optical/near-IR data. EAZY fits the observed SED by linearly combining a number of template SEDs. It thus calculates the best-fit coefficients of the linear combination together with the photometric redshift. We used the EAZY version ``v1.3'' which comprises 9 (5+1+1+1+1) templates, shown in Figure \ref{figeazysed}.  The original ``v1.0'' version of EAZY comprises 5 ``principal component'' templates obtained from \cite{grazian_goods-music_2006} using the \cite{blanton_k-corrections_2007} algorithm. One dusty starburst template ($t=50$ Myr, $A_v=2.75$) was added to account for the extremely dusty galaxies. The new adopted version has these 6 templates modified to include line emissions. It also includes one dust template, a template taken from \cite{erb_physical_2010} and the evolved SSP by \cite{maraston_evolutionary_2005} to account for massive old galaxies at $z<1$.
Dust absorption from the intergalactic medium (IGM), in accordance with \cite{madau_radiative_1995}, is also included by EAZY during its calculations. We selected a redshift range from $z=1$ to 8 in steps of $\Delta z=0.01$. 

\subsubsection{Redshifts from the sub-mm photometry}

Another redshift estimate was derived from the {\it Herschel}/SPIRE photometry at $250\,\mu$m, $350\,\mu$m and $500\,\mu$m. We fitted it with the empirical SED of \cite{pearson_h-atlas_2013}, which is the sum of two modified blackbody spectra:
\begin{equation}
    F_{\nu} = N[\nu^{\beta}B_{\nu}(T_1)+\xi\nu^{\beta}B_{\nu}(T_2)]
\end{equation}
where, $F_{\nu}$ is the flux density at the restframe frequency $\nu$, $N$ is the normalisation factor, $B_{\nu}$ is the Planck function, $\beta$ is the dust emissivity index, $T_1$ is the hot dust temperature, $T_2$ is the cold dust temperature and $\xi$ is the ratio of the mass of the cold dust to that of the hot dust. We adopt $T_1=46.9\,$K, $T_2=23.9$\,K, $\beta=2$ and $\xi=30.1$ \citep{pearson_h-atlas_2013,negrello_herschel-atlas_2017}. The template SED was redshifted between $0\leq z\leq8$ in steps of $0.01$ and a $\chi^2$ minimisation was performed to estimate the photometric redshift.


\subsection{Estimating the galaxies properties}

\begin{table*}
\caption{Parameter values given as input to CIGALE for SED fitting}
\label{tabcigale}
\begin{tabular}{|ll|}
\hline\hline
\multicolumn{2}{|l|}{\bf{SFH : sfhdelayed - delayed SFH with optional exponential burst}}                                                                                                \\ \hline\hline
\multicolumn{1}{|l|}{e-folding time of the main stellar population (Myr)}    & 1000, 2000, 5000, 6000, 7000                                    \\ 
\multicolumn{1}{|l|}{e-folding time of the late starburst population (Myr)}  & 5000, 10000                                          \\ 
\multicolumn{1}{|l|}{mass fraction of the late starburst population}   & 0.0, 0.1, 0.15, 0.30                  \\ 
\multicolumn{1}{|l|}{age of the main stellar population (Myr)}               & 500, 1000, 2000, 3000, 5000, 6000, 7000, 8000, 9000, 10000 \\ 
\multicolumn{1}{|l|}{age of the late starburst (Myr)}                        & 1, 10, 30, 50, 70, 100, 150, 300                                   \\ \hline\hline
\multicolumn{2}{|l|}{\bf{SSP : bc03} \citep{bruzual_stellar_2003}}                                                                                                   \\ \hline\hline
\multicolumn{1}{|l|}{Initial mass function (IMF)}                            & Chabrier \citep{chabrier_galactic_2003}                                                 \\ 
\multicolumn{1}{|l|}{Metallicity}                                      & 0.02 ($Z_{\odot}$)                                                     \\ \hline\hline
\multicolumn{2}{|l|}{\bf{Dust attenuation :  dustatt\_modified\_CF00} \citep{charlot_simple_2000}}                                                                  \\ \hline\hline
\multicolumn{1}{|l|}{V-band attenuation in ISM ($A_{V}^{\rm ISM}$)}                        & 0.3, 1.7, 2.8                                \\ 
\multicolumn{1}{|l|}{$\mu$}                                            & 0.44, 0.5                                                      \\ 
\multicolumn{1}{|l|}{power law slope of attenuation in the ISM}        & -0.7                                                     \\ 
\multicolumn{1}{|l|}{power law slope of attenuation in the BCs}         & -0.7                                                     \\ \hline\hline
\multicolumn{2}{|l|}{\bf{Dust emission : dl2014} \citep{draine_andromedas_2014}}                                                                                       \\ \hline\hline
\multicolumn{1}{|l|}{PAH mass fraction ($q_{\rm \rm PAH}$)}                             & 2.5, 3.9, 4.58, 5.26                                         \\ 
\multicolumn{1}{|l|}{minimum radiation field ($U_{\rm min}$) (Habing)}                          & 5, 10, 15, 25, 40                                             \\ 
\multicolumn{1}{|l|}{dust emission power law slope ($\alpha$)}                          & 2                                                       \\ 
\multicolumn{1}{|l|}{fraction illuminated from $U_{\rm min}$ to $U_{\rm max}$ ($\gamma$)} & 0.02                                                       \\ \hline\hline
\multicolumn{2}{|l|}{\bf{AGN : fritz2006} \citep{fritz_revisiting_2006}}                                                                                       \\ \hline\hline
\multicolumn{1}{|l|}{ratio of the maximum to minimum radii of the dusty torus ($r_{ratio}$)}                             & 60, 150                                          \\ 
\multicolumn{1}{|l|}{equatorial optical depth at $9.7$ $\mu$m ($\tau$)}                          & 0.6, 1.0, 3.0, 6.0,10.0                                             \\ 
\multicolumn{1}{|l|}{radial dust distribution within the torus ($\beta$)}                          & -0.5, 0.0                                                       \\ 
\multicolumn{1}{|l|}{angular dust distribution within the torus ($\gamma$)} & 0, 6                                                       \\
\multicolumn{1}{|l|}{full opening angle of the dusty torus (Opening angle)} & 100, 140                                                       \\
\multicolumn{1}{|l|}{Angle between the equatorial axis and line of sight ($\Psi$)} & 0.001, 30.1, 60.1, 89.990                                                       \\
\multicolumn{1}{|l|}{AGN fraction ($f_{\rm AGN}$)} & 0.0, 0.1, 0.15, 0.25, 0.5                                                       \\\hline\hline\end{tabular}
\end{table*}

Once the photo-$z$ is determined, we estimate the physical properties of the simulated galaxies using the SED fitting code \textbf{CIGALE}  \citep[\textbf{C}ode \textbf{I}nvestigating \textbf{GAL}axy \textbf{E}mission;][]{burgarella_star_2005, noll_analysis_2009, boquien_cigale_2019}. \textbf{CIGALE} is a code written in \textit{Python} which is used to efficiently model the multi-wavelength (UV to radio) spectrum of galaxies and to estimate the physical properties of the galaxies. It is based on the principle of ``energy balance'', where the stellar energy absorbed by dust in the UV-optical is entirely re-emitted in the far-IR/sub-mm/mm.\\
CIGALE creates model SEDs based on the modules and parameter values given as input by the user. Once an SFH is chosen, CIGALE couples it with the stellar population synthesis models to create the stellar spectrum of the galaxies. Several SFH functions are available depending on the physics of star formation and on the morphology of the galaxies. In selecting the modules of CIGALE, we took help from \cite{traina_3cosmos_2023}, who studied 1620 sub-mm galaxies from $0.5\leq z\leq6$ by using the data from the largest available Atacama Large Millimeter Array (ALMA)\footnote{\url{https://www.almaobservatory.org/en/home/}} survey in the COSMOS field \citep{scoville_cosmic_2007}, known as $A^3$COSMOS \citep{liu_automated_2019}. We chose a delayed SFH with an additional late burst of star formation. For modelling the stellar spectrum, we adopted the Chabrier IMF \citep{chabrier_galactic_2003} and chose the SSP models by BC03. The metallicity was set at the solar value ($Z_\odot=0.02$). The dust attenuation was modelled using the attenuation law by \cite{charlot_simple_2000}. To model the dust emission, we adopted the templates from \cite{draine_andromedas_2014}. Finally, the AGN component was modelled using the templates from \cite{fritz_revisiting_2006}.\\ 
The chosen modules and parameter values, which are summarized in Table \ref{tabcigale}, produced a total of 39\,029\,760 SED templates. CIGALE considered all these templates to perform SED fitting by implementing a $\chi^2$ minimisation process \citep{noll_analysis_2009}. 

\begin{figure}
    \centering
      \includegraphics[width=.47\textwidth]{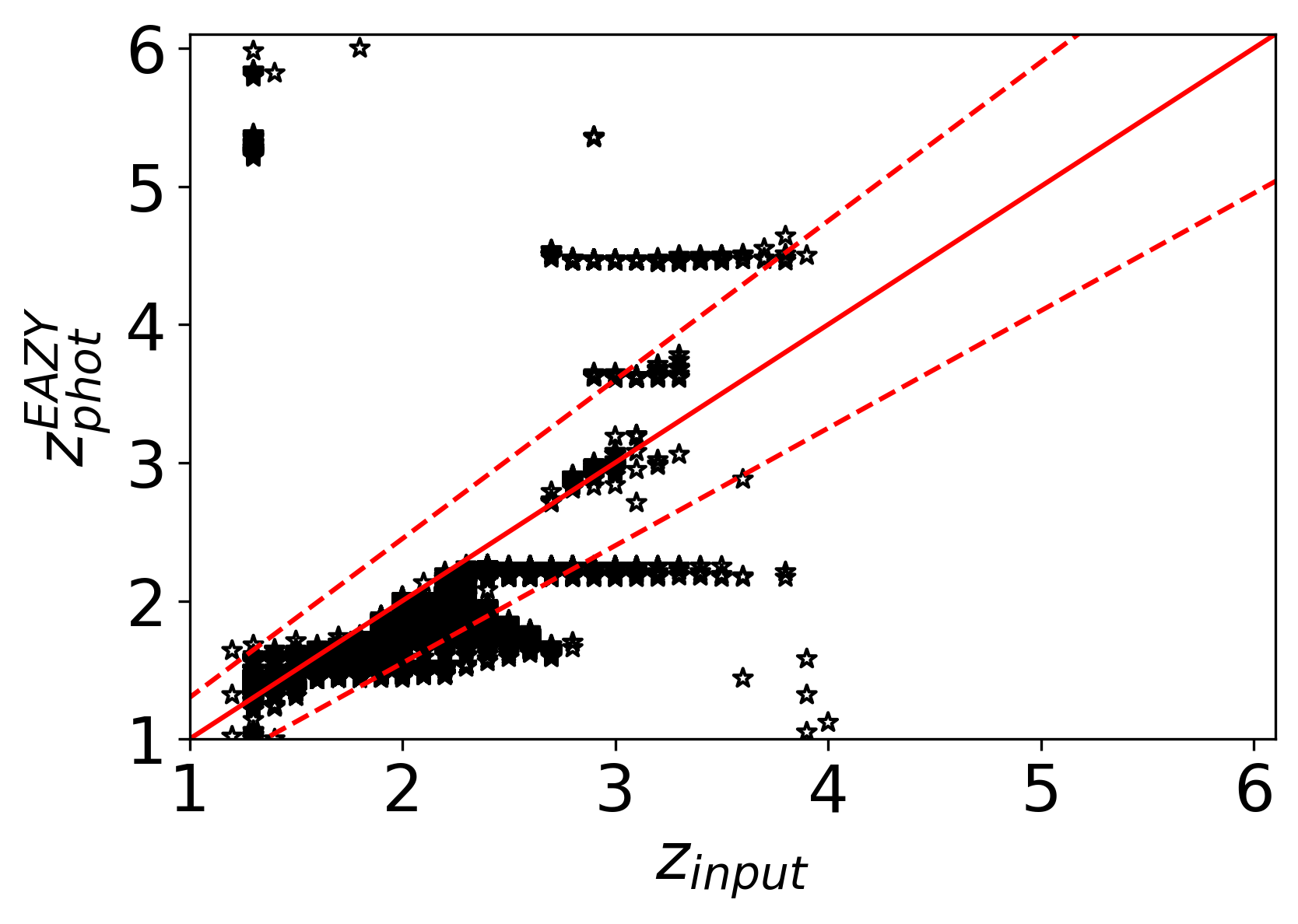}
      \caption{Derived photometric redshift using EAZY vs the input redshift of the galaxies detected by {\it Herschel} which are also detected by all the four bands of \textit{Euclid} at $\ge 3\,\sigma$. The solid red line denotes $z_{\rm input}=z_{\rm phot}^{\rm EAZY}$ while the dashed red lines define the region where $|\Delta z| \leq 0.15(1+z_{\rm input})$.}
      \label{figeazyphot}
\end{figure}


\section{Results}
\label{sec4}

This section presents the results obtained for the estimation of photometric redshift, stellar mass and SFR for the simulated catalogue for \textit{Euclid}, using the SED fitting technique.

\subsection{Derived photometric redshift}

\begin{figure*}

\centering
\centering
\includegraphics[width=.33\textwidth]{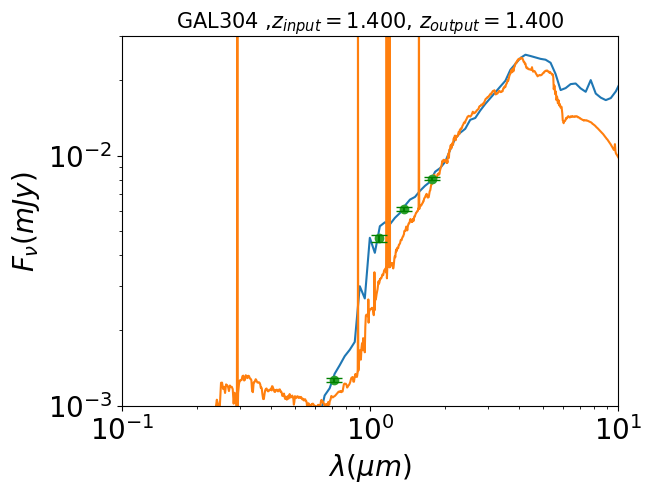}\hfill
\includegraphics[width=.33\textwidth]{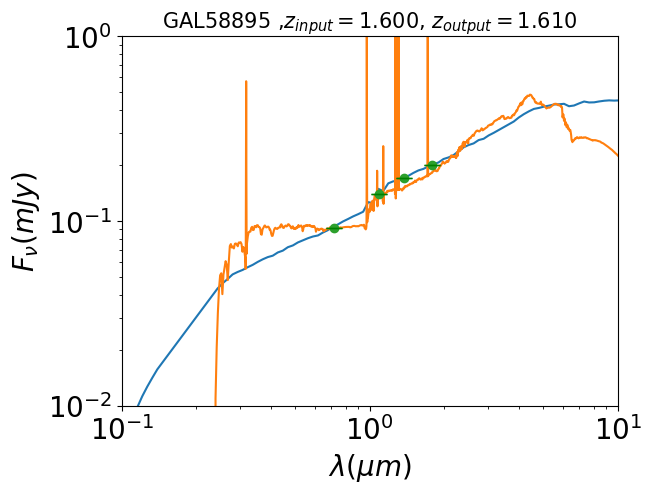}\hfill
\includegraphics[width=.33\textwidth]{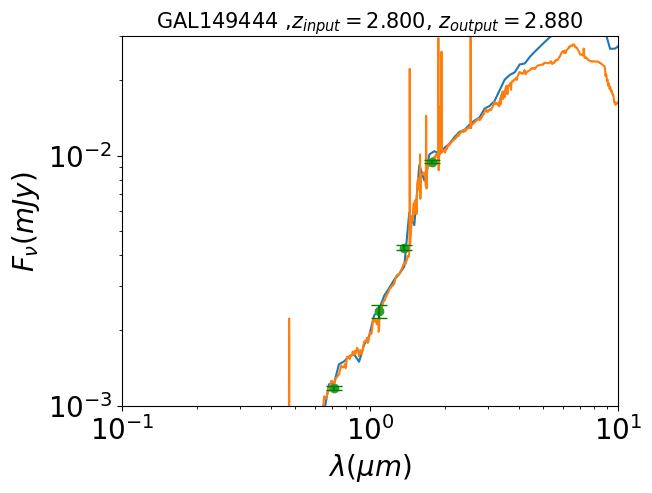}\hfill
\includegraphics[width=.33\textwidth]{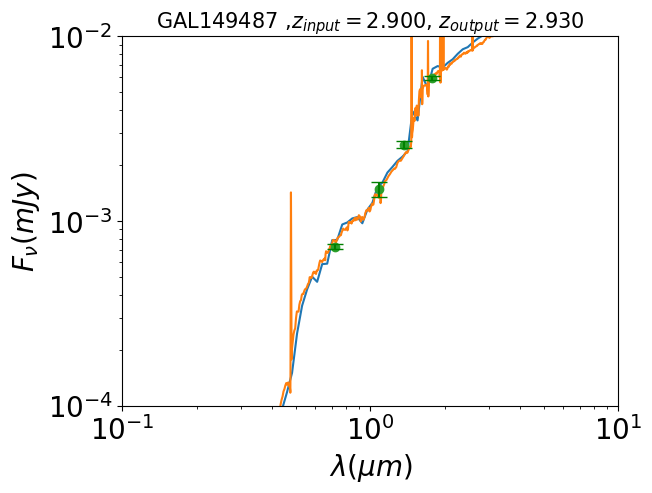}\hfill
\includegraphics[width=.33\textwidth]{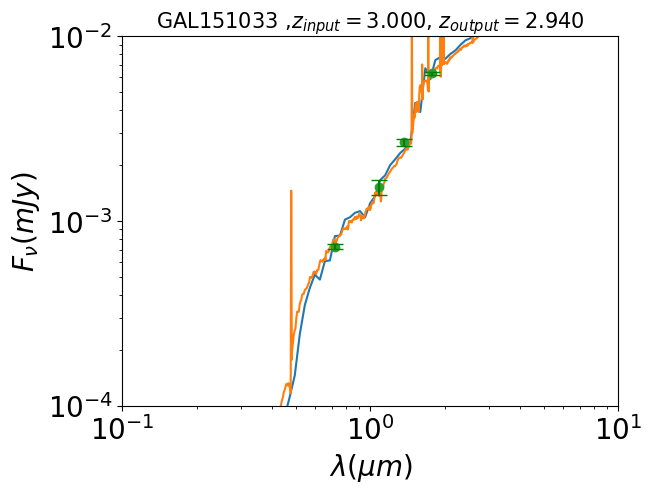}\hfill
\includegraphics[width=.33\textwidth]{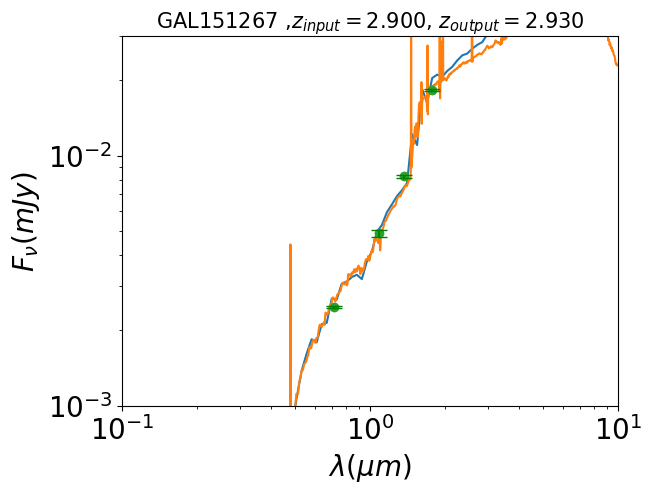}\hfill
\includegraphics[width=.33\textwidth]{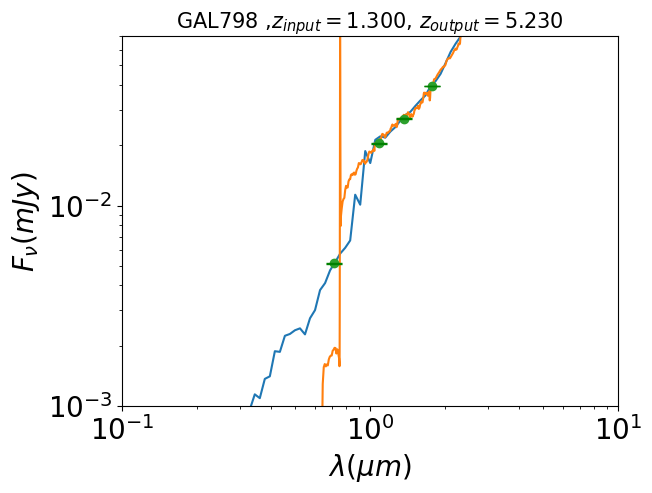}\hfill
\includegraphics[width=.33\textwidth]{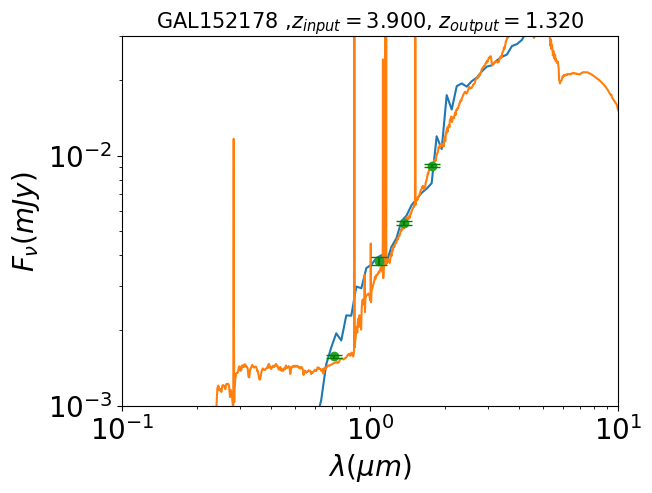}\hfill
\includegraphics[width=.33\textwidth]{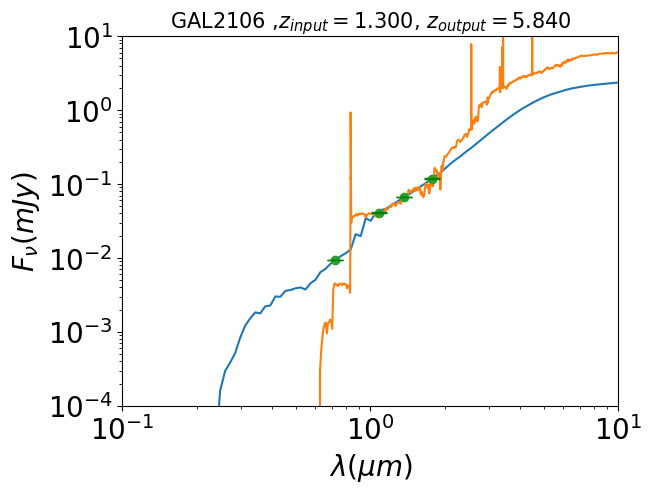}

\caption{Examples of simulated near-IR SEDs fitted with EAZY, illustrating the cases of both good and catastrophic redshift estimates. The blue curve denotes the simulated SED while the orange curve is the best-fit SED produced by EAZY. The green points denote the ``observed'' flux density values at the four {\it Euclid} bands.}
\label{figeazyplot}
\end{figure*}

\begin{figure}
    \centering
      \includegraphics[width=.47\textwidth]{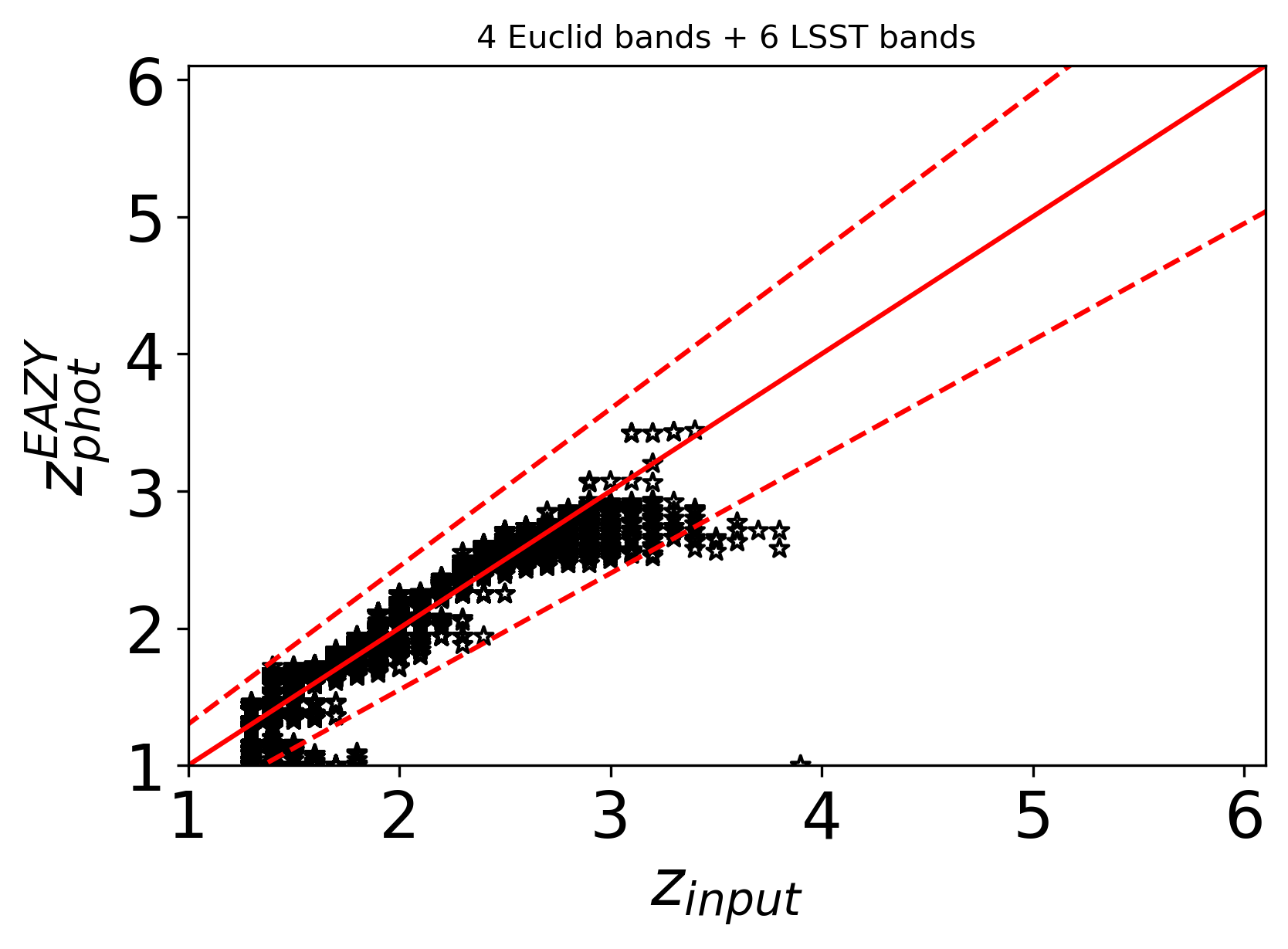}
      \caption{Derived photometric redshift using EAZY vs the input redshift of the galaxies detected by {\it Herschel} which are also detected by all the four bands of \textit{Euclid} and any four out of six bands of \textit{LSST} at $\ge 3\,\sigma$. The solid red line denotes $z_{\rm input}=z_{\rm phot}^{\rm EAZY}$ while the dashed red lines define the region where $|\Delta z| \leq 0.15(1+z_{\rm input})$.}
      \label{figeazyEL}
\end{figure}

In Figure \ref{figeazyphot}, we show the photometric redshift estimated with EASY versus the input redshift for the detected galaxies. There is a good level of agreement between the photometric redshift and the input redshifts for $z_{\rm input}\leq 3$.  The fraction of outliers, i.e. of objects with $|\Delta z|/(1+z_{\rm input}) > 0.15$ where $|\Delta z| = |z_{\rm input}-z_{\rm phot}^{\rm EAZY}|$, as commonly defined in the literature \citep[e.g.,][]{laigle_cosmos2015_2016}, is $f_{\rm outlier} = 0.128$. Most of the disastrous estimates of the photometric redshift are associated with galaxies at $z_{\rm input}\gtrsim3.5$, which are wrongly assigned a redshift in the range $z_{\rm phot}^{\rm EAZY}\sim 1-1.5$. Also, there are some galaxies with input redshift $z_{\rm input}\sim1-2$ which are estimated to be at much higher redshift, i.e. $z_{\rm EAZY}\sim5-6$. 
Figure \ref{figeazyplot} shows some of the best-fit SEDs produced by EAZY along with the corresponding estimated redshift. 
In order to achieve a good photometric redshift, EAZY mainly exploits the 4000\,Å break. At low redshifts (i.e. $z_{\rm input}\lesssim 3$), such feature falls within the wavelength range sampled by {\it Euclid}, thus leading to photometric redshifts that agree with the input values.
However, at higher redshifts, the 4000\,Å break is missed by {\it Euclid} thus causing catastrophic redshift estimates. Moreover, due to the limited wavelength coverage, the 4000 Å break is wrongly identified as the Lyman$-\alpha$ break at $912$ Å, which also leads to a wrong redshift estimate. 
The addition of filters blueward of \textit{Euclid} can help to identify the Lyman$-\alpha$ break, thus leading to a more accurate redshift estimate. This is observed when the \textit{Euclid} photometry is complemented with the \textit{LSST} data. Figure \ref{figeazyEL} shows a plot of the input redshift vs the photometric redshift obtained by EAZY from the \textit{Euclid} +\textit{LSST} photometry. Out of the 113\,897 \textit{Euclid} sources that are also detected by {\it LSST} in at least four bands, 112\,831 (approximately $99\%$) have $|\Delta z|/(1+z_{\rm input}) \leq 0.15$. The outlier fraction $f_{\rm outlier}$ is thus down to just $1\%$.\\

\begin{figure}
    \centering
      \includegraphics[width=.47\textwidth]{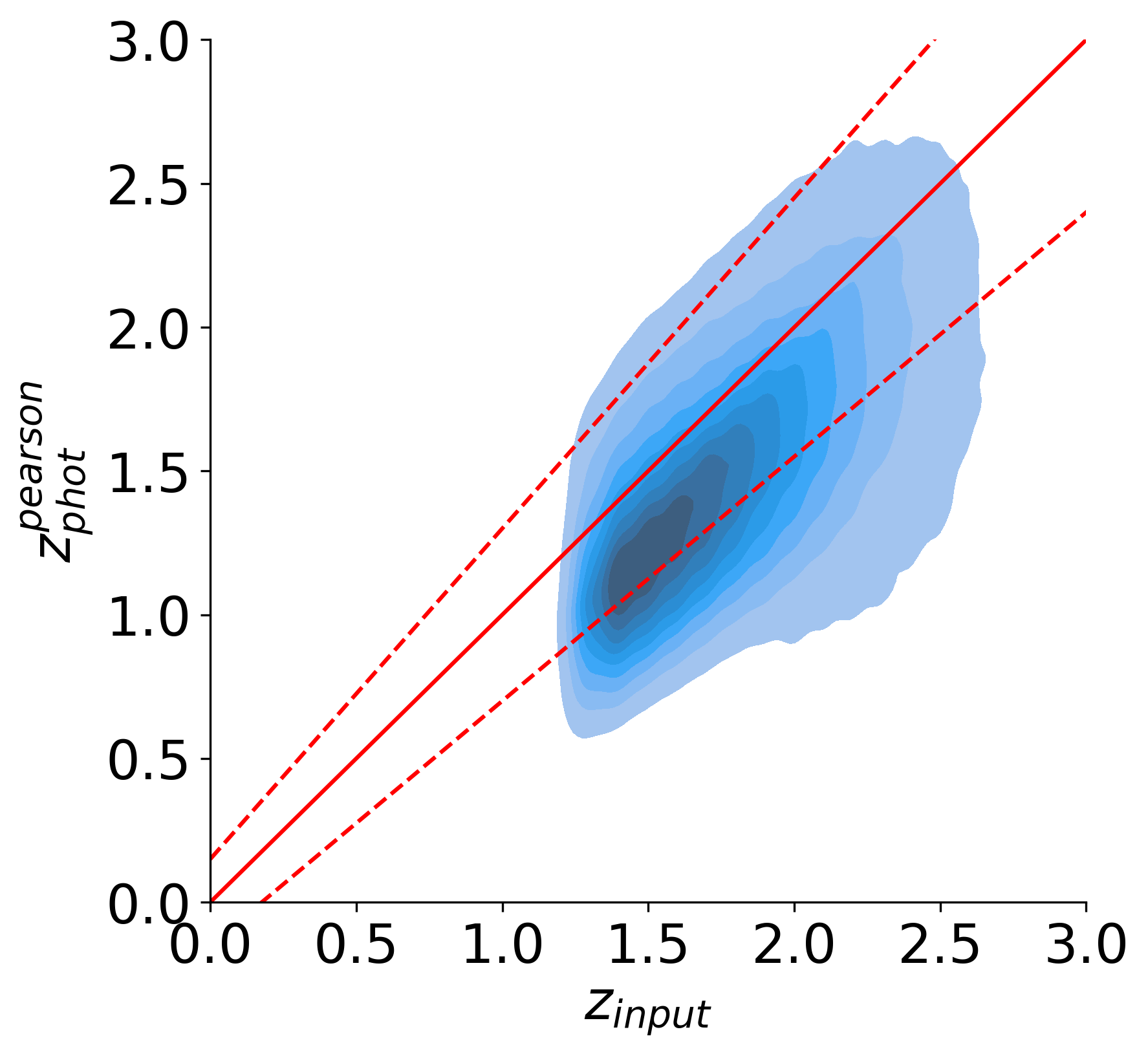}
      \caption{Contour plot of the derived photometric redshifts using Pearson templates on \textit{Herschel} photometry vs the input redshift of the \textit{Euclid} detected \textit{Herschel} galaxies. The red solid line denotes $z_{\rm input}=z_{\rm phot}^{\rm pearson}$ while the red dashed lines bound the region where $|\Delta z| \leq 0.15(1+z_{\rm input})$.}
      \label{figpearphot}
\end{figure}

\begin{figure}
    \centering
      \includegraphics[width=.47\textwidth]{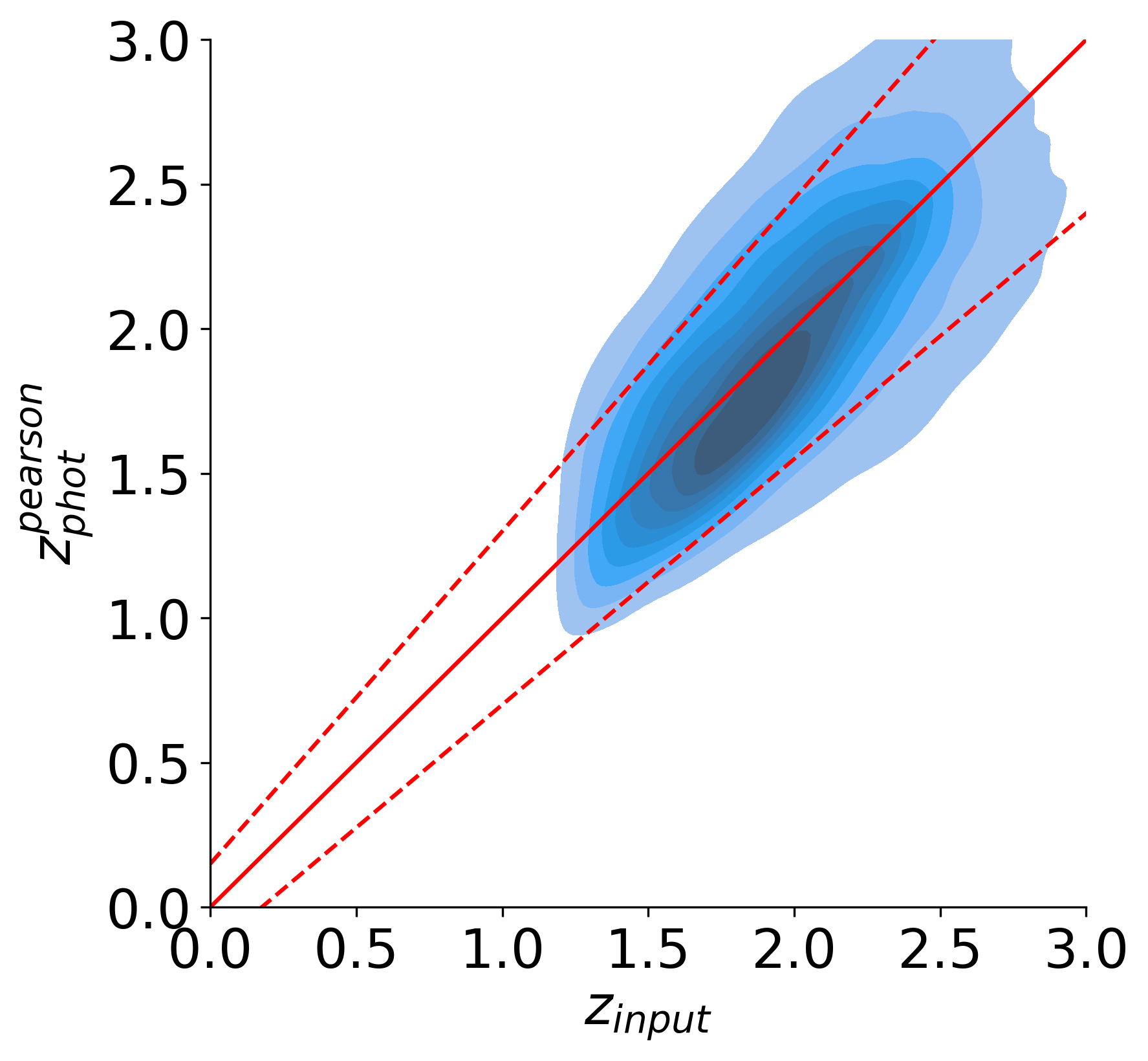}
      \caption{Contour plot of the derived photometric redshifts  using Pearson templates on \textit{Herschel} photometry vs the input redshift of the \textit{Herschel}/\textit{Euclid} galaxies detected above $3\,\sigma$ at both $350\,\mu$m and $500\,\mu$m. The solid red line denotes $z_{\rm input}=z_{\rm phot}^{\rm pearson}$ while the dashed  red lines bound the region where $|\Delta z| \leq 0.15(1+z_{\rm input})$.}
      \label{figpearphot1}
\end{figure}

 Figure \ref{figpearphot} shows the contour plot of the derived photometric redshift of the same galaxies using the \textit{Herschel}/SPIRE photometry alone. The fraction of outlier is $f_{\rm outlier} = 0.37$, more than twice the one obtained from the {\it Euclid} photometry. Moreover, there are 47\,618 sources for which the Pearson template gives ${|\Delta z|}/{(1+z_{\rm input})}> 0.15$ while EAZY produces ${|\Delta z|}/{(1+z_{\rm input})}\leq 0.15$. At the same time, there are 10\,623 sources with ${|\Delta z|}/{(1+z_{\rm input})}\leq 0.15$ when using the Pearson template while EAZY produces ${|\Delta z|}/{(1+z_{\rm input})}> 0.15$. Therefore, in general, the near-IR data seem to be more successful at estimating the redshift of the simulated galaxies compared to the \textit{Herschel} photometry.  However, it is worth pointing out that 
 for the objects which are also detected above $3\,\sigma$ at both $350\,\mu$m and $500\,\mu$m (only 20\% of the sample), the sub-mm photometry provides a better estimate of the photometric redshift as illustrated in Figure \ref{figpearphot1}. 
 Therefore, the inaccuracy of the sub-mm photometry in estimating the redshift for the bulk of the objects can be attributed to the absence of robust flux density measurements at $350\,\mu$m and $500\,\mu$m. 
 
 In the following analysis, we will rely on the photometric redshift estimated by EAZY from the \textit{Euclid} photometry.

\subsection{Derived physical properties}

In order to extract the physical properties of the simulated galaxies, such as the stellar mass, $M_{\star}$, and the star formation rate, SFR, we fitted the SED of the simulated {\it Herschel} galaxies detected with {\it Euclid} by using CIGALE and adopting the photometric redshift produced by EAZY. In the following sections, we discuss the results obtained using the {\it Euclid} photometry alone and the {\it Euclid} plus {\it Herschel}/SPIRE photometry combined. We also analyse the improvement of the above when including the \textit{LSST} photometry. 

\subsubsection{Stellar Mass}

\begin{figure*}

\centering
\centering
\includegraphics[width=.3\textwidth]{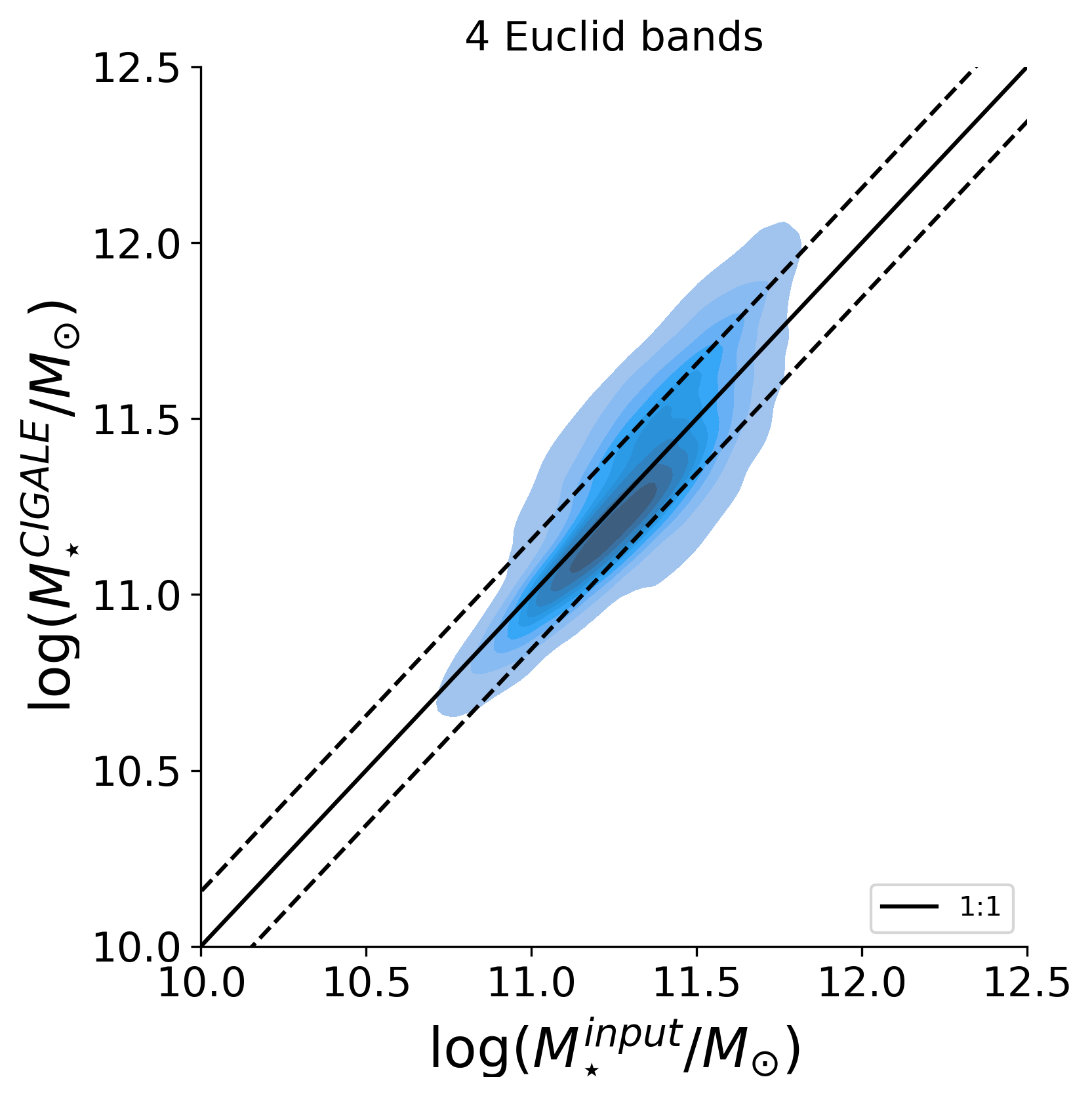}\hfill
\includegraphics[width=.3\textwidth]{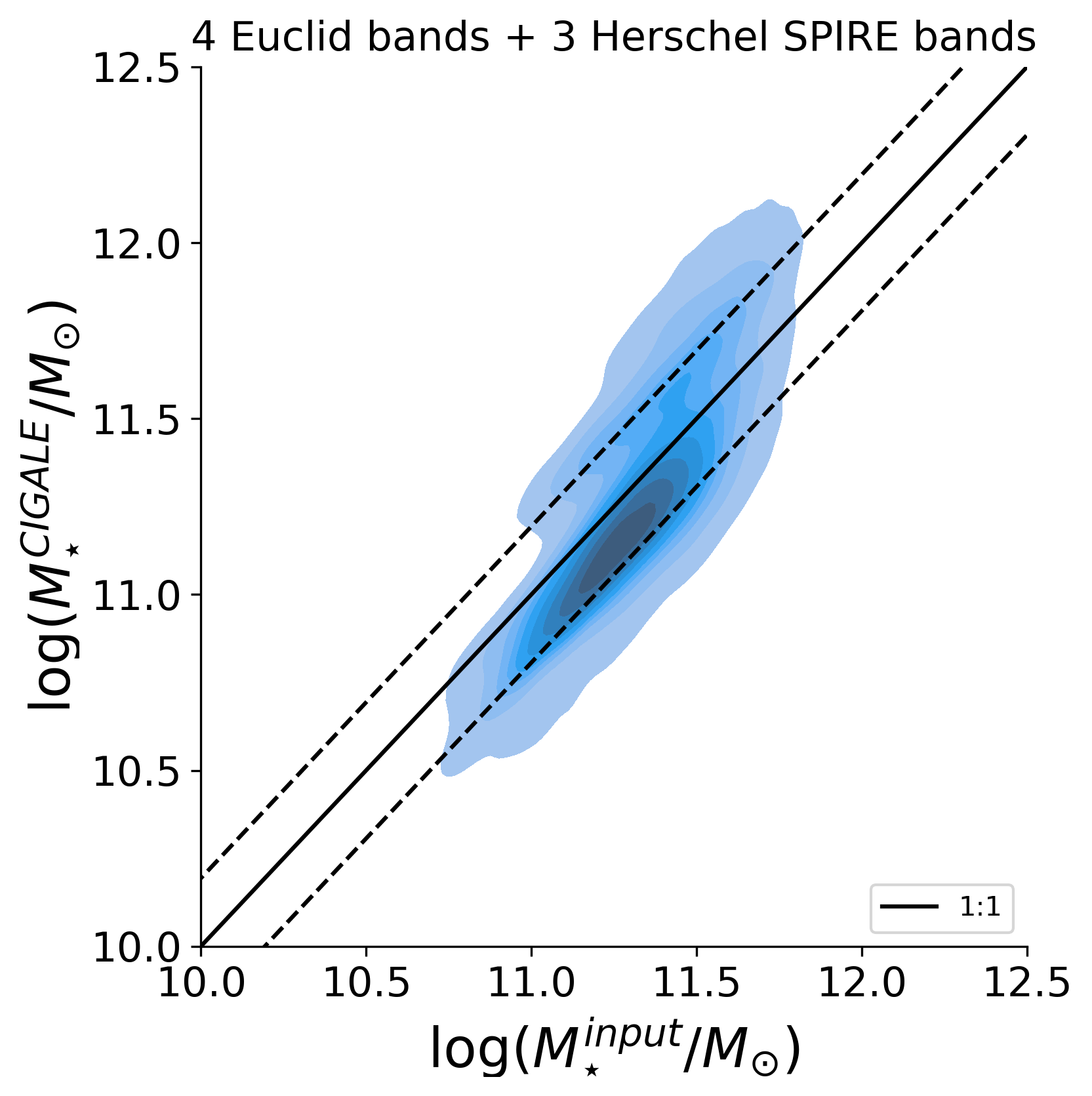}\hfill
\includegraphics[width=.3\textwidth]{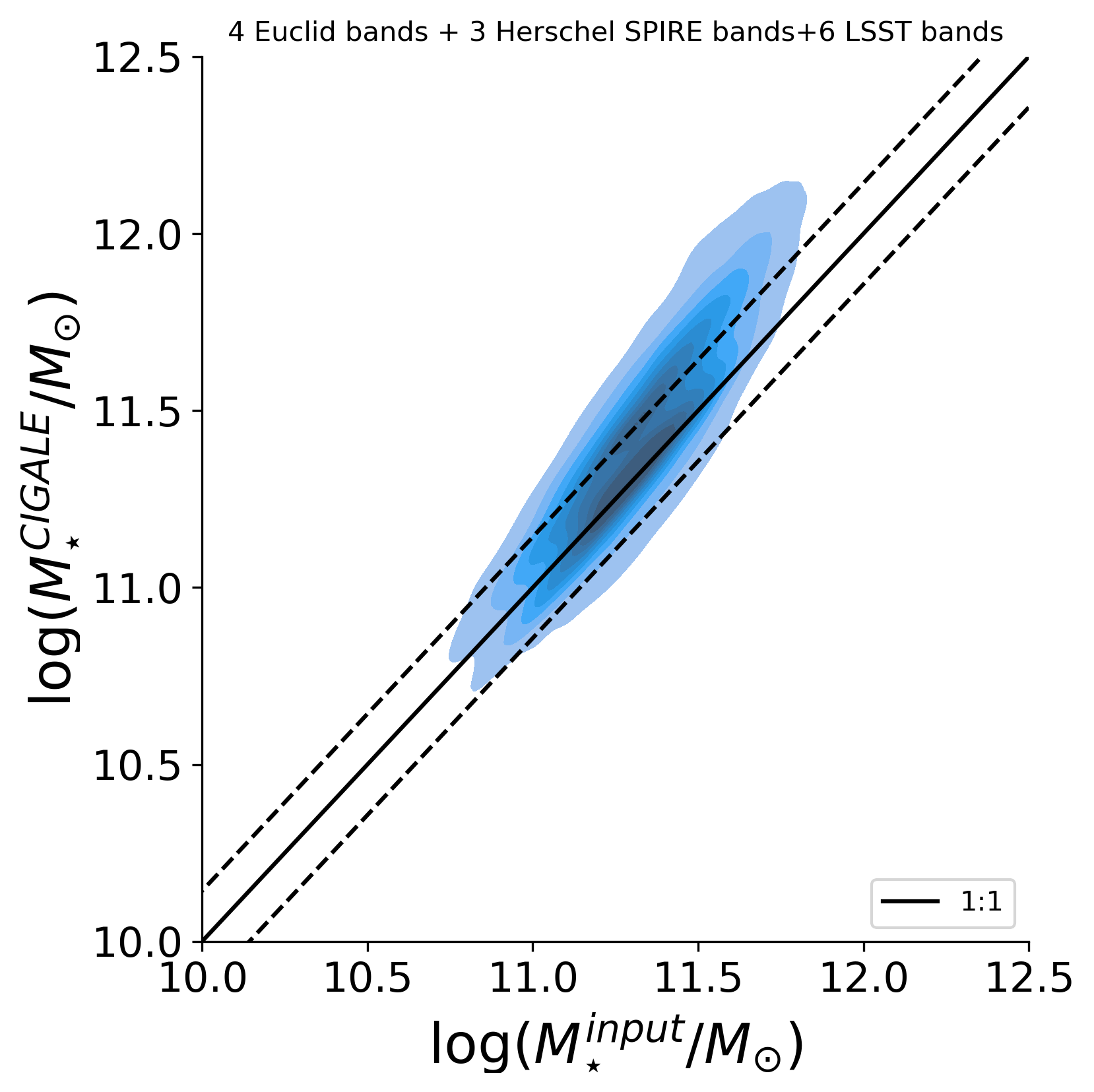}\hfill
\includegraphics[width=.3\textwidth]{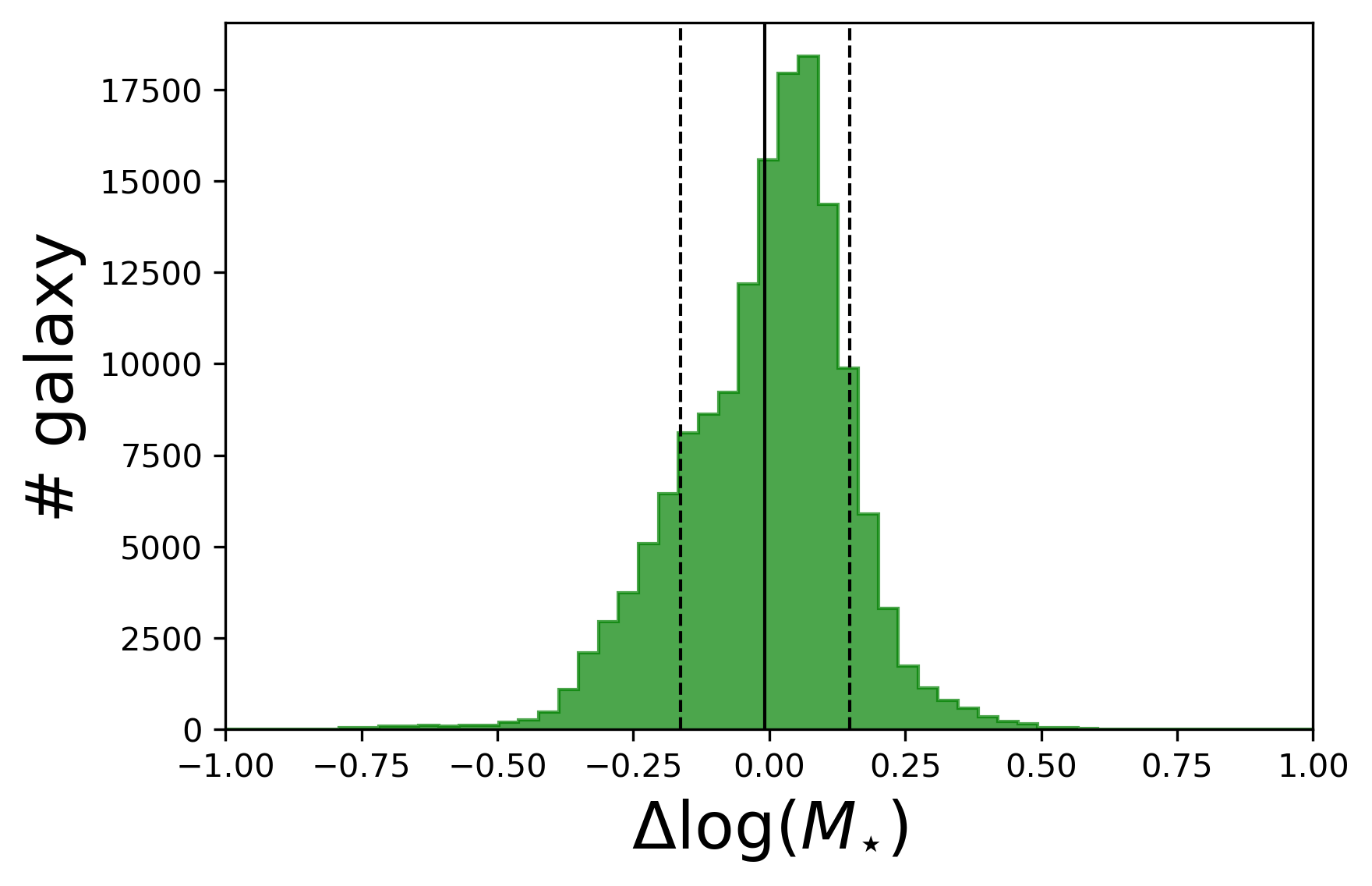}\hfill
\includegraphics[width=.3\textwidth]{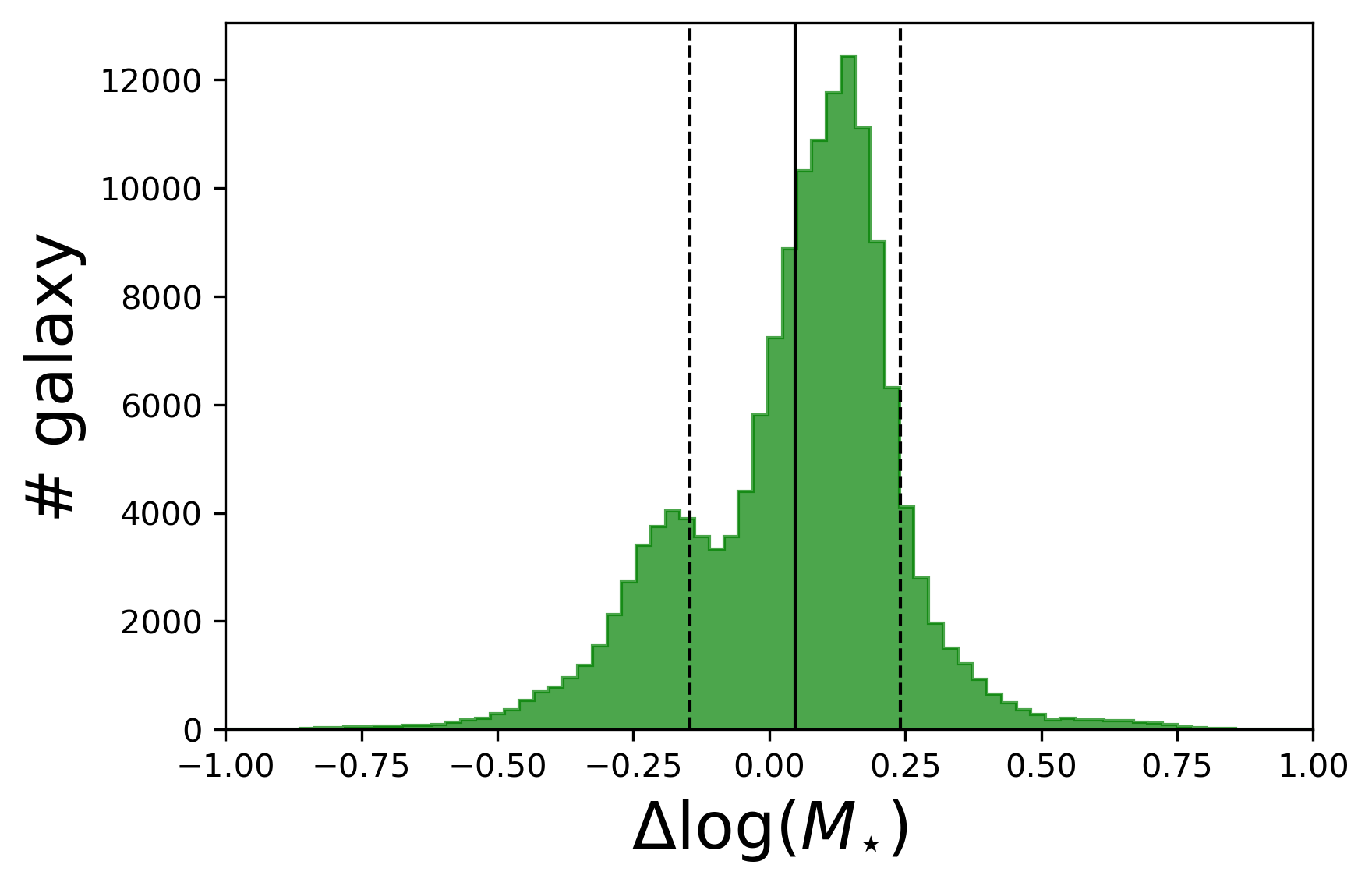}\hfill
\includegraphics[width=.3\textwidth]{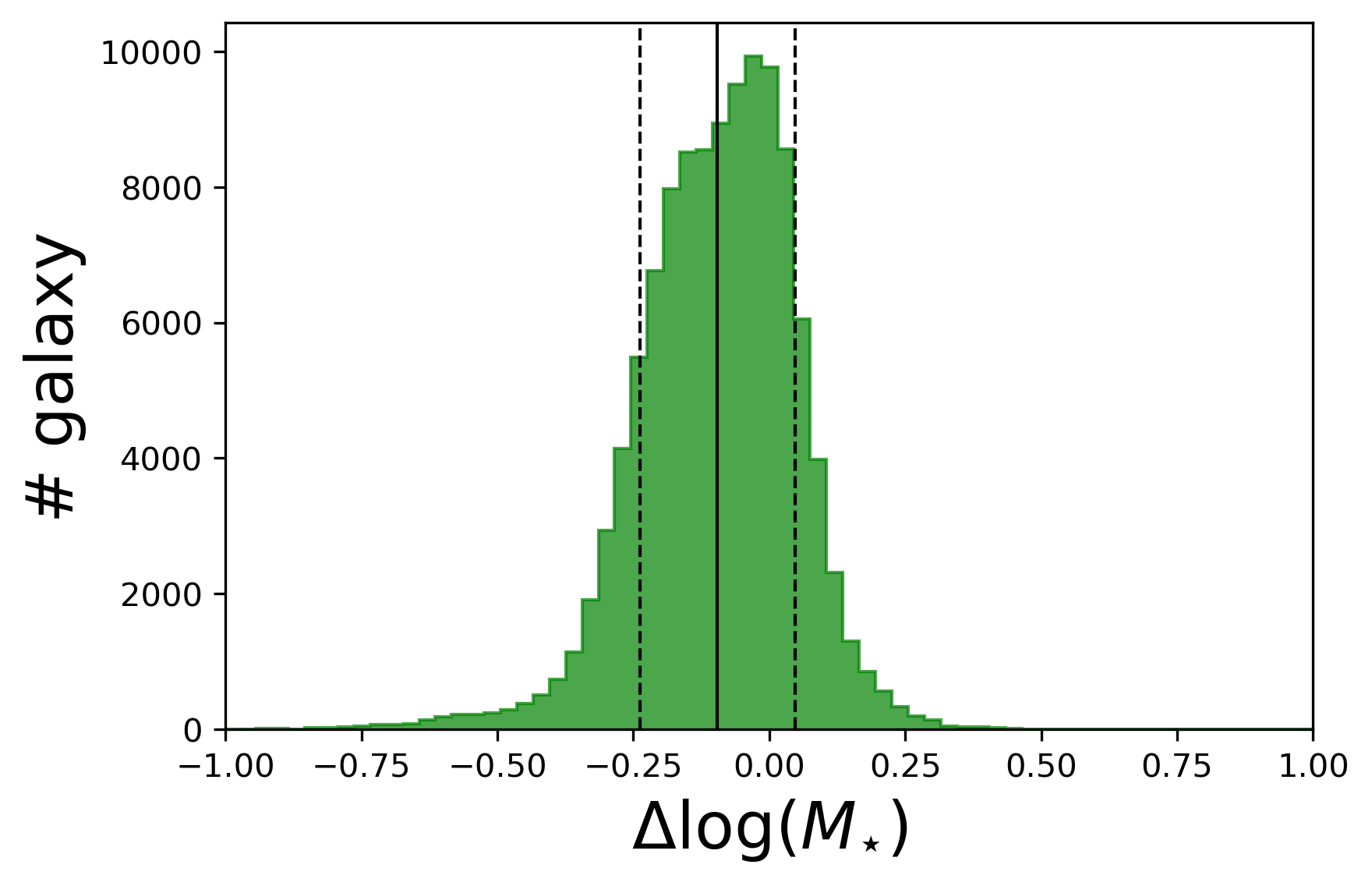}
\caption{Comparison between the input and the estimated values of the stellar masses obtained with CIGALE from \textit{Euclid} photometry alone (left panels), {\it Euclid} + {\it Herschel} photometry (middle panels) and {\it Euclid} + {\it Herschel}+ \textit{LSST} photometry (right panels). The top panels show the contour of the estimated vs the true stellar masses for both cases. The black straight line marks the 1:1 relation. The black dashed line marks the boundaries of the $1\,\sigma$ dispersion around the mean. The bottom panels show the distribution of differences between the logarithms of the input stellar mass and of the estimated stellar mass. The black solid line denotes the mean value while the dashed lines mark the $1\,\sigma$ dispersion around the mean.}
\label{figcigalesm}
\end{figure*}

Figure \ref{figcigalesm} shows the comparison between the derived stellar masses, $M_{\star}^{\rm CIGALE}$, and the input ones, $M_{\star}^{\rm input}$. 
Also shown in the figure is the histogram of the difference between the logarithm of the stellar masses defined as $\Delta\log (M_{\star}) \equiv \log(M_{\star}^{\rm input})-\log (M_{\star}^{\rm CIGALE})$. 
Overall, the retrieved stellar mass is in good agreement with the input values. The median value of the stellar mass is found to be $\log(M_{\star}/M_{\odot})=11.29\pm0.181$. The 1\,$\sigma$ dispersion in 
$\Delta\log (M_{\star})$ are $0.16$ and $0.19$ from the \textit{Euclid} and \textit{Euclid}+{\it Herschel} photometry, respectively. These dispersions are mainly a reflection of the uncertainties in the derived photometric redshifts. 
As expected, because the evolved stellar populations $-$ which contribute most of the stellar mass in a galaxy $-$ dominate the rest-frame optical/near-IR part of the spectrum, the stellar mass estimates produced by CIGALE do not improve when the SED fitting also includes the sub-mm {\it Herschel} data. However, the addition of the \textit{LSST} data does improve the stellar mass estimate, by reducing the   
$1\,\sigma$ dispersion in 
$\Delta\log (M_{\star})$ to $0.14$.

We observe that, on average, CIGALE overestimates the stellar mass, the effect being particularly relevant for the higher masses. This offset can be attributed to the choice of the SFH and its parameterisation. Here, we choose a multi-component SFH, which, in general, has a tendency to yield a higher mass-to-light ratio, hence higher stellar mass estimates, as pointed out by \cite{michalowski_stellar_2012}. The discrepancy in stellar mass estimates could also be 
explained by the difference in the value of the dust attenuation law in birth clouds used in CIGALE versus the one adopted here, i.e. -0.7 versus -1.3, respectively, and therefore to the extension of the extinction curve to the near-IR, where it directly impacts the stellar mass.




\subsubsection{SFR}

\begin{figure*}
    \centering
      \centering
\includegraphics[width=.5\textwidth]{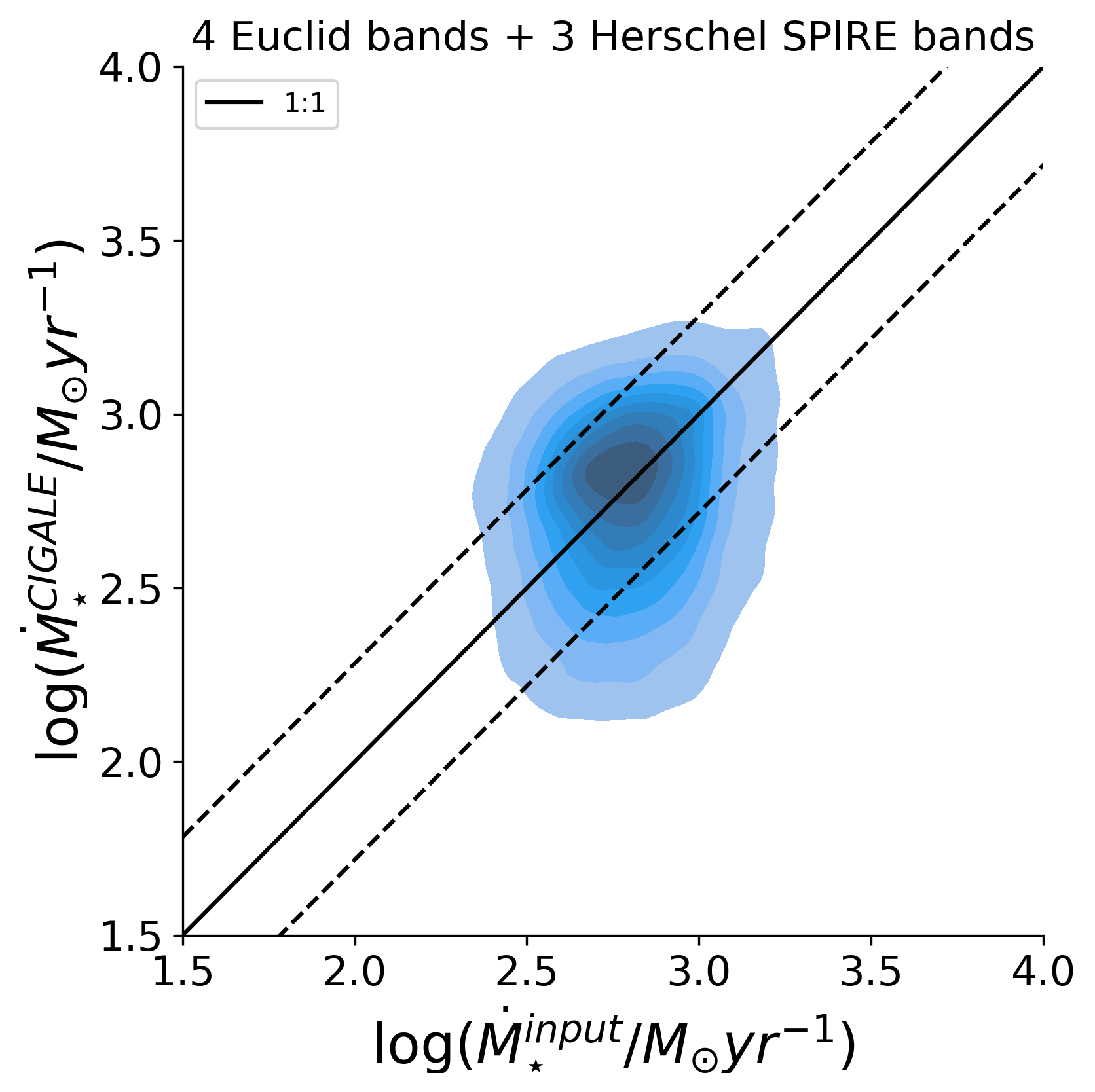}\hfill
\includegraphics[width=.5\textwidth]{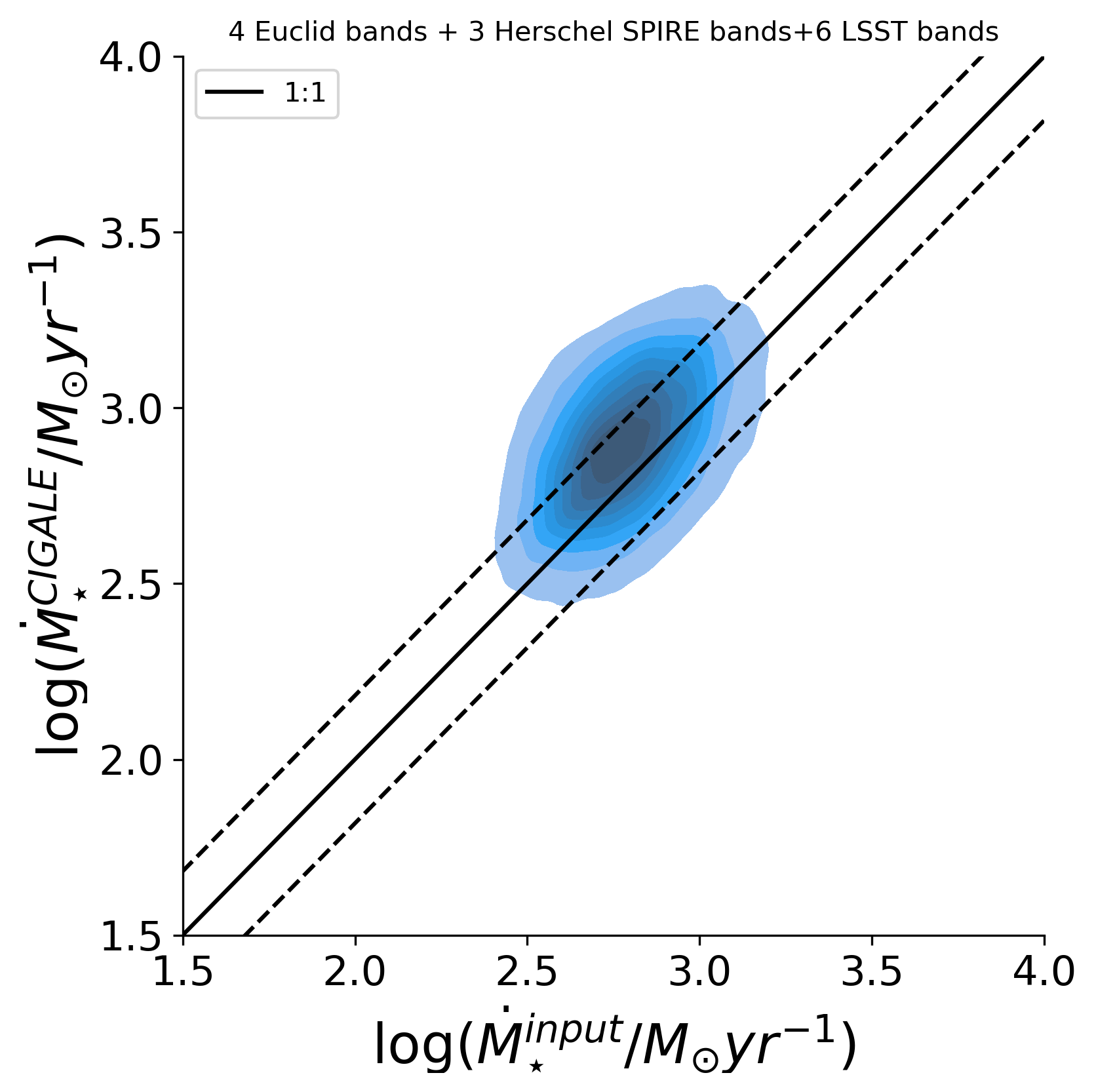}\hfill
\includegraphics[width=.5\textwidth]{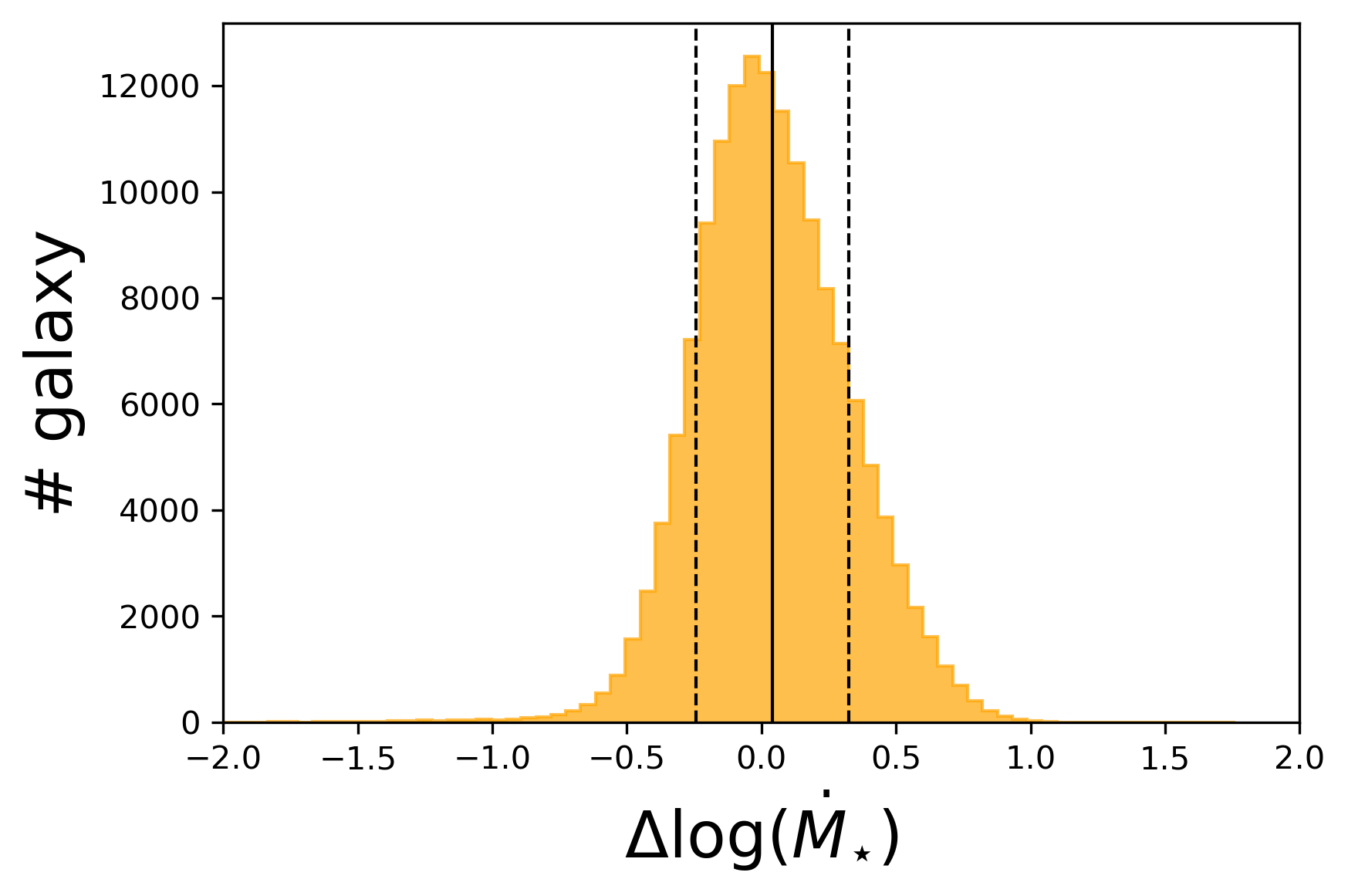}\hfill
\includegraphics[width=.5\textwidth]{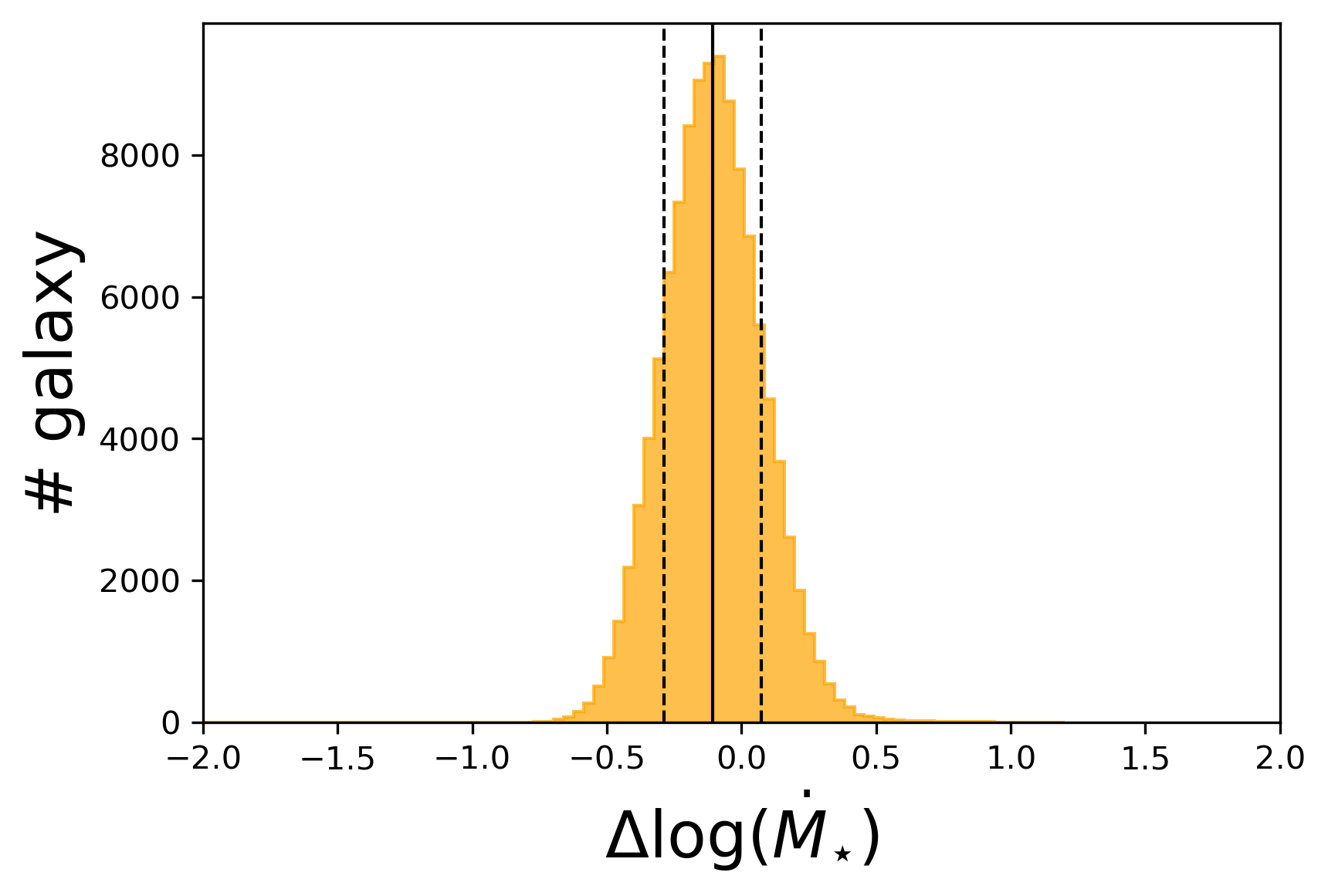}\hfill

      \caption{Same as Figure \ref{figcigalesm}, but for the SFR, with the SED fit being performed on \textit{Euclid}+\textit{Herschel}/SPIRE photometry (left) and \textit{Euclid}+\textit{Herschel}/SPIRE+\textit{LSST} photometry (right).}
      \label{figcigsfr}
\end{figure*}

\begin{figure*}
    \centering
      \centering
\includegraphics[width=.25\textwidth]{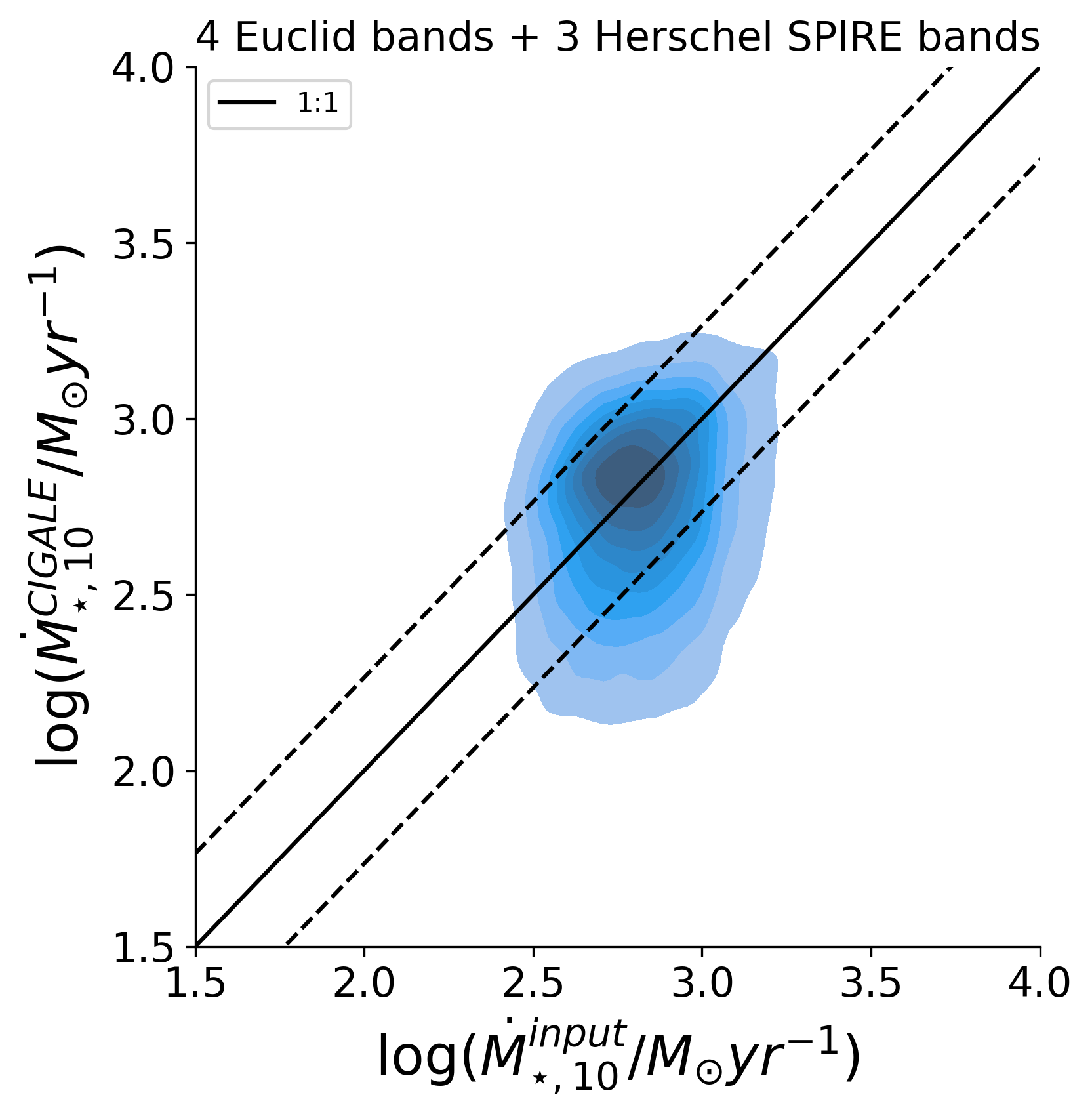}\hfill
\includegraphics[width=.25\textwidth]{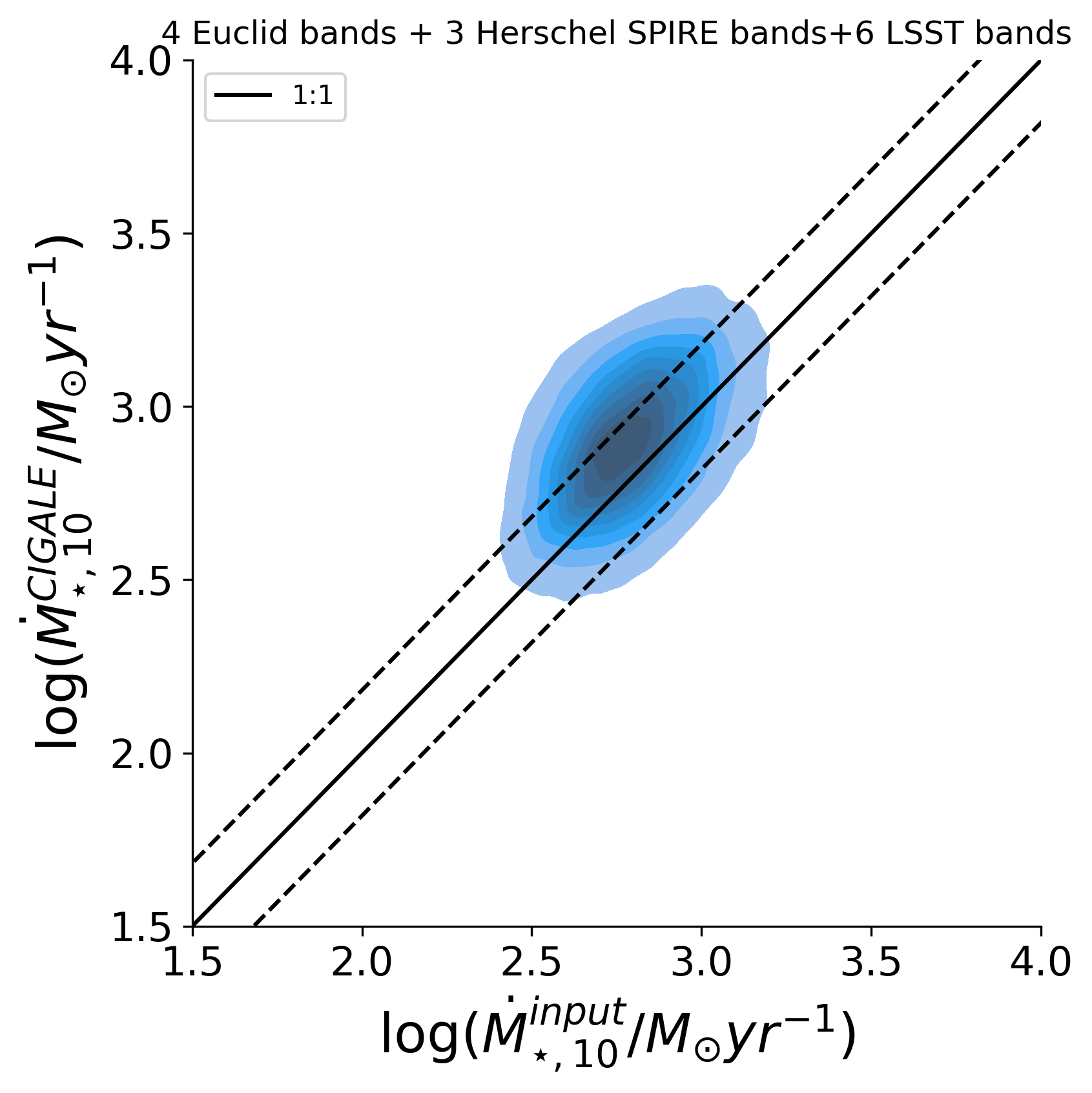}\hfill
\includegraphics[width=.25\textwidth]{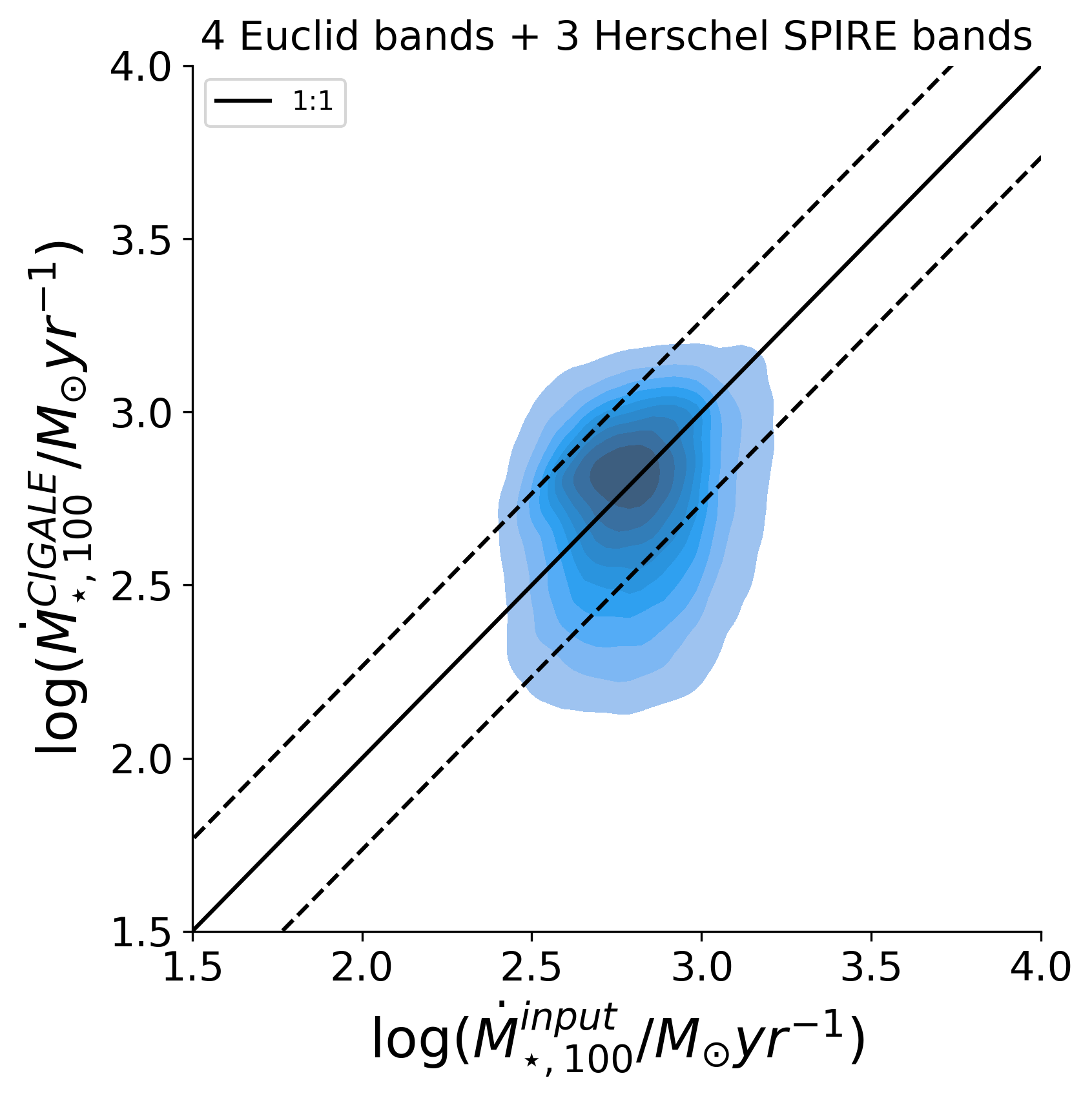}\hfill
\includegraphics[width=.25\textwidth]{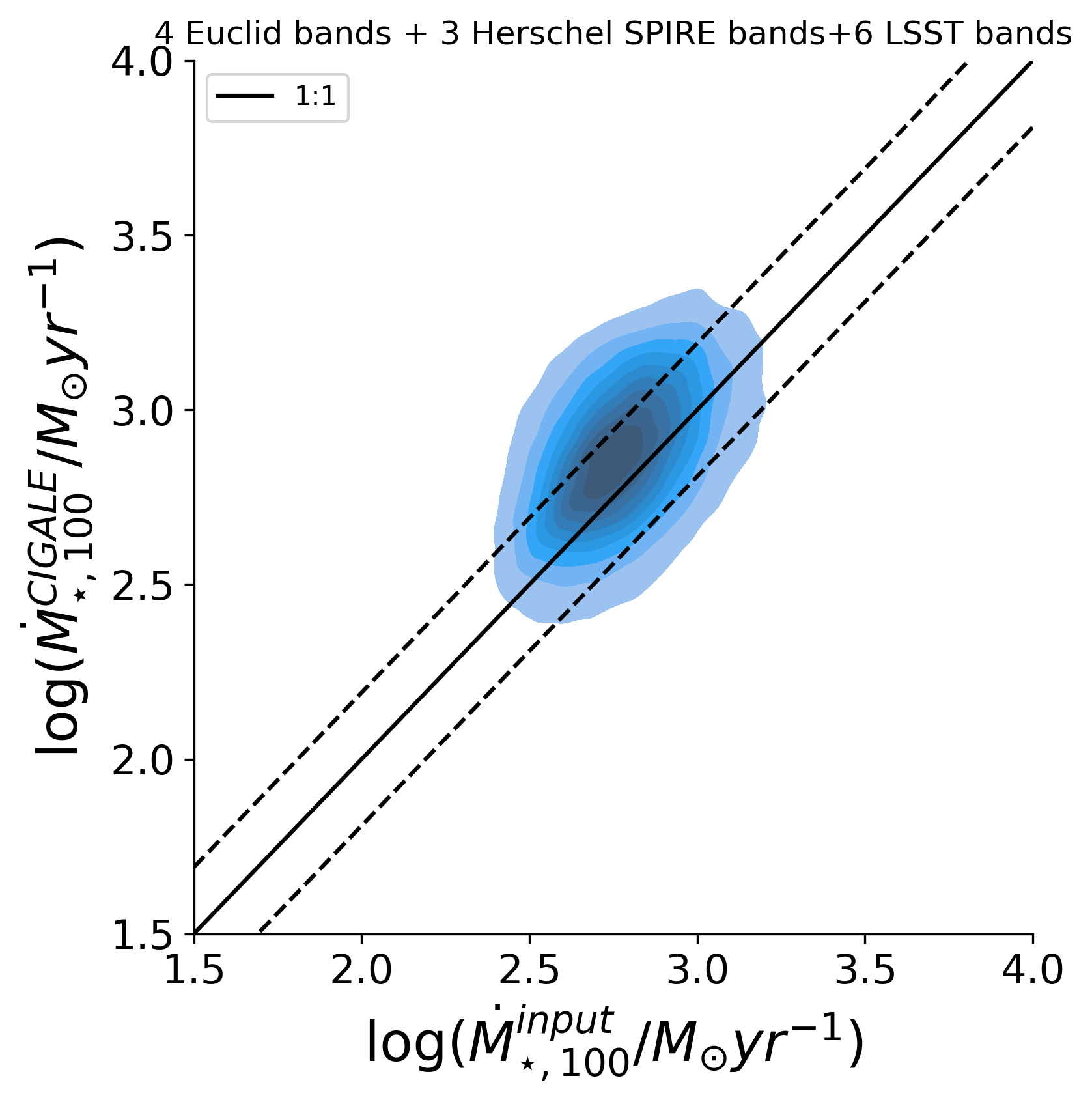}\hfill
\includegraphics[width=.25\textwidth]{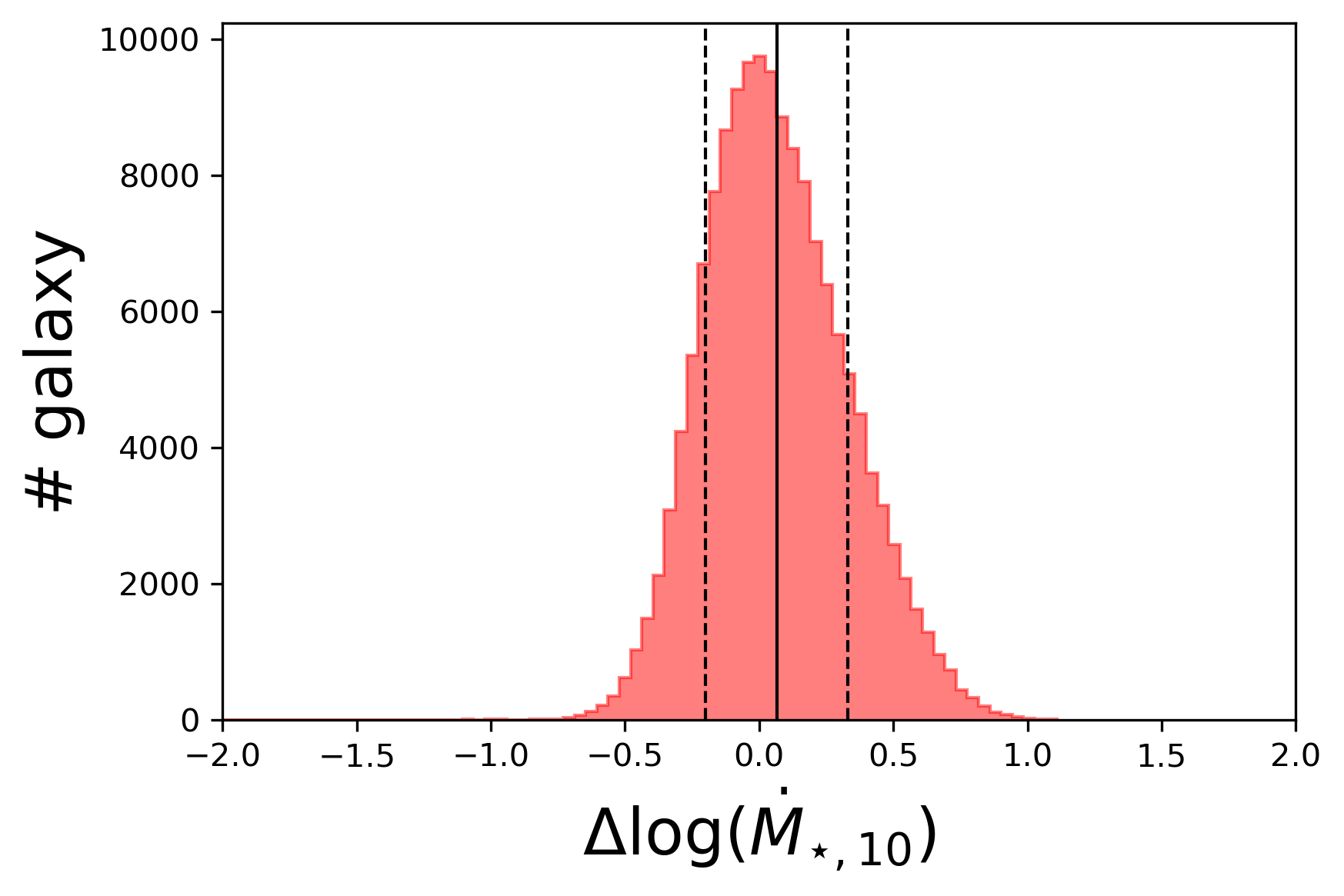}\hfill
\includegraphics[width=.25\textwidth]{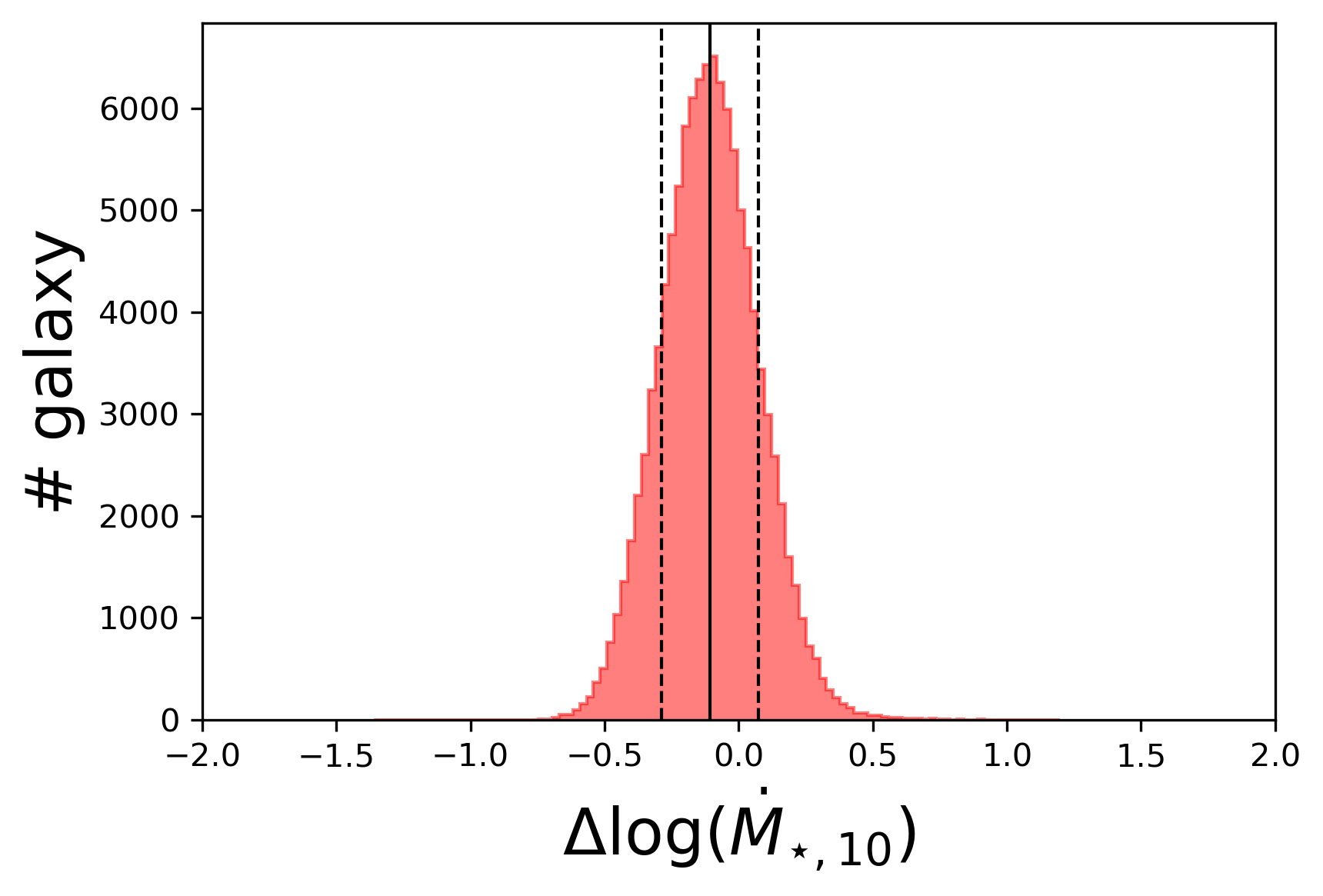}\hfill
\includegraphics[width=.25\textwidth]{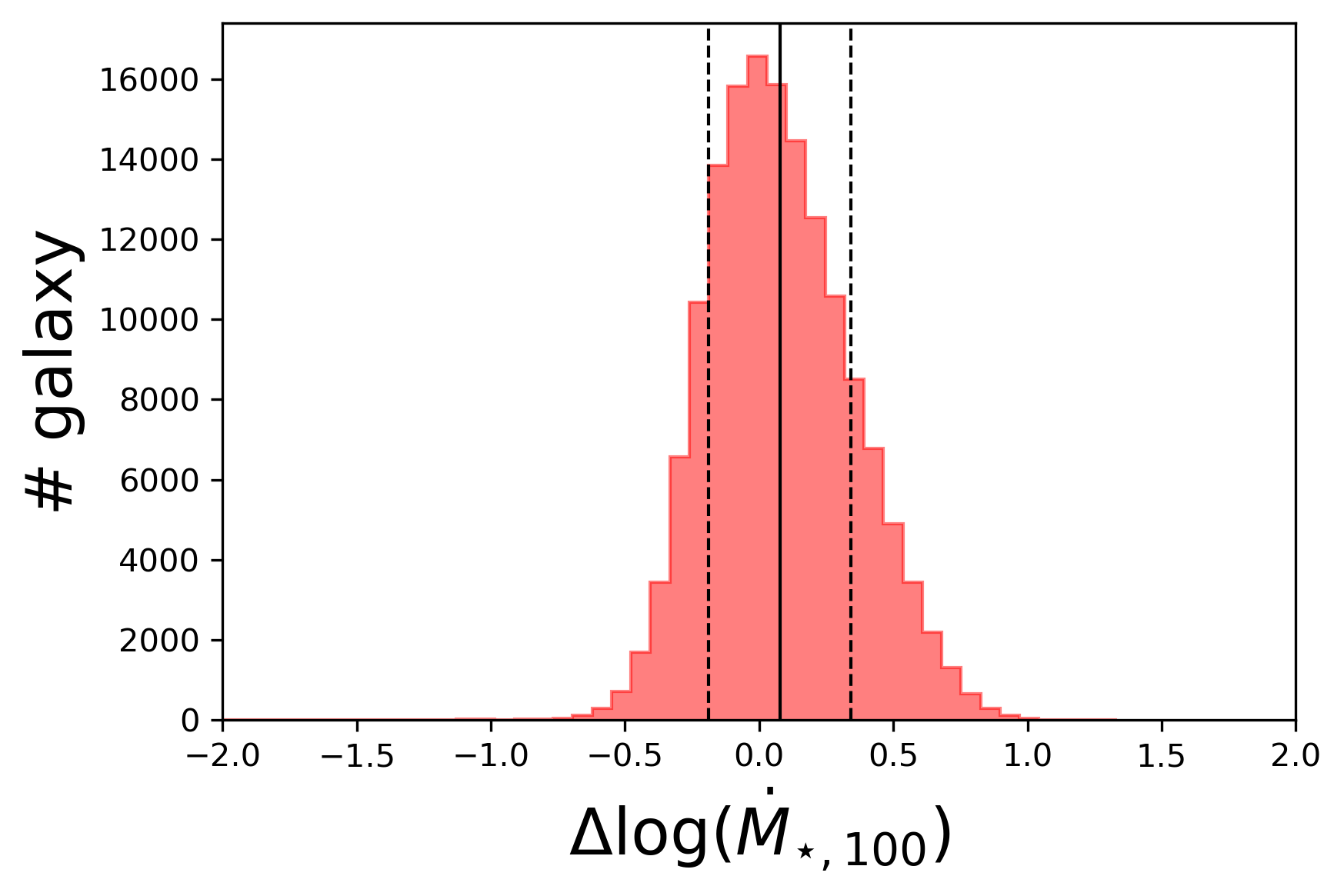}\hfill
\includegraphics[width=.25\textwidth]{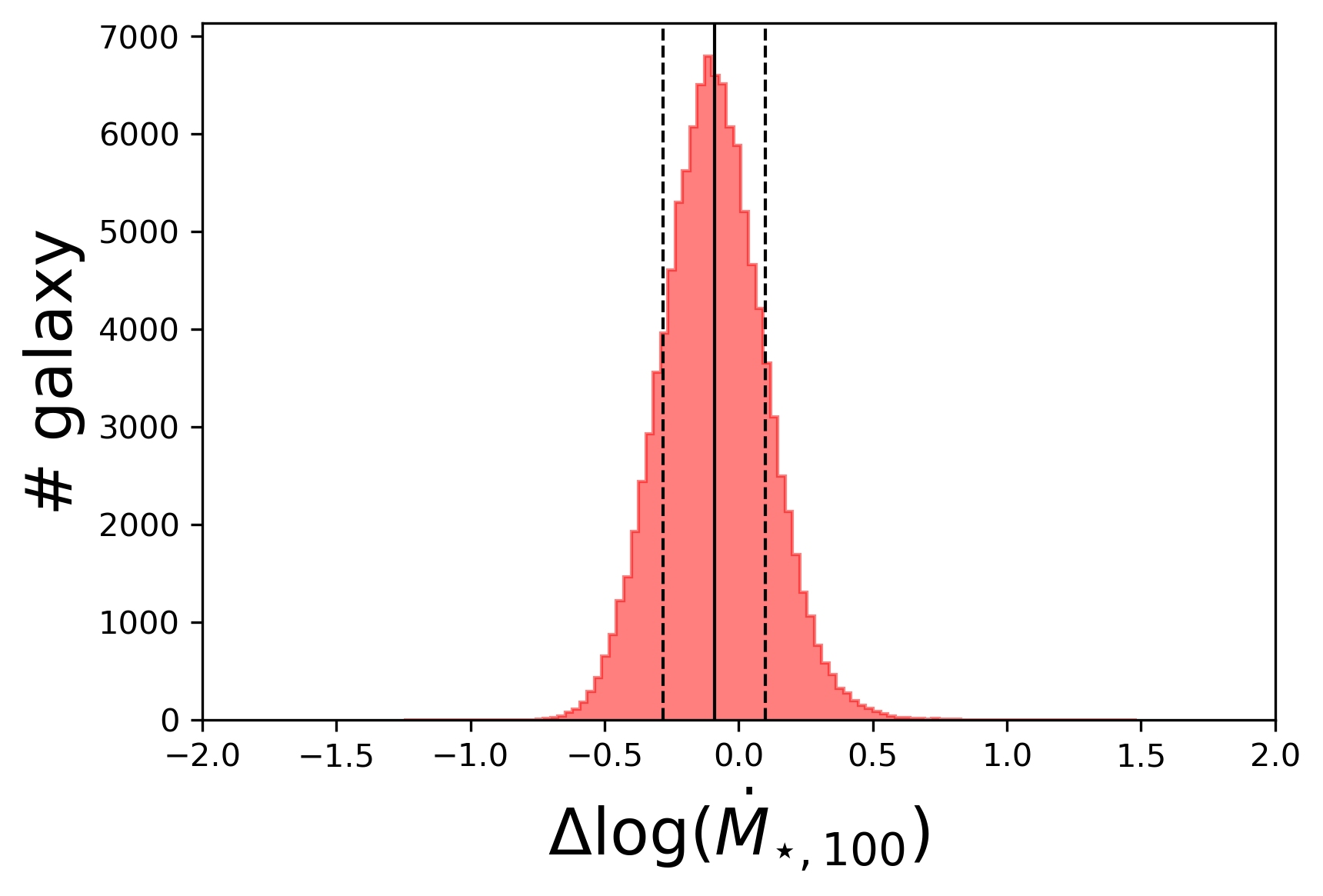}

      \caption{Same as Figure \ref{figcigalesm}, but for the SFR averaged over 10 Myrs (first two top and bottom figures) and 100 Myrs (last two top and bottom figures) respectively, with the SED fit being performed on \textit{Euclid}+\textit{Herschel}/SPIRE photometry and \textit{Euclid}+\textit{Herschel}/SPIRE+\textit{LSST} photometry.}
      \label{figcigsfravg}
\end{figure*}

The comparison between the input values of the SFR and the estimated ones is shown in Figure \ref{figcigsfr}. In this case, the SED fit has been performed using both the {\it Euclid} and the {\it Herschel} photometry. In fact, for dusty objects, the NIR wavelengths are not good tracers of the star formation \citep{pforr_recovering_2012,pforr_recovering_2013, euclid_collaboration_euclid_2023-1} and the use of sub-mm data becomes crucial to reliably infer the SFR. 
The difference in logarithm between the input and the output values of the SFR has a dispersion of $0.26$. As the LSST filters cover the UV/optical wavelengths, adding the LSST data to the NIR and FIR photometry can help to trace the SFR more accurately as can be seen in Figure \ref{figcigsfr} (right panels). In fact, when using the \textit{Euclid}+\textit{Herschel}+{\it LSST} photometry, the $1\,\sigma$ dispersion 
in $\Delta\log (\dot{M_{\star}})$ drops to $0.18$. However, we observe a systematic overestimation of the SFR by CIGALE when adding the \textit{LSST} photometric data. As for the stellar mass estimates, the discrepancy in the SFR measurements can also be explained in terms of differences in the adopted values of the slope of the dust attenuation law between CIGALE and our formalism. Overall, estimating the SFR seems to be more challenging than inferring the stellar mass. This may be due to the more sensitive dependence of the latter on the SFH compared to the former. In fact, the SFR is an {\it instantaneous} quantity while $M_{\star}$ is an integrated quantity. To test this, we compared the values of the SFR averaged over a defined time scale. We introduce two quantities, $\dot{M}_{\star,10}$ and $\dot{M}_{\star,100}$, which represent the SFR averaged over the 10 Myr and the 100 Myr prior to the time of observation, respectively, i.e.:
\begin{equation}
    \dot{M}_{\star,10 } = \frac{1}{10\rm\,Myr}\int_{t_{obs}-10 \rm \,Myr}^{t_{obs}}\dot{M}_{\star}(t)dt
\end{equation}
and 
\begin{equation}
    \dot{M}_{\star,100 } = \frac{1}{100\rm\,Myr}\int_{t_{obs}-100 \rm \,Myr}^{t_{obs}}\dot{M}_{\star}(t)dt
\end{equation}
Figure \ref{figcigsfravg} shows the distribution of the input values versus the measured values of $\dot{M}_{\star,10}$ and $\dot{M}_{\star,100}$ along with the difference in logarithm between the true and the estimated values of the SFR from the \textit{Euclid}+\textit{Herschel} and the \textit{Euclid}+\textit{Herschel}+{\it LSST} photometry. Both $\Delta\log(\dot{M}_{\star,10})$ and $\log(\Delta\dot{M}_{\star,100})$ have a $1\,\sigma$ dispersion of 0.26 when estimated from the \textit{Euclid}+\textit{Herschel} photometry. Upon adding the LSST data to the former, the $1\,\sigma$ dispersion reduces to 0.18. Even though $\dot{M}_{\star,10}$ and $\dot{M}_{\star,100}$ are integrated properties, we get similar dispersion to what is derived for the instantaneous SFR. 
Therefore, we can conclude that the use of an instantaneous SFR cannot explain the larger dispersion in the recovered values of SFR compared to what was observed for the stellar mass. 
However, the galaxies we are dealing with are dust-obscured; therefore, the accuracy in the estimate of the SFR is also dependent on the availability of data at far-IR/sub-mm/mm wavelengths. Only 20\% of the simulated DSFGs detected by Herschel above 5\,$\sigma$ at 250\,$\mu$m are also detected above 3\,$\sigma$ at 350\,$\mu$m and at 500\,$\mu$m. Indeed, if we restrict the calculation of the SFR to that subsample we find a $1\,\sigma$ dispersion of 0.23 and 0.14 for \textit{Euclid}+\textit{Herschel} photometry and \textit{Euclid}+\textit{Herschel}+\textit{LSST} photometry, respectively, which clearly shows an improvement in the estimation of the SFR. The median SFR of the sources in our sample is $\log(\dot{M_{\star}}/M_{\odot}\hbox{yr}^{-1})=2.77\pm0.162$.

%

\subsubsection{SFR-$M_{\star}$ relation}

\begin{figure*}
    \centering
\includegraphics[width=.5\textwidth]{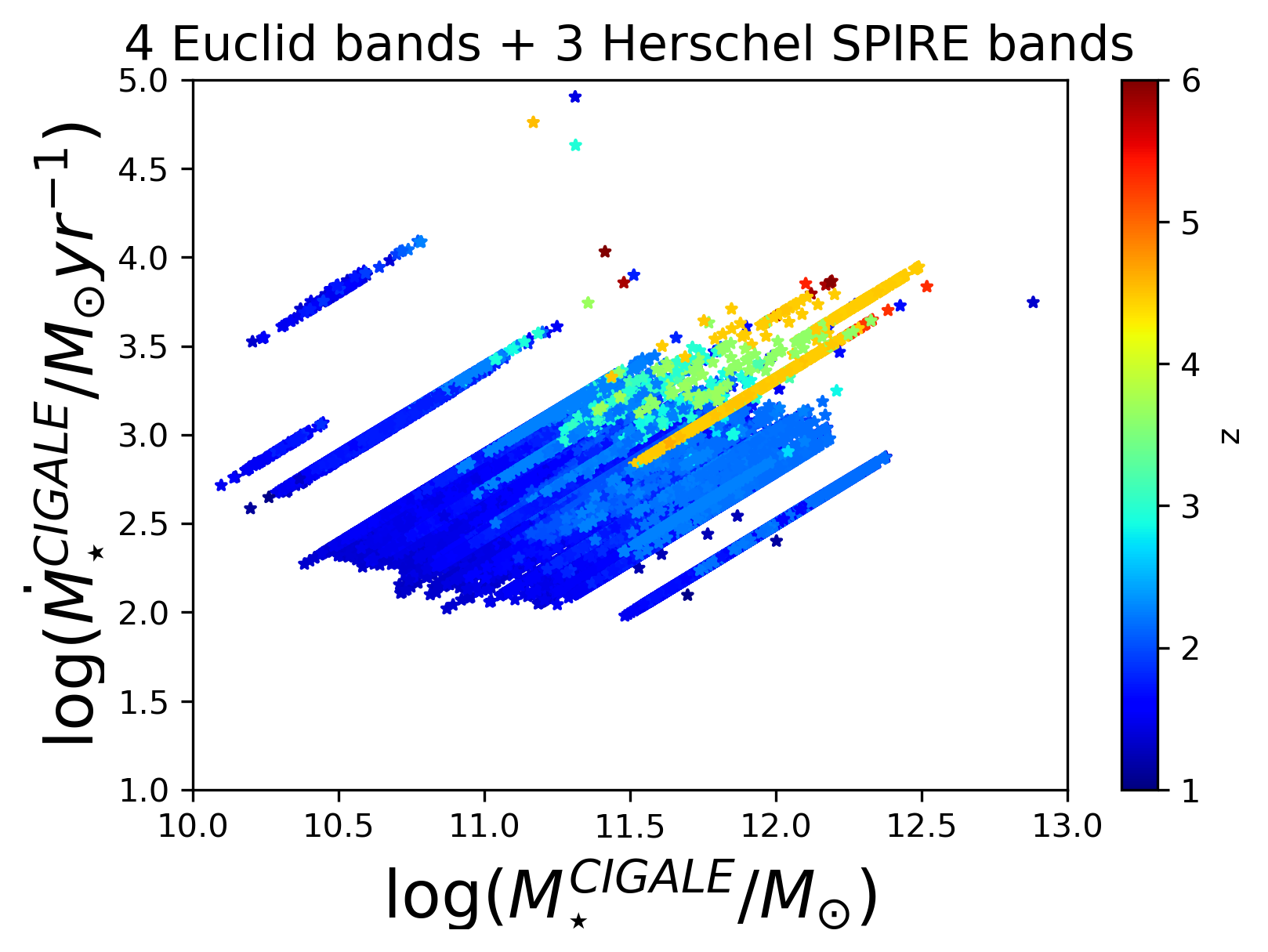}\hfill
\includegraphics[width=.5\textwidth]{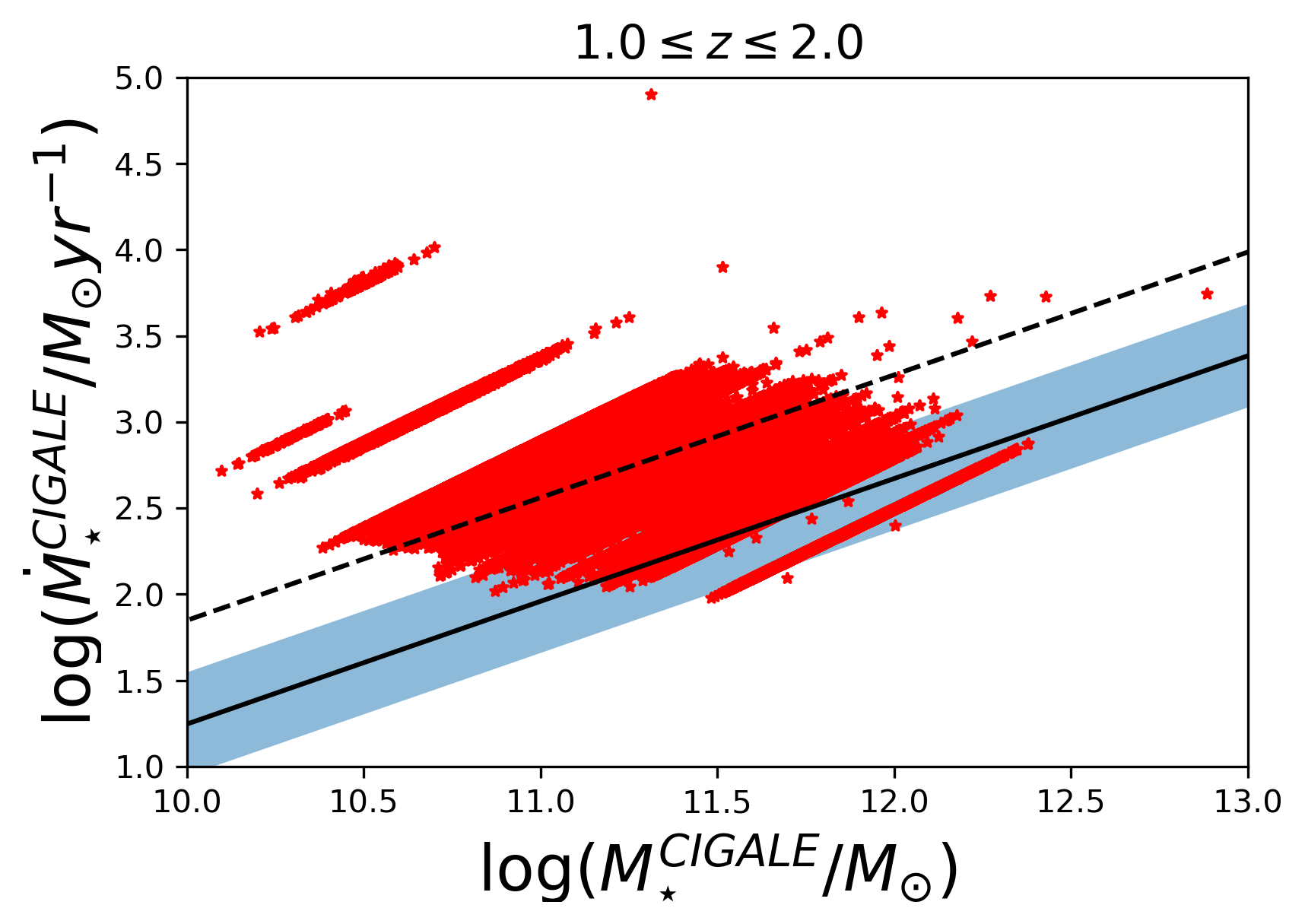}\hfill
\includegraphics[width=.5\textwidth]{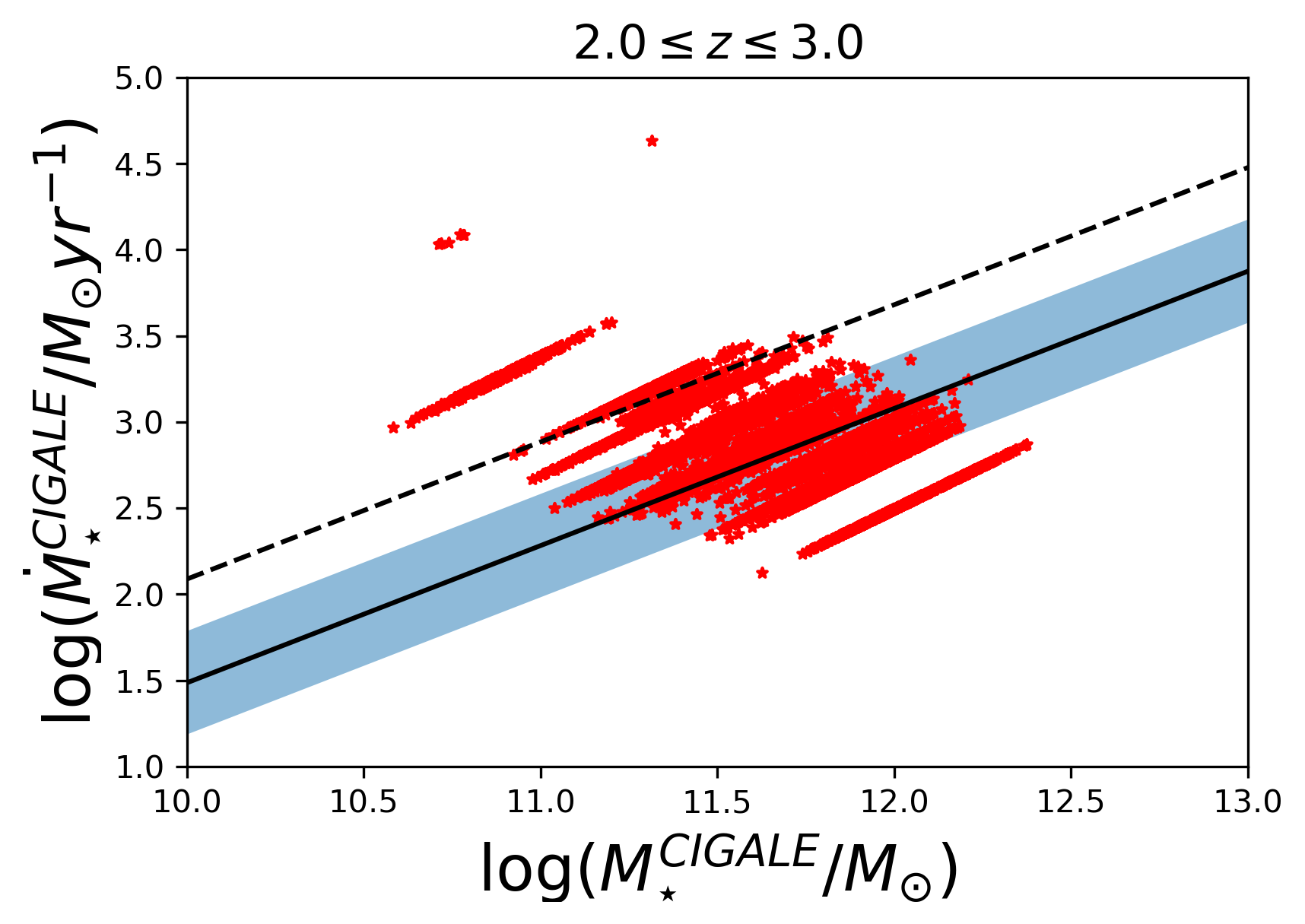}\hfill
\includegraphics[width=.5\textwidth]{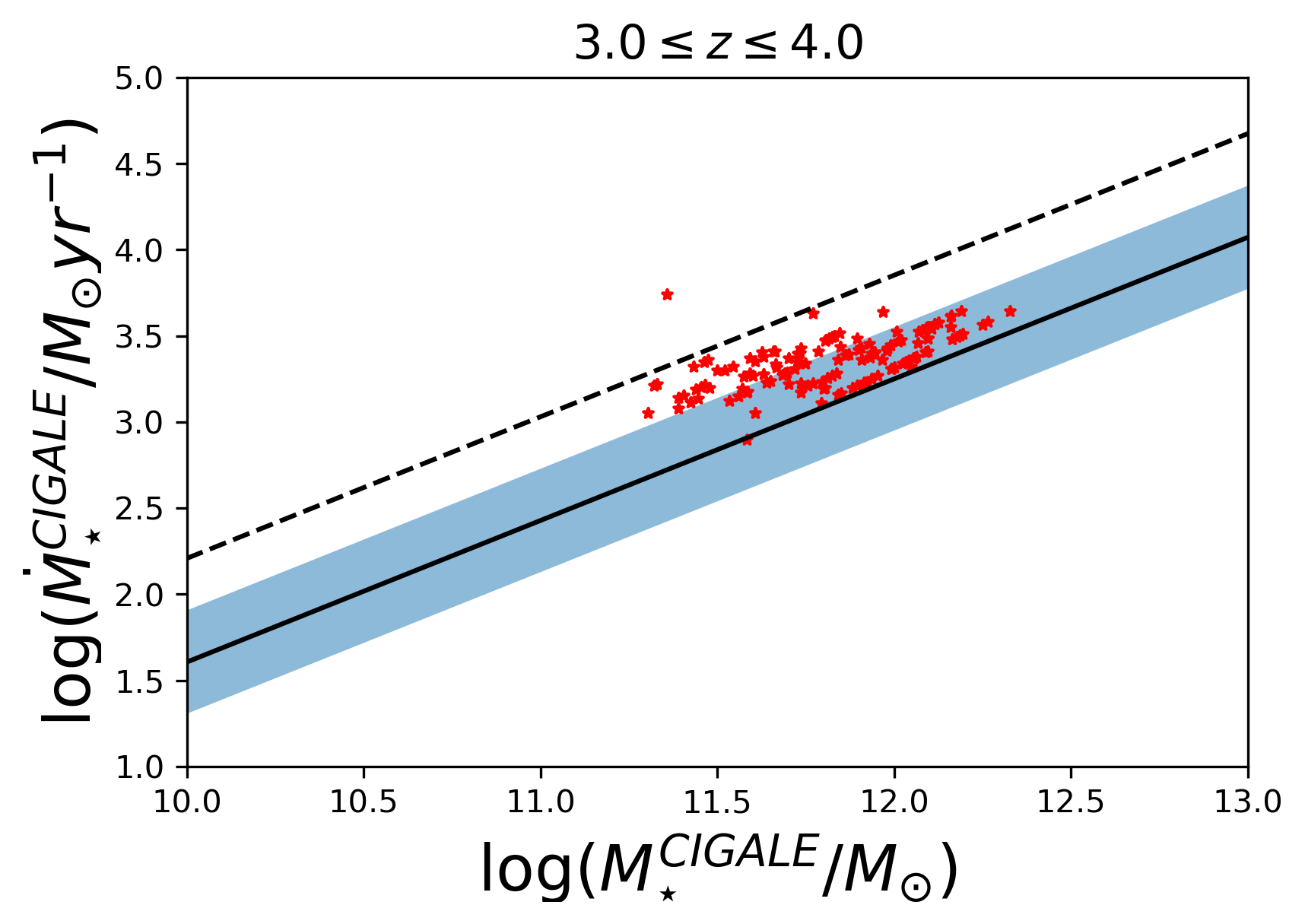}\hfill
\includegraphics[width=.5\textwidth]{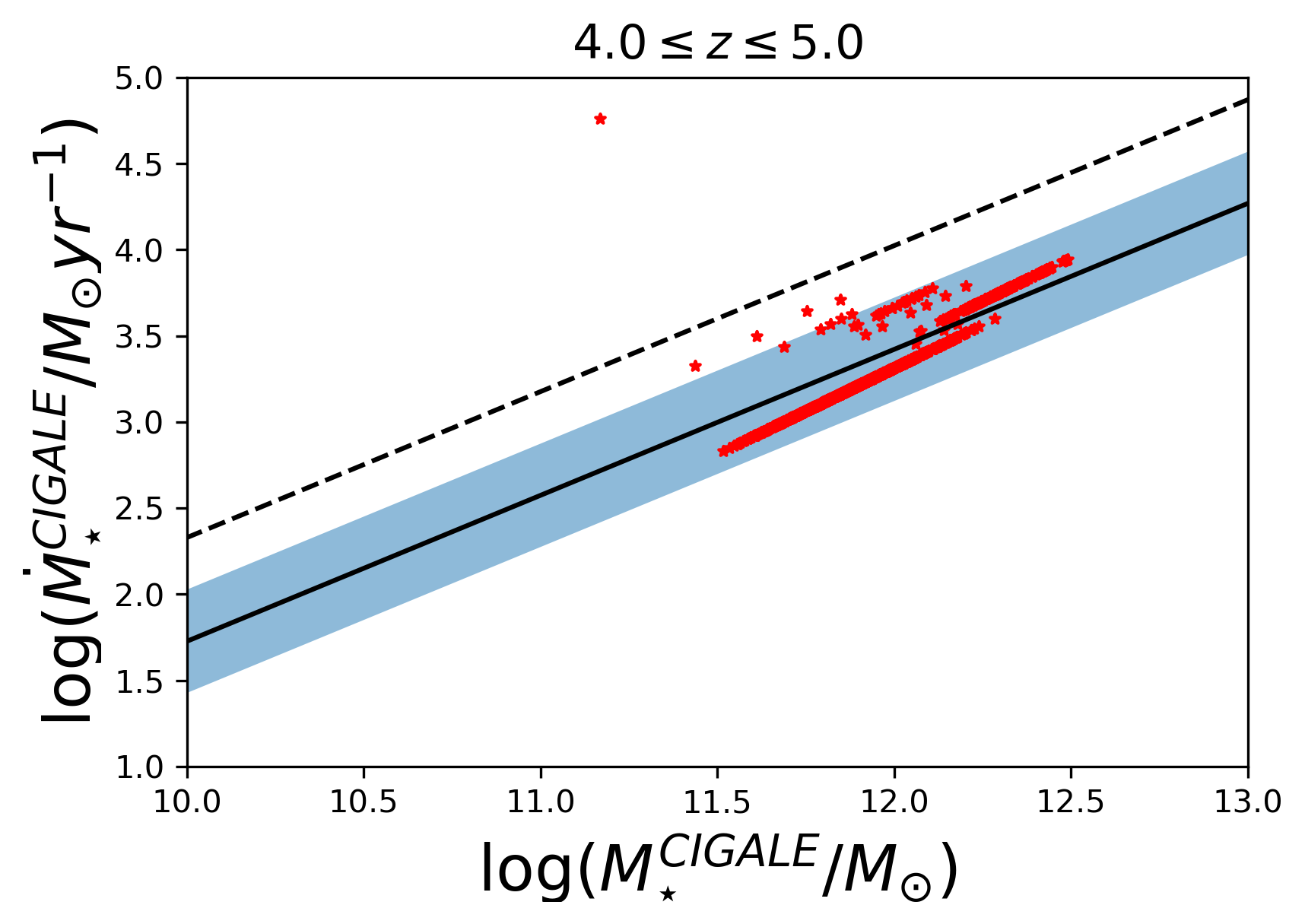}\hfill
\includegraphics[width=.5\textwidth]{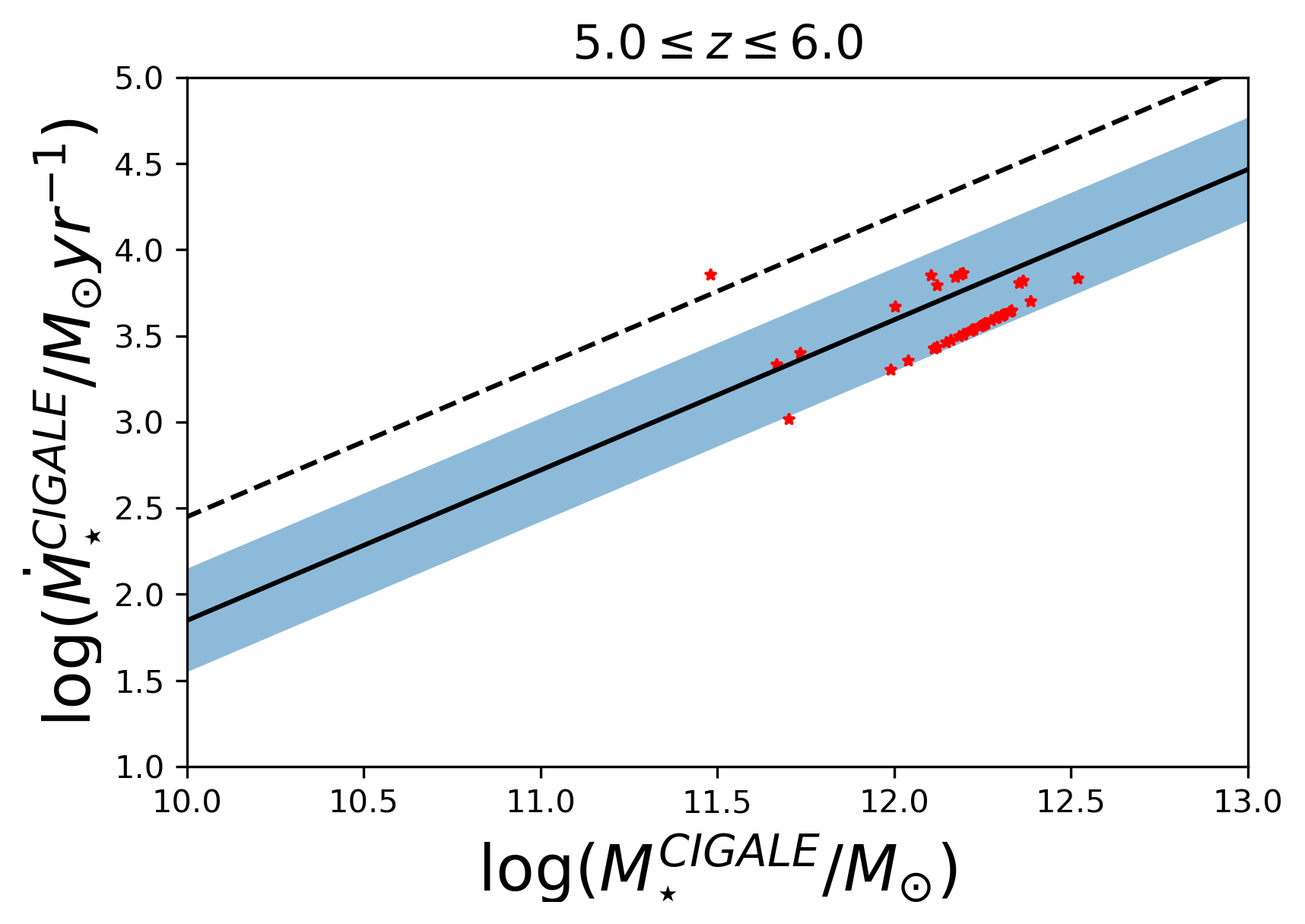}
      \caption{Star formation versus stellar mass for the simulated sample of galaxies detected by {\it Herschel} and {\it Euclid}. The top left panel shows the correlation for the entire catalogue of galaxies. The points are colour-coded according to their photometric redshift, $z_{\rm phot}^{\rm EAZY}$. The remaining plots show the SFR versus $M_{\star}$ for different redshift bins as follows: $1\leq z\leq2$ (top right), $2<z\leq3$ (middle left), $3<z\leq4$ (middle right), $4<z\leq5$ (bottom left), and $5\leq z\leq 6$ (bottom right). The black line in each of the plots represents the galaxy-MS relation from  calculated at the central redshift of the interval. The blue-filled region denotes the 0.3 dex scatter around that relation. The dashed black line shows the MS relation scaled up by a factor of four, the conventional boundary between MS and starburst galaxies. }
      \label{figsfrsm}
\end{figure*}

Studies of large galaxy samples have shown that most of the star-forming galaxies in the Universe follow an SFR-$M_{\star}$ correlation, called the galaxy main sequence (MS), while a fraction of galaxies lie above that correlation and are regarded as {\it starburst} (SB) galaxies \citep{speagle_highly_2014-1,mancuso_main_2016,popesso_main_2022}.  
Figure \ref{figsfrsm} shows the SFR-$M_{\star}$ plot of our simulated galaxies for several bins of redshifts, along with the redshift-dependent galaxy-MS relation derived by \cite{speagle_highly_2014-1}. Our sample comprises both MS and starburst galaxies. The bulk of the galaxies, approximately $57\%$, with $z\sim1-2$, are above the MS relation and they are in their starburst stage, forming stars at an average rate of $\sim300-3000\,M_{\odot}$ per year. There are very few cases (11) where the estimated SFR is $>10^4\,M_{\odot}\hbox{yr}^{-1}$. This is due to the wrong redshift estimation by EAZY, leading to the wrong estimation of SFR by CIGALE. Around $15\%$ of the galaxies at $z\sim2-3$ are above the MS relation and in their starburst stage. On the other hand, at high redshifts $z\sim3.5-5.0$, we find that the bulk of the sources are MS galaxies. 

\subsubsection{Dust Luminosity and Dust Mass}

\begin{figure*}
    \centering
      \centering
\includegraphics[width=.5\textwidth]{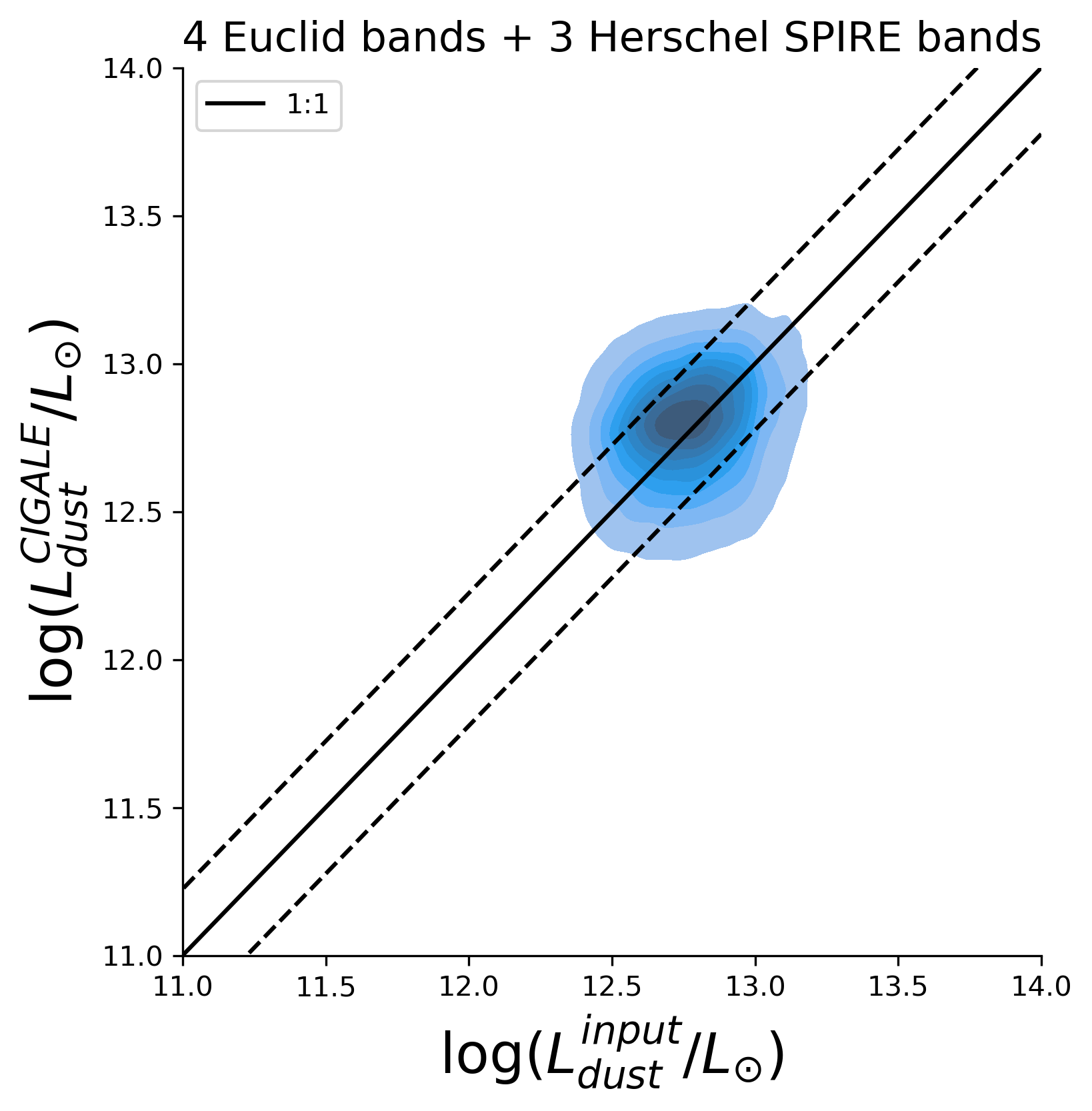}\hfill
\includegraphics[width=.5\textwidth]{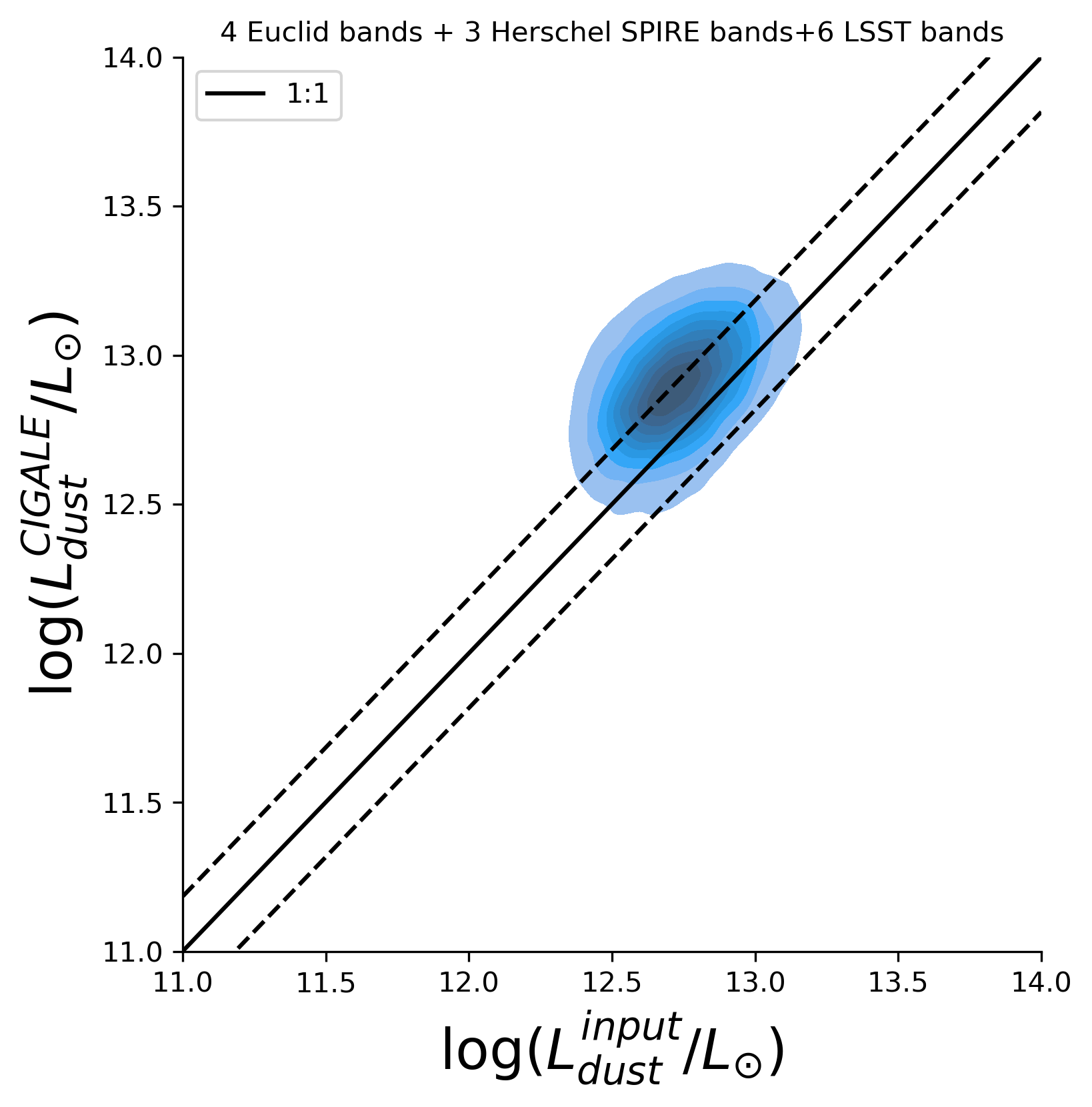}\hfill
\includegraphics[width=.5\textwidth]{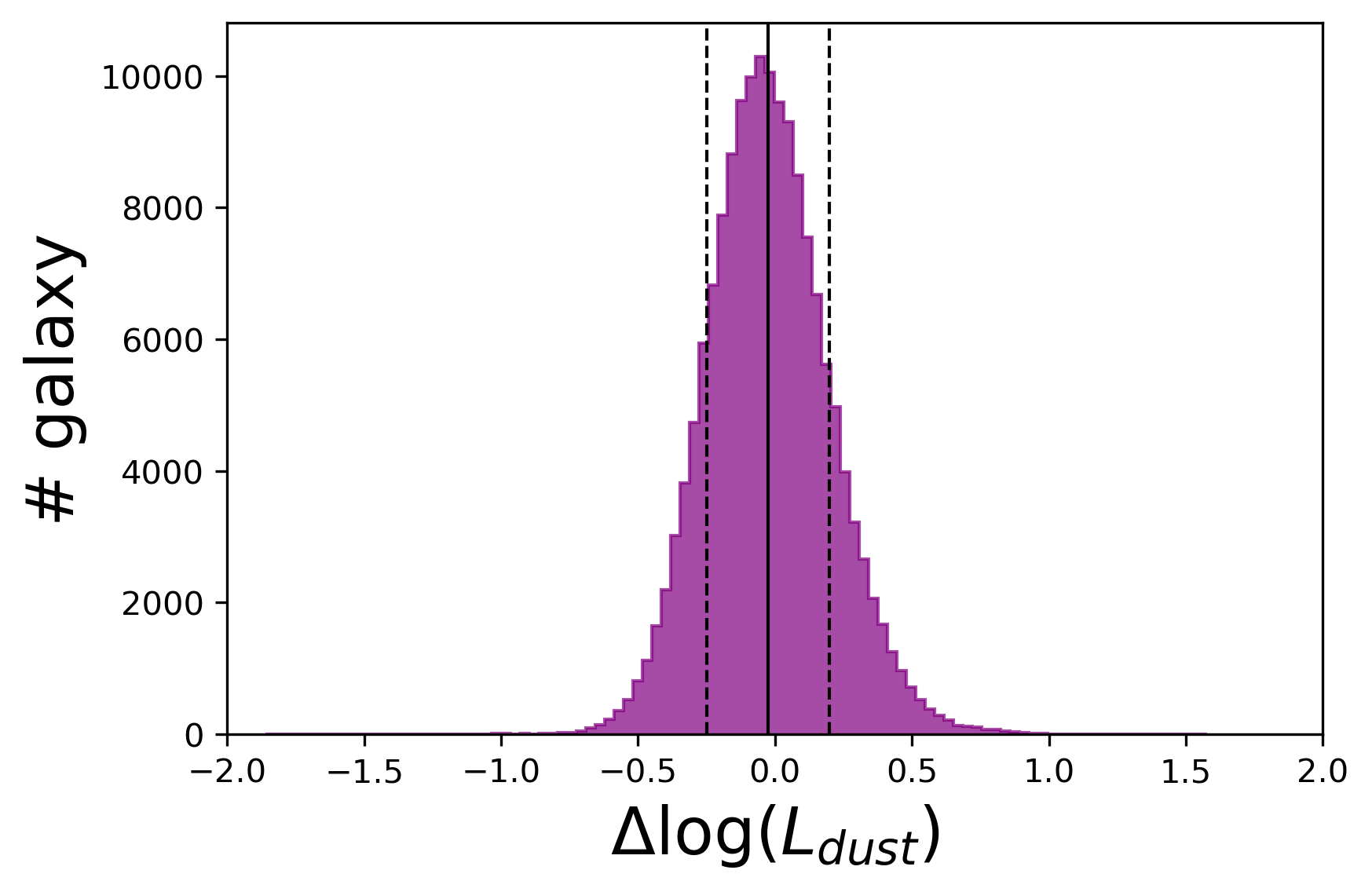}\hfill
\includegraphics[width=.5\textwidth]{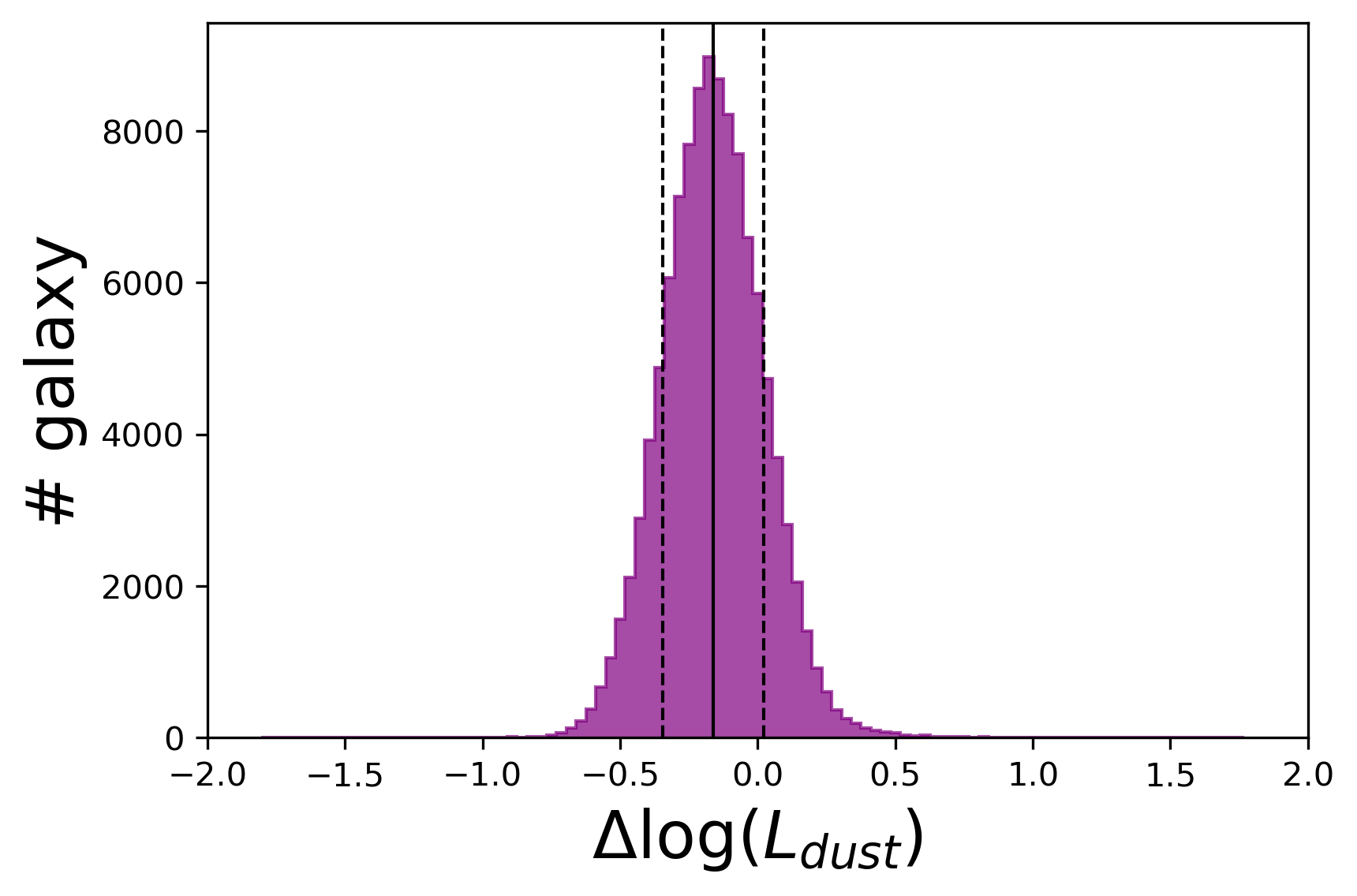}

      \caption{Same as Figure \ref{figcigalesm}, but for the dust luminosity ($L_{\rm dust}$), with the SED fit being performed on \textit{Euclid}+\textit{Herschel}/SPIRE photometry (left-hand panels) and \textit{Euclid}+\textit{Herschel}/SPIRE+\textit{LSST} photometry (right-hand panels)}.
      \label{figcigdl}
\end{figure*}

The comparison between the measured values of the dust luminosity and the input values is shown in Figure \ref{figcigdl}. We considered again the results obtained by using the \textit{Euclid}+\textit{Herschel} photometry (left-hand panels) and the \textit{Euclid}+\textit{Herschel}+\textit{LSST} photometry (right-hand panels). The histogram of the differences between the input values and the measured values is shown in the bottom panels.
We found that, when fitting the \textit{Euclid} and the \textit{Herschel} data together, we could estimate the dust luminosity ($L_{\rm dust}$) accurately, with a $1\,\sigma$ dispersion in $\Delta\log(L_{\rm dust})$ of about 0.22. On adding the \textit{LSST} photometry to the fit, the constraint on $L_{\rm dust}$ becomes tighter and the $1\,\sigma$ dispersion in $\Delta\log(L_{\rm dust})$ reduces to 0.17. In fact, by tracing the UV/optical wavelengths, the LSST data help to better constrain the extinction of starlight by dust, thus leading to a more precise estimate of $L_{\rm dust}$. It is worth noticing, though, that in this case CIGALE systematically overestimates the dust luminosity by a factor of about 1.07. This is due to a higher standard value of the slope of the dust-attenuation law adopted by CIGALE. 
Overall, with a median dust luminosity of $\log(L_{\rm dust}/L_{\odot})=12.8\pm0.12$, these galaxies sample the bright tail of the population of ultra-luminous infrared galaxies (ULIRGs).

\begin{figure*}
    \centering
      \centering
\includegraphics[width=.5\textwidth]{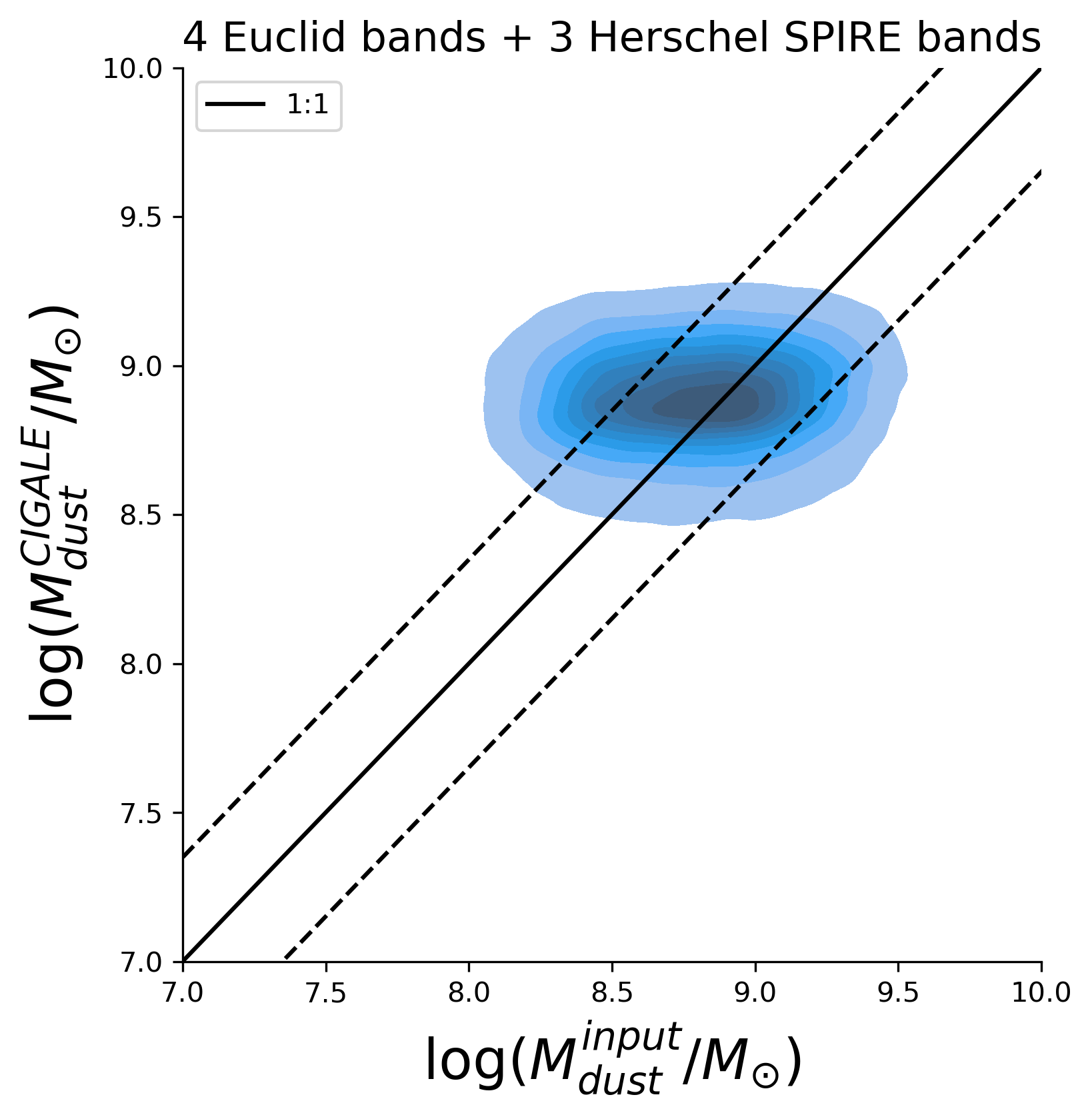}\hfill
\includegraphics[width=.5\textwidth]{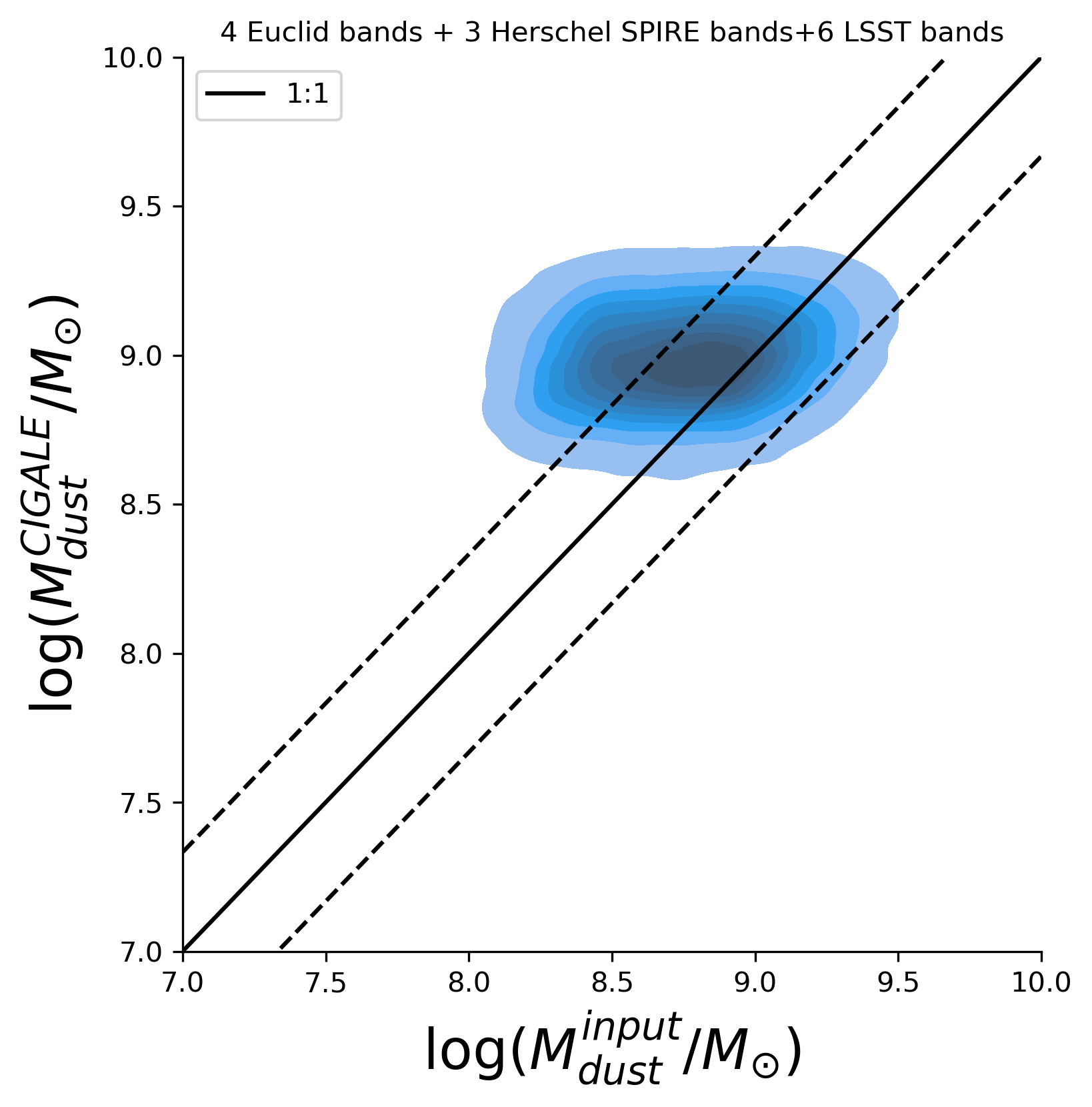}\hfill
\includegraphics[width=.5\textwidth]{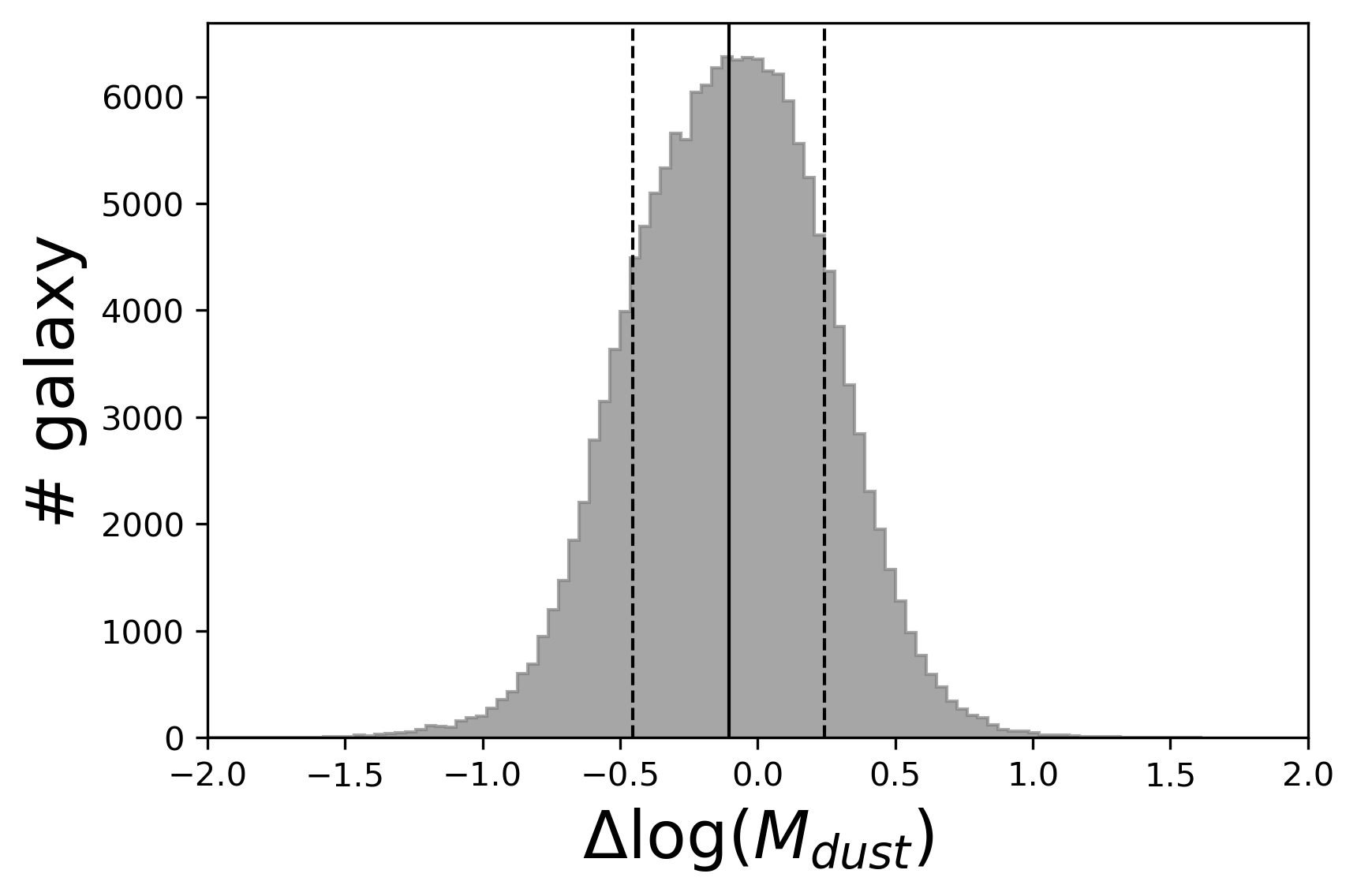}\hfill
\includegraphics[width=.5\textwidth]{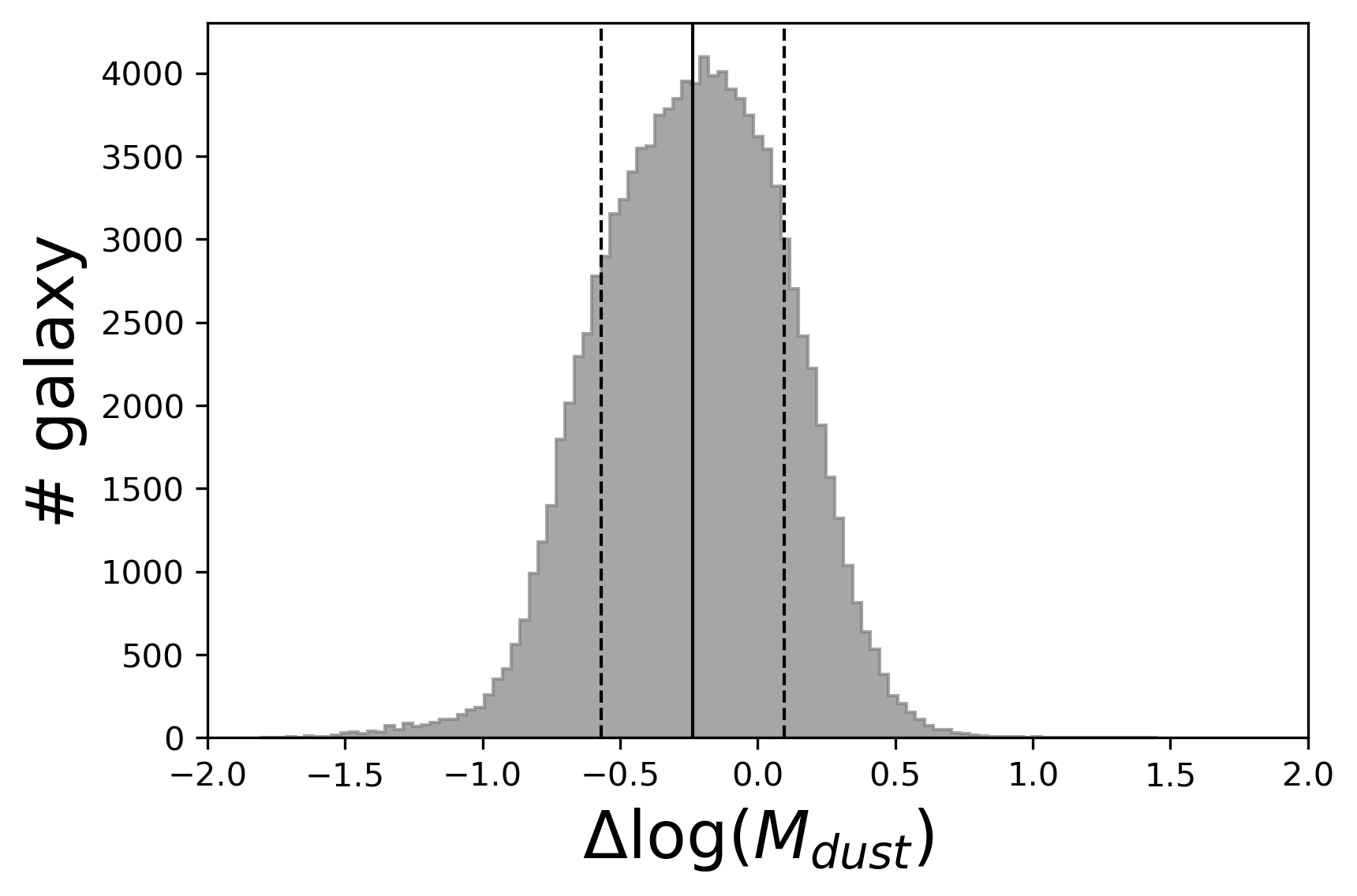}

      \caption{Same as Figure \ref{figcigalesm}, but for the dust mass ($M_{\rm dust}$), with the SED fit being performed on \textit{Euclid}+\textit{Herschel}/SPIRE photometry (left-hand panels) and \textit{Euclid}+\textit{Herschel}/SPIRE+\textit{LSST} photometry (right-hand panels).}
      \label{figcigdm}
\end{figure*}

In Figure \ref{figcigdm} we show the results for the dust mass estimates.The dust mass has a median value of $\log(M_{\rm dust}/M_{\odot})=8.9\pm0.10$. This is consistent with the values of $M_{\rm dust}$ of other ULIRGs reported by \cite{dudzeviciute_alma_2020} and \cite{pantoni_unveiling_2021-1}, and shows that these DSFGs have higher dust content than local ($z<1$) star forming galaxies selected with Herschel (e.g.: R14). The dust mass is affected by larger uncertainties compared to the dust luminosity and the use of the LSST photometry does not help to reduce that uncertainty. We find, indeed, that the $1\,\sigma$ dispersion in $\Delta\log(M_{\rm dust})$ is 0.35 and 0.32 for the \textit{Euclid}+\textit{Herschel} photometry and the \textit{Euclid}+\textit{Herschel}+LSST photometry, respectively. We also observe that CIGALE systematically overestimates $M_{\rm dust}$ by a factor of $1.2$. Interestingly, \cite{liao_alma_2024} fit the SED of a sample of 18 bright ($S_{870}=12.4-19.4\,$mJy) $870\,\mu$m-selected DSFGs from AS2COSMOS \citep{simpson_alma_2020} using both CIGALE and MAGPHYS, the latter being based on the same \cite{da_cunha_simple_2008} formalism we have used in this work. They found that CIGALE produced values of $M_{\rm dust}$ higher by a factor 1.5 than those estimated with MAGPHYS. An overestimation factor of 1.3 is also quoted by \cite{birkin_almanoema_2021} when comparing dust mass estimates by CIGALE and MAGPHYS for a sample 61 sources selected from ALMA-identified $870$ $\mu$m-selected DSFGs from AS2COSMOS, AS2UDS \citep{stach_alma_2019}, and ALESS \citep{hodge_alma_2013} surveys. These discrepancies in $M_{\rm dust}$ can be mainly explained by the difference in the standard values of $\kappa_0$ adopted by the two SED fitting codes \citep{liao_alma_2024}.

\section{Summary and Conclusions}
\label{sec5}
We have presented predictions for how well the recently launched space telescope \textit{Euclid} can study the $z\gtrsim1.5$ DSFGs detected by the \textit{Herschel}-ATLAS survey. To this end, we have built a simulated catalogue of galaxies exploiting the C13 model for the evolution of proto-spheroidal galaxies. We combined the output of model equations with the SED formalism of D08, which is based on the principle of energy balance in the UV-optical and IR. 

By setting the values of the main parameters of the model to those derived by R14 from the study of a sample of DSFGs we obtained a good agreement between the predicted infrared luminosity function of DSFGs and the measured one in the redshift range $z=2 - 4$.
However, compared with the prescriptions of R14, we fine-tuned the values of both the dust emissivity index and the temperature of the warm dust in the BCs and of the cold dust in the ISM to reproduce the observed number counts of at 850\,$\mu$m, 870\,$\mu$m and 1.1\,mm.



After these preliminary checks, 
we simulated a catalogue of 458\,994 galaxies with $L_{\rm IR}\gtrsim10^{12}\,L_{\odot}$ within a survey area of 100$\,\hbox{deg}^2$, which is of the same order of magnitude of the total area surveyed by H-ATLAS, and applied a $5\,\sigma$ detection condition at 250\,$\mu$m for \textit{Herschel} and a $3\,\sigma$ detection condition for all the 4 \textit{Euclid} bands. All the 151\,720 sources detected by {\it Herschel} at 250\,$\mu$m are also detected in all the {\it Euclid} bands. We used EAZY on the \textit{Euclid} photometry to estimate the photometric redshifts of the above sources. 
We then exploited CIGALE to extract the physical properties of the galaxies from the SED fitting, assuming the previously derived photometric redshift.

Our main findings are as follows
\begin{enumerate}
    \item For 87\,\% of the sample, the {\it Euclid} photometry alone provides a good redshift estimate up to $z\sim3$, with a discrepancy $\leq15\%$ from the true value of $1+z$. For input redshifts $z\gtrsim3$, we observe some catastrophic outliers, where the estimated redshift is significantly lower than the input one, i.e. $z_{\rm phot}\sim1-1.5$. This discrepancy is attributed to the fact that the 4000\,{\AA} break, which is used as the main feature to constrain the redshift, is not sampled by \textit{Euclid} above $z\sim3$. Also, there are some galaxies at $z\sim1-2$ which are estimated to be at $z\sim4-6$, due to the Balmer break (4000\,{\AA}) being mistaken as the Lyman break (912\,{\AA}). Adding the shorter wavelength data from LSST does help to break some degeneracies. In fact, the outlier fraction reduces to just 1\% for the 75\% of the {\it Euclid} detected galaxies that are also detected above 3$\,\sigma$ in at least 4 LSST bands.
    
    We also investigated the use of the sub-mm {\it Herschel} photometry to derive the photometric redshift by fitting those data with the empirical template of \cite{pearson_h-atlas_2013}. However, we 
    found a higher fraction of outliers compared to what was obtained with the {\it Euclid} photometry, i.e. $f_{\rm outlier}=0.35$ compared to $f_{\rm outlier}=0.13$. 
    This is mainly due to the lack of sufficient sub-mm data for most of the sources in the sample. In fact, only 20\% of the {\it Herschel} detected galaxies (i.e. those above $5\,\sigma$ at 250 $\mu$m) are also detected at both 350$\,\mu$m and $500\,\mu$m above $3\,\sigma$. For this sub-sample the outlier fraction is reduced to 13$\%$. 
    In summary, for the vast majority of the simulated DSFGs, we cannot gain any significant improvement on the photo-z from the sub-mm measurements and that is why we have adopted the photometric redshift estimates based on the {\it Euclid} data alone. 


    \item The distribution of the stellar masses, as recovered from the {\it Euclid} and {\it Euclid+Herschel} photometry, has a dispersion of 0.16\,dex and 0.19\,dex respectively, around the true value. As expected, adding the {\it Herschel} photometry to the SED fit does not significantly improve the results, as the stellar mass is sensitive to the rest-frame optical/near-IR portion of the SED. The median value of the stellar mass is found to be 
    $\log(M_{\star}/M_{\odot})=11.29\pm0.181$, while the range of values of $\log(M_{\star}/M_{\odot})$ is $10.5-12.9$. 

    \item The recovered SFR shows a wider distribution around the true value compared to the stellar mass, with a dispersion 0.26\,dex.  Overall, the sources in our sample have a median SFR of $\log(\dot{M_{\star}}/M_{\odot}\hbox{yr}^{-1})=2.77\pm0.162$, while the probed range of values of $\log(\dot{M_{\star}}/M_{\odot}\hbox{yr}^{-1})$ is $1.51-3.99$. When the SFRs are averaged over 10 Myr and 100 Myr respectively, we obtain a similar $1\,\sigma$ dispersion of 0.26 in both cases. The median value is $\log(\dot{M}_{\star,10 }/M_{\odot}yr^{-1})=2.77\pm0.16$ and $\log(\dot{M}_{\star,100 }/M_{\odot}yr^{-1})=2.74\pm0.16$.

    \item From the SFR$-M_{\odot}$ relation, we found that our simulated catalogue comprised both MS galaxies and starburst galaxies. Approximately $55\%$ of the galaxies in our sample lie above the MS relation and are starbursting, with a median SFR of $\log(SFR/M_{\odot}\hbox{yr}^{-1})=2.87$.

    \item The median dust luminosity obtained for our simulated DSFG sample is $\log(L_{\rm dust}/L_{\odot})=12.8\pm0.12$ and the dust mass has a median value of $\log(M_{\rm dust}/M_{\odot})=8.9\pm0.10$. These DSFGs have a dust luminosity in the range $\log(L_{\rm dust}/L_{\odot})\sim12.2-13.3$ and dust mass in the range $\log(M_{\rm dust}/M_{\odot})\sim7.5-9.9$.

    \item The photometric redshift can be improved by complementing the \textit{Euclid} data with the \textit{LSST} observations. From the combined \textit{Euclid+LSST} photometry, EAZY provides an accurate redshift, i.e. $\Delta z/(1 + z)\leq 0.15$, for $99\%$ of the $87\%$ {\it Euclid} objects that are also detected in at least 4 of the 6 {\it LSST} bands. For this subsample of galaxies that have both Euclid and LSST photometry, the dispersion on the recovered stellar mass and star formation rate reduces to $0.14$\,dex and $0.18$\,dex respectively. By covering the UV/optical wavelengths using \textit{LSST} filters, we could constrain the extinction of starlight by dust in the UV, which reduced the dispersion in the estimation of $L_{\rm dust}$. However, the dispersion in dust mass shows no significant improvement with the inclusion of \textit{LSST} photometry.


\end{enumerate}

\section*{Acknowledgements}

DM acknowledges the postgraduate studentship provided by the UK Science and Technology Facilities Council (STFC). 
DM would also like to thank Dr Denis Burgarella for his help while running CIGALE and Dr M.W.L. Smith for valuable discussions on Python programming. The authors wish to thank the anonymous referee for the valuable comments provided, which helped to improve the manuscript.
In this work, we have used the following Python packages: \textit{Astropy}\footnote{\url{https://www.astropy.org/}}, \textit{Scipy}\footnote{\url{https://scipy.org/index.html}}, \textit{Numpy}\footnote{\url{https://numpy.org/}}, \textit{Joblib}\footnote{\url{https://joblib.readthedocs.io/en/stable/}}, \textit{COLOSSUS}\footnote{\url{https://pypi.org/project/colossus/}} \citep{diemer_colossus_2018}, \textit{Seaborn}\footnote{\url{https://seaborn.pydata.org/index.html}} and \textit{Matplotlib}\footnote{\url{https://matplotlib.org/}}.

\section*{Data Availability}

Data in this paper will be made available upon request.







\appendix

\section{Equations Governing the Evolution of Proto-Spheroidal Galaxies}
\label{appendixA}

\subsection{The Model}
\label{model}

Massive proto-spheroidal galaxies are high-$z$ submm galaxies with high SFRs \citep{lapi_quasar_2006, lapi_lessigreaterherschellessigreater-atlas_2011}. The model depicts two ways of DM halo formation as follows,
\begin{enumerate}
    \item fast collapse of the halo bulk triggering star formation, and 
    \item slow growth of the halo outskirts (accretion flows) having little effect on the inner portion.
\end{enumerate}

The star formation takes place due to the cooling of gas in the region of radius
\begin{equation}
\approx 70\left(\frac{M_{\rm vir}}{10^{13}M_{\odot}}\right)^{\frac{1}{3}}\left[\frac{(1+z_{\rm vir})}{3}\right]^{-1}\ \hbox{kpc},
\end{equation}
where, $M_{\rm vir}$ and $z_{\rm vir}$ are the halo mass and the virialisation redshift respectively. This region is approximately $30\%$ of the halo virial radius, and comprises $\approx40\%$ of the total mass (C13). Both the star formation and the growth of BH at the centre are controlled by feedback mechanisms from the supernovae (SNe) and the AGN. AGN feedback is responsible for sweeping out the residual gas in the most massive proto-spheroids, while the SFR is mostly controlled by SN feedback in the less massive ones. 

Consider a galactic halo mass $M_{\rm vir}$. The initial gas mass fraction in the halo has the cosmological value $f_b={M_{\rm gas}}/{M_{\rm vir}}=0.165$. The gas is heated to the virial temperature at the virialization redshift $z_{\rm vir}$. There are 3 gas phases:
\begin{enumerate}
    \item hot dense medium infalling and/or cooling towards the centre,$M_{\rm inf}$;
    \item cold gas condensing into stars, $M_{\rm cold}$;
    \item gas stored in a reservoir around the SMBH, $M_{\rm res}$ (low angular momentum gas).
\end{enumerate}

\noindent In addition, there are two condensed phases:

\begin{enumerate}
    \item total mass of stars, $M_{\star}$;
    \item black hole mass, $M_{\bullet}$.
\end{enumerate}

\noindent The equations governing the evolution of the gas phases are

\begin{equation}
    \dot{M}_{\rm inf}=-\dot{M}_{\rm cond}-\dot{M}_{\rm inf}^{\rm QSO},
\end{equation}

\begin{equation}
    \dot{M}_{\rm cold}=\dot{M}_{\rm cond}-[1-R(t)]\dot{M}_{\star}-\dot{M}_{\rm cold}^{SN}-\dot{M}_{\rm cold}^{\rm QSO},
\end{equation}

\begin{equation}
    \dot{M}_{\rm res}=\dot{M}_{\rm inflow}-\dot{M}_{\rm BH}.
\end{equation}

\noindent The cooling of hot gas and its flow to the centre take place at a rate 
\begin{equation}
    \dot{M}_{\rm cond}=\frac{\dot{M}_{\rm inf}}{t_{\rm cond}},
\end{equation}
where
\begin{equation}
    t_{\rm cond}\simeq8\times10^8\left(\frac{1+z}{4}\right)^{-1.5}\left(\frac{M_{\rm vir}}{10^{12}M_{\odot}}\right)^{0.2}\ \hbox{yr},
\end{equation}

\noindent and we take the initial value of $M_{\rm inf}$  as $M_{\rm inf}^0=f_b M_{\rm vir}$. The SFR is (for the Chabrier IMF, $R=0.54$)
\begin{equation}
    \dot{M}_{\star}=\frac{\dot{M}_{\rm cold}}{t_{\star}}=\frac{\dot{M}_{\rm cold}}{t_{\rm cond}/s}=\frac{5\dot{M}_{\rm cold}}{t_{\rm cond}}.
    \label{eqnsfr}
\end{equation}

\noindent The rate of gas loss due to SN feedback is
\begin{equation}
\begin{split}
    \dot{M}_{\rm cold}^{SN}=\beta_{\rm SN}\dot{M}_{\star}=\frac{N_{\rm SN}\epsilon_{\rm SN}E_{\rm SN}}{E_{\rm bind}}\dot{M}_{\star}
    \simeq0.6\left(\frac{N_{\rm SN}}{8\times10^{-3}/M_{\odot}}\right)\left(\frac{\epsilon_{\rm SN}}{0.05}\right)\\ \left(\frac{E_{\rm SN}}{10^{51} \hbox{erg}}\right)\times
    \left(\frac{M_{\rm vir}}{10^{12}M_{\odot}}\right)^{-\frac{2}{3}}\left(\frac{1+z}{4}\right)^{-1}\dot{M}_{\star}.
\end{split}
\end{equation}

\noindent The following values are adopted: 
\begin{enumerate}
    \item number of SNe $/M_{\odot}$, $N_{\rm SN}\simeq1.4\times10^{-2}/M_{\odot}$;
    \item fraction of energy released used to heat the gas, $\epsilon_{\rm SN}=0.05$;
    \item kinetic energy/SN, $E_{\rm SN}\simeq10^{51}$ erg;
    \item binding energy of the halo per unit stellar mass, $$E_{\rm bind}\simeq3.2\times10^{14}\left(\frac{M_{\rm vir}}{10^{12}M_{\odot}}\right)^{\frac{2}{3}}\left(\frac{1+z}{4}\right)\hspace{3ex}\hbox{cm}^2\,\hbox{s}^{-2}.$$ 
\end{enumerate}

\noindent The IR luminosity related to the dust-enveloped SF is given by
\begin{equation}
    L_{\star,IR}(t)=k_{\star,IR}\times10^{43}\left(\frac{\dot{M}_{\star}}{M_{\odot}\hbox{yr}^{-1}}\right)\ \hbox{erg}\,\hbox{s}^{-1},
\label{eqnIRlum}
\end{equation}
where $k_{\star,IR}$ is SED dependent.\\

\noindent The rate of cold gas inflow into the reservoir around the SMBH is 
\begin{equation}
    \dot{M}_{\rm inflow}\simeq\frac{L_{\star}}{c^2}(1-e^{-\tau_{RD}})\simeq\alpha_{RD}\times10^{-3}\dot{M}_{\star}(1-e^{-\tau_{RD}}),
\end{equation}
with 
\begin{equation}
    \tau_{RD}=\tau^0_{RD}\left(\frac{Z_{\rm cold}(t)}{Z_{\odot}}\right)\left(\frac{M_{\rm cold}}{10^{12}M_{\odot}}\right)\left(\frac{M_{\rm vir}}{10^{13}M_{\odot}}\right)^{-\frac{2}{3}}.
\end{equation}

\noindent The cold gas metallicity evolves according to the equation
\begin{equation}
\begin{split}
    Z_{\rm cold}(t)=Z^0_{\rm inf}+\frac{s}{s\gamma-1}\varepsilon_Z(t)-\frac{st/t_{\rm cond}}{e^{(s\gamma-1)t/t_{\rm cond}}-1}\\
    \left\{\varepsilon_Z(t)+B_Z\sum_{i=2}^{\infty}\frac{1}{i.i!}\left[(s\gamma-1)\frac{\min(t,t_Z)}{t_{\rm cond}}\right]^{i-1}\right\},
\end{split}
\end{equation}

\noindent where $\gamma=1-R-\beta_{\rm SN}$.

\vspace{2ex}

\noindent The mass fraction of metals formed in stars is given by
\begin{equation}
\begin{split}
   \varepsilon_Z(t)&= A_z+B_z\left[\frac{\min(t,t_{\rm saturated})}{t_z}\right]\\&=0.03+0.02\left[\frac{\min(t,t_{\rm saturated})}{t_z}\right],
   \end{split}
\end{equation}

\noindent where $t_z=20$ Myr, $t_{\rm saturated}=40$ Myr (for a Chabrier IMF) and $Z_{\odot}=0.02$.\\

\noindent The above equation accounts for the fact that after star formation begins, contributions to the metal yield take place
\begin{enumerate}
    \item initially from massive stars with masses $\geq 20\, M_{\odot}$ and lifetime $\leq 20$ Myrs;
    \item afterwards from moderate/intermediate stars having masses $\sim9-20\, M_{\odot}$ and lifetime $\sim20-40$ Myrs;
    \item finally, shifting to low mass stars $(M\leq9\,M_{\odot})$ with lifetime of approximately $\geq 40$ Myrs.
\end{enumerate}

\noindent The black hole accretion rate is given by 
\begin{equation}
    \dot{M}_{\rm BH}=\min(\dot{M}^{\rm visc}_{\rm BH},\lambda_{\rm Edd}\dot{M}_{\rm Edd}).
\end{equation}

\noindent The accretion rate into the reservoir around the SMBH  of gas which has dissipated its angular momentum is
\begin{equation}
    \dot{M}^{\rm visc}_{\rm BH}=\frac{M_{\rm res}}{\tau_{\rm visc}}=\kappa_{\rm acc}5\times10^3\left(\frac{V_{\rm vir}}{500\hspace{1ex}\mbox{kms}^{-1}}\right)^3\left(\frac{M_{\rm res}}{M_{\bullet}}\right)^{\frac{3}{2}}
     \left(1+\frac{M_{\bullet}}{M_{\rm res}}\right)^{\frac{1}{2}},
\end{equation}

\noindent where
\begin{enumerate}
    \item $\kappa_{\rm acc}=10^{-2}$;
    \item $V_{\rm vir}={GM_{\rm vir}^{2/3}}{\left[(4/3)\pi\triangle_{\rm vir}(z)\bar{\rho}_m(z)\right]^{-1/3}}$;
    \item $\dot{M}_{\rm Edd}={M_{\bullet}}/({\epsilon t_{\rm Edd}})$; for a mass-to-light conversion efficiency $\epsilon=0.1$ the Salpeter time is $\epsilon t_{\rm Edd}=4.5\times10^7$ yr;
    \item $\lambda_{\rm Edd}\simeq0.1(z-1.5)^2+1.0$ for $z\gtrsim1.5$ ($\lambda_{\rm max,Edd}=4$).
\end{enumerate}

\noindent The BH grows according to the equation
\begin{equation}
    \dot{M_{\bullet}}(t)=(1-\epsilon)\dot{M}_{\rm BH}.
\end{equation}

\noindent The AGN luminosity is given by
\begin{equation}
 \label{eqnagn}   L_{\bullet}=\epsilon\dot{M}_{\rm BH}c^2=5.67\times10^{45}\left(\frac{\epsilon}{0.1}\right)\left(\frac{\dot{M}_{\rm BH}}{M_{\odot}\hbox{yr}^{-1}}\right)\ \hbox{erg}\,\hbox{s}^{-1}.
\end{equation}

\noindent The feedback from the AGN is given by
\begin{equation}
    \dot{M}_{\rm inf,cold}^{\rm QSO}=\dot{M}_{\rm wind}\frac{M_{\rm inf,cold}}{M_{\rm inf}+M_{\rm cold}}=\frac{L_{\rm ISM}^{\rm QSO}}{E_{\rm bind}}\frac{M_{\rm inf,cold}}{M_{\rm inf}+M_{\rm cold}},
\end{equation}
where
\begin{equation}
    L_{\rm ISM}^{\rm QSO}=2\times10^{44}\epsilon_{\rm QSO}\left(\frac{\dot{M}_{\rm BH}}{M_{\odot}\hbox{yr}^{-1}}\right)^{3/2}\ \hbox{erg}\,\hbox{s}^{-1}.
\end{equation}

\subsection{Halo Mass Function and Halo Formation Rate Function}

In cosmology, the halo mass function specifies the mass distribution of dark matter halos. Technically speaking, it gives the number density of halos per unit interval of mass. We adopt the analytical expression given by \citet{sheth_large-scale_1999}, 
\begin{equation}
    N_{\rm ST}(M_{\rm vir},z)dM_{\rm vir}=\frac{\bar{\rho}_{m,0}}{M_{\rm vir}^2}f_{\rm ST}(\nu)\frac{d\ln\nu}{d\ln M_{\rm vir}}dM_{\rm vir},
\label{eqnHMF}
\end{equation}
where $\bar{\rho}_{m,0}=\Omega_{m,0}\rho_{c,0}$ is the present day mean matter density of the Universe, $\nu\equiv\left({\delta_{c}(z)}/{\sigma(M_{\rm vir})}\right)^2$ with $\delta_c(z)=\delta_0(z)({D(0)}/{D(z)})$ where 

\begin{equation}
    \delta_0(z)\approx1.6865[1+0.0123\log\Omega_m(z)] 
\end{equation}
and
\begin{equation}
    D(z)=\frac{5\Omega_m(z)/[2(1+z)]}{\frac{1}{70}+\frac{209}{140}\Omega_m(z)-\frac{1}{140}\Omega_m^2(z)+\Omega_m^{4/7}(z)}
\end{equation}
from \cite{carroll_cosmological_1992}.

\noindent The mass variance is given by \citep{bardeen_statistics_1986, hu_anisotropies_1995}
\begin{equation}
\begin{split}
    \sigma(M_{\rm vir})=\frac{0.8}{0.84}[14.110393-1.1605397x-0.0022104939x^2\\
    +0.0013317476x^3-2.1049631\times10^{-6}x^4],
    \end{split}
\end{equation}
%
where $x=\log(M_{\rm vir}/M_{\odot})$ and $10^6<{M_{\rm vir}}/{M_{\odot}}<10^{16}$.
Furthermore,
\begin{equation}
    f_{\rm ST}(\nu)=A[1+(a\nu)^{-p}]\left(\frac{a\nu}{2}\right)^{1/2}\frac{e^{-a\nu/2}}{\pi^{1/2}},
\end{equation}
where $A=0.322$, $p=0.3$ and $a=0.707$.
%
The halo formation rate (HFR) is approximated by the positive cosmic time derivative of the halo mass function, $N_{\rm ST}$, valid for $z\gtrsim1.5$). It is given by
\begin{equation}
\begin{split}
\frac{dN_{\rm ST}}{dt} & =  N_{\rm ST}\frac{d \ln f_{\rm ST}(\nu)}{dt}\\
& \approx N_{\rm ST}\left[\frac{a\nu}{2}+\frac{p}{1+(a\nu)^p}\right]\frac{d \ln\nu}{dz}\left|\frac{dz}{dt}\right|,
\end{split} 
\label{eqnHFR}
\end{equation}
where ${dz}/{dt}=-H_0(1+z)E(z)$ with $E(z)=\sqrt{\Omega_{\Lambda,0}+\Omega_{m,0}(1+z)^3}$.


\bsp	
\label{lastpage}
\end{document}